\documentclass[12pt]{article}
\usepackage{amssymb,amsfonts,times}
\usepackage[mathscr]{eucal}
\DeclareFontFamily{U}{rsf}{}
\DeclareFontShape{U}{rsf}{m}{n}{
  <5> <6> rsfs5 <7> <8> <9> rsfs7 <10-> rsfs10}{}
\DeclareMathAlphabet\Scr{U}{rsf}{m}{n}
\makeatletter

\@addtoreset{equation}{section}
\renewcommand\tableofcontents{
 \begin{center}\bf\Large\contentsname\end{center}
  \@starttoc{toc}
}
\makeatother

\newcommand{\nn}{\nonumber}
\newcommand{\ts}{\textstyle}
\newcommand{\ds}{\displaystyle}
\newcommand{\tfrac}[2]{{\textstyle\frac{#1}{#2}}}
\newcommand{\tsum}{{\textstyle\sum}}

\newcommand{\ket}[1]{\vert  #1\rangle}
\newcommand{\bra}[1]{\langle #1 \vert}
\newcommand{\vev}[1]{\langle{#1}\rangle}
\newcommand{\kket}[1]{\,\vert  #1\rangle\!\rangle}
\newcommand{\bbra}[1]{\langle\!\langle #1 \vert\,}

\newcommand{\wtilde}[1]{\widetilde{#1}}

\newcommand{\bi}{{\bar\imath}}
\newcommand{\bj}{{\bar\jmath}}
\newcommand{\ti}{{\tilde\imath}}
\newcommand{\tj}{{\tilde\jmath}}

\newcommand{\SR}{\mathscr{R}}

\newcommand{\SF}{\mathscr{F}}
\newcommand{\SM}{\mathscr{M}}

\newcommand{\PP}{{\mathbb P}}
\newcommand{\ZZ}{{\mathbb Z}}
\newcommand{\CC}{{\mathbb C}}
\newcommand{\RR}{{\mathbb R}}

\newcommand{\ssub}[1]{_{_{{#1}}}}
\newcommand{\SEC}{{\rm Y}}
\newcommand{\SB}{\mathscr{B}}
\newcommand{\SC}{\mathscr{C}}
\newcommand{\ST}{\mathscr{T}}
\newcommand{\nc}[1]{{ \left[#1\right]}}
\newcommand{\lc}[1]{{ |\!|#1|\!|}}
\newcommand{\NSNS}{{\rm NSNS}}
\newcommand{\RARA}{{\rm RR}}
\newcommand{\NS}{{\rm NS}}
\newcommand{\RA}{{\rm R}}
\newcommand{\npm}{{\hskip9pt}}
\newcommand{\gammaA}{\gamma_{_{(A)}}}
\newcommand{\gammaB}{\breve\gamma}
\newtheorem{PIB}{Condition for Parity Invariant Branes (PIB)}

\newcommand{\SBPLM}[3]
{\SB_{{}^{{\bf L}\;=({#2})}_{{\bf M}=({#3})}}^{{}_{#1}}}
\newcommand{\charge}[1]{{[#1]}}
\begin{document}

\pagestyle{empty}
\baselineskip 5mm
\hfill
\hbox to 3cm
{\parbox[t]{5cm}{hep-th/0612109\\ SPhT-T06/170} \hss}\\

\baselineskip0.8cm\vskip2cm
\vskip10mm
\begin{center}

 {\Large\bf Permutation Orientifolds of Gepner Models}

\end{center}
\vskip10mm
\baselineskip0.6cm
\begin{center}

    Kazuo Hosomichi
 \\ \vskip2mm
{\it Service de Physique Th\'eorique, CEA Saclay\\
     F-91191 Gif sur Yvette, France} \vskip3mm

\end{center}
\vskip8mm\baselineskip=3.5ex
\begin{center}{\bf Abstract}\end{center}\par\smallskip

In tensor products of a left-right symmetric CFT, one can define
permutation orientifolds by combining orientation reversal
with involutive permutation symmetries.
We construct the corresponding crosscap states in general
rational CFTs and their orbifolds, and study in detail those in
products of affine $U(1)_2$ models or $N=2$ minimal models.
The results are used to construct permutation orientifolds
of Gepner models.
We list the permutation orientifolds in a few simple
Gepner models, and study some of their physical properties ---
supersymmetry, tension and RR charges.
We also study the action of corresponding parity
on D-branes, and determine the gauge group on a stack of
parity-invariant D-branes.
Tadpole cancellation condition and some of its solutions are
also presented.

\vspace*{\fill}
\noindent December~~2006
\newpage
\setcounter{page}{0}
\tableofcontents\newpage
\pagestyle{plain}
\pagenumbering{arabic}

\section{Introduction}\label{sec:Intro}

In the construction of type II string vacua with ${\cal N}=1$
supersymmetry in four dimensions, orientifolds play
an important role along with branes and fluxes.
While we wish to obtain a global picture for the whole variety of such
vacua, it would be desirable to understand better each of the
ingredients at different vacua.
At one regime of vacua where the compactification manifold has
large volume, the supergravity and Dirac Born-Infeld theory will
give a reliable geometric description of the system.
On the other hand, in a different regime where the size of the
compactification manifold is very small, there are vacua admitting
an exactly solvable worldsheet description.
The worldsheet theory describing such vacua was found by Gepner \cite{Gepner}
and involves an orbifold of products of $N=2$ minimal models, which
are very well-understood rational CFTs.

D-branes and orientifolds in Gepner models were studied in
many papers.
A class of D-branes were first constructed in \cite{Recknagel-S}
using Cardy's boundary states\cite{Cardy} in $N=2$ minimal models.
Since then, different aspects of them were studied including how they
continue in moduli space to the large volume \cite{Brunner-DLR}.
Similar analyses for orientifolds were first made in
\cite{Angelantonj-BPSS, Blumenhagen-W} and then in
\cite{Brunner-H2}--\cite{Aldazabal-AJ} using the standard crosscap states
in $N=2$ minimal models, and provided us with a large number
of tadpole-free backgrounds where the particle spectra are explicitly
computable \cite{Dijkstra-HS}.

The D-branes and orientifolds studied in those works are
mostly made from products of boundary or crosscap states in minimal models.
On the other hand, in Gepner models containing products of minimal
models of the same level, there are also D-branes and orientifolds
corresponding to boundary conditions on fields twisted by permutation
symmetries.
Permutation branes in general CFTs were first constructed by Recknagel
\cite{Recknagel} by generalizing Cardy's standard prescription
\cite{Cardy} (see also \cite{Gaberdiel-S}).
Some generalizations of it have been discussed
in \cite{Sarkissian-Z, Fredenhagen-Q, Fredenhagen-G}.
There have also been many work on permutation branes in Gepner models
\cite{Ashok-DD}--\cite{Brunner-GK},
some of which employ the description in terms of
{\it matrix factorization} of Landau-Ginzburg superpotential
\cite{Kapustin-L, Brunner-HLS}.
A natural extension of these developments will be to construct
permutation orientifolds in a similar manner.

One of our goal in this paper is to give a general prescription to
construct permutation orientifolds in tensor product CFTs as well as
their orbifolds, generalizing the standard construction of
crosscap states in RCFT given by \cite{Pradisi-SS} and
developed further by \cite{Bantay}--\cite{Brunner-H1}.
The other goal is to apply it to Gepner models and study type II
string vacua made of permutation orientifolds.
Accordingly, the paper is organized into two parts.

In Section \ref{sec:PBCRCFT} we present our general construction
of permutation orientifolds in RCFTs and orbifolds thereof.
In Section \ref{sec:U1_2} we apply our prescription to the theory
of $n$ Dirac fermions, using the fact that the theory is
related to the affine $U(1)_2^{\otimes n}$ model by orbifolding.
We pay particular attention to assigning Grassmann parity to
states and operators so that the acnticommutativity of fermions
is correctly reproduced.
In a similar manner, we construct in Section \ref{sec:MM} the boundary and
crosscap states in $N=2$ minimal models preserving an $N=2$
superconformal symmetry.

In Section \ref{sec:GM} we classify permutation orientifolds in Gepner
models and write down their explicit form.
The construction of permutation D-branes will also be
given here although there have been a lot of works on it; in
particular we discuss in full detail the properties of short
orbit branes, i.e. branes in orbifolds which are not simply
the sum over orbifold images.
In Section \ref{sec:SSTP} we analyze further some physical properties of
permutation orientifolds in Gepner models.
We will find out how various orientifolds act on D-branes, and
determine the gauge group on a stack of parity-invariant
D-branes.
We also analyze the condition of tadpole cancellation and some of its
solutions.
We conclude in Section \ref{sec:CONC} with some brief remarks.

\paragraph{Note added.} A part of the results presented in this paper
was obtained independently by Brunner and Mitev\cite{Brunner-M}.
We were informed of their work in progress at an early stage of our work.

\subsection*{Rudiments of one-loop amplitudes}

Here we collect our convention for various one-loop
amplitudes in string theory.

\paragraph{Cylinder.}

The one-loop of open string stretching between two D-branes is a cylinder.
We parametrize the worldsheet by $(\sigma,t)$ with
$0\le\sigma\le\pi,~t\sim t+2\pi l$ or a complex coordinate
$z=\sigma+it$.
The endpoints $\sigma=0$ and $\pi$ are on D-branes $\bra{\SB_0}$
and $\ket{\SB_\pi}$ respectively.
The D-branes are characterized by different boundary conditions
on fields.
We assume the worldsheet conformal field theory to have a symmetry
generated by holomorphic currents $W(z),\wtilde W(\bar z)$ with spin
$S_W\in\frac12\ZZ$, and assume that the currents with integer
(half-odd-integer) spins are bosonic (resp. fermionic).
We restrict our interest to the boundary states
satisfying
\begin{equation}
  \bra{\SB_0}\left(\wtilde W(\bar z)- e^{-i\pi S_W}W(z)\right)_{\sigma=0}
  ~=~ 0 ~=~
  \left(\wtilde W(\bar z)- e^{ i\pi S_W}W(z)\right)_{\sigma=\pi}\ket{\SB_\pi}.
\end{equation}
Let $X$ be a symmetry of the theory.
The open closed duality relates the overlap of boundary states
in $X$-twisted sector and the trace over open string Hilbert space
with weight $X$,
\begin{equation}
   ^X\!\bra{\SB_0}e^{-\pi H_{\rm c}/l}\ket{\SB_\pi}^X
   ~=~
  {\rm Tr}_{\SB_0,\SB_\pi}\left[(-)^Fe^{-2\pi H_{\rm o}l}X\right].
\label{cyl}
\end{equation}
The right hand side is formally calculated as the path integral
on the cylinder with the fields $\phi(\sigma,t)$ obeying boundary
conditions specified by D-branes and the periodicity along time,
\begin{equation}
 \phi(\sigma,t) ~=~ X^{-1}\phi(\sigma,t+2\pi l)X.
\end{equation}
If one is interested in summing over spin structures, it is convenient
to introduce the indices $\NSNS\pm,\RARA\pm$ to label different
boundary conditions for fermionic currents $W$ and $\wtilde W$,
\begin{equation}
  \ssub{\SEC\!\pm}\bra{\SB_0}
  \left(\wtilde W(\bar z)\mp e^{-i\pi S_W}W(z)\right)
  ~=~ 0 ~=~
  \left(\wtilde W(\bar z)\mp e^{ i\pi S_W}W(z)\right)
  \ket{\SB_\pi}\ssub{\SEC\!\pm}.
\end{equation}
$\SEC=\NSNS~(\RARA)$ indicates that the fermionic fields are
anti-periodic (periodic) along time $t$.

\paragraph{M\"obius strip.}

If the theory on a strip has a parity symmetry
exchanging fields at $\sigma$ and $\pi-\sigma$, the one-loop of
open string of width $\pi$ and the periodicity along time
($t\sim t+2\pi l$) twisted by the parity is a M\"obius strip.
The boundary states $\bra{\SB_0}$ and $\ket{\SB_\pi}$ then
have to be parity images of each other.
We assume there is a ``basic'' involutive parity $P$
acting on the currents as
\begin{equation}
 PW(\sigma,t)P ~=~ e^{-i\pi S_W}\wtilde W(\pi-\sigma,t),~~~~~
 P\wtilde W(\sigma,t)P ~=~ e^{ i\pi S_W}W(\pi-\sigma,t),
\label{pbas}
\end{equation}
and consider parities of the form $gP$, defined by combining
$P$ with various symmetries $g$ acting locally on fields
and symmetry currents.
The M\"obius strip amplitude associated to the parity $gP$ is a
trace over open string Hilbert space (\ref{cyl}) with $X=gP$.
Alternatively, it is given by a path integral on a strip of width $\pi/2$
and period $4\pi l$ bounded by a boundary and a crosscap states.
The fields satisfy twisted periodicity along time,
\[
 \phi(\sigma,t)~=~ X^{-1}\phi(\sigma,t+4\pi l)X,~~~~X\equiv(gP)^2.
\]
The fields obey the boundary condition specified by $\bra{\SB_0}$
at $\sigma=0$, and the crosscap condition at $\sigma=\pi/2$,
\begin{equation}
  \phi(\tfrac\pi2,t) ~=~  gP\phi(\tfrac\pi2,t-2\pi l)Pg^{-1},
\end{equation}
The corresponding crosscap state is denoted by $\ket{gP}$.
The open-closed duality then tells
\begin{equation}
 {\rm Tr}_{\SB_0,\SB_\pi}[(-)^Fe^{-2\pi H_{\rm o}l}gP]
 ~=~ ^X\!\bra{\SB_0} e^{-\pi H_{\rm c}/4l}\ket{gP}^X
 ~=~ ^X\!\bra{(-)^FgP} e^{-\pi H_{\rm c}/4l}\ket{\SB_\pi}^X.
\label{MS}
\end{equation}
The second equality tells how the boundary states are transformed
under the parity.
The additional factor $(-)^F$ in the definition of crosscap bra-state
is because we define the bra and ket states to satisfy the crosscap conditions
\begin{eqnarray}
0 &=&
  \bra{gP}\left(
  \wtilde W(t)- e^{-i\pi S_W}gW(t-2\pi l)g^{-1}\right)_{\sigma=\frac\pi2}
\nn\\ &=&
  \left(\wtilde W(t)- e^{ i\pi S_W}gW(t-2\pi l)g^{-1}
  \right)_{\sigma=\frac\pi2}\ket{gP},
\end{eqnarray}
so that (i) the conditions on bra and ket states are related by rotation
by 180 degrees, and (ii) the bra and ket states are related by
the dagger operation.

Different spin structures give a pair of NSNS crosscaps
$\ket{(-)^{F_L}P},\ket{(-)^{F_R}P}$ and a pair
of RR crosscaps $\ket{(\pm)^FP}$ for each involutive parity
symmetry $P$.
In general the NSNS parity maps a boundary state
$\ssub{\NSNS\pm}\bra\SB$ to $\ket{\SB'}\ssub{\NSNS\pm}$ by (\ref{MS}),
while the RR parity maps $\ssub{\RARA\pm}\bra\SB$ to
$\ket{\SB'}\ssub{\RARA\mp}$.

If $\ket P$ is the crosscap state corresponding to the parity $P$
of (\ref{pbas}), then $g\ket P$ satisfies
\[
  \left(g\wtilde W(t)g^{-1}- e^{ i\pi S_W}gW(t-2\pi l)g^{-1}
  \right)_{\sigma=\frac\pi2}g\ket{P} ~=~ 0.
\]
We can therefore put
\begin{equation}
  g\ket P ~=~ \ket{gPg^{-1}}.
\label{gketP}
\end{equation}

\paragraph{Klein bottle.} Let us next consider a closed string
with spatial coordinate $\sigma\sim\sigma+2\pi$.
The one-loop of closed string with the periodicity
along time ($t\sim t+2\pi l$) twisted by parity is a Klein bottle.
If the parity maps $\sigma$ to $-\sigma$ modulo $2\pi$,
then the Klein bottle is equivalent to a periodic strip of width $\pi$,
period $t\sim t+4\pi l$ bounded by two crosscap states at
$\sigma=0$ and $\pi$.
If the two crosscaps correspond to different parities $g_0P$ and
$g_\pi P$, then the fields obey
\begin{equation}
 \phi(\sigma,t)
 ~=~ g_0   P_{(0  )}\phi(\sigma,t-2\pi l)P_{(0  )}g_0  ^{-1}
 ~=~ g_\pi P_{(\pi)}\phi(\sigma,t-2\pi l)P_{(\pi)}g_\pi^{-1},
\end{equation}
where the suffix $(0)$ or $(\pi)$ indicates the fixed point of
the parity.
Therefore the closed string is in the sector twisted by
$g\equiv(g_0 g_\pi^{-1})$.
The open-closed duality then tells that
\begin{equation}
  {\rm Tr}_g[(-)^Fe^{-2\pi H_{\rm c}l}g_0P_{(0)}]
  ~=~ {\rm Tr}_g[(-)^Fe^{-2\pi H_{\rm c}l}g_\pi P_{(\pi)}]
  ~=~ \bra{(-)^Fg_0P}e^{-\frac{\pi H_c}{2l}}\ket{g_\pi P}.
\end{equation}

The closed string states form a representation of the symmetry
algebra of the currents $W(z)$ and $\wtilde W(\bar z)$.
The action of parity $P_{(0)},P_{(\pi)}$ on the currents
are given by (\ref{pbas}) with modified fixed points.
Introducing the coordinate $\zeta\equiv \pm e^{-iz}$ and expanding
the currents in standard power series, one finds these parities
act on the modes as $W_n \leftrightarrow \wtilde W_n$, as
expected.

\section{Permutation Branes and Crosscaps in RCFT}\label{sec:PBCRCFT}%

In this section we present the construction of permutation branes and
orientifolds in tensor products of general rational CFTs,
and then extend it to their simple current orbifolds.
The argument follows that of \cite{Recknagel}.

Let ${\cal X}$ be a general left-right symmetric RCFT
with chiral symmetry algebra ${\cal A}\otimes{\cal A}$,
and denote the tensor product of $N$ copies of it by ${\cal X}^N$.
The D-branes or orientifolds in $\cal X$ are described by the
states $\ket\SB,~\ket\SC$ satisfying the boundary or crosscap
conditions on currents generating two copies of ${\cal A}$:
\begin{equation}
\begin{array}{rcl}
  (\wtilde W_{n}-e^{-i\pi S_W}    W_{-n})\ket{\SB}^{\!\cal X} &=& 0,\\
  (\wtilde W_{n}-e^{-i\pi (S_W-n)}W_{-n})\ket{\SC}^{\!\cal X} &=& 0.
\end{array}
\label{bcWX}
\end{equation}
Here $S_W$ is the spin of the current $W$.
Any product of states $\ket{\SB}$ or $\ket{\SC}$ of ${\cal X}$
gives a state of ${\cal X}^N$ satisfying the boundary or crosscap conditions
\begin{equation}
\begin{array}{rcl}
  (\wtilde W_{n}^a-e^{-i\pi S_W}    W_{-n}^a)\ket{\SB}^{\!{\cal X}^N} &=& 0,\\
  (\wtilde W_{n}^a-e^{-i\pi (S_W-n)}W_{-n}^a)\ket{\SC}^{\!{\cal X}^N} &=& 0.
\end{array}
\end{equation}
Here the suffix $a$ is for operators in the $a$-th copy of ${\cal X}$.
Permutation branes and permutation orientifolds in ${\cal X}^N$
are characterized by the conditions on currents twisted by
permutations $\pi\in S_N$:
\begin{equation}
\begin{array}{rcl}
  (\wtilde W_{n}^{\pi(a)}-e^{-i\pi S_W}   W_{-n}^a)
  \ket{\SB^\pi}^{\!{\cal X}^N} &=& 0,\\
  (\wtilde W_{n}^{\pi(a)}-e^{-i\pi(S_W-n)}W_{-n}^a)
  \ket{\SC^\pi}^{\!{\cal X}^N} &=& 0.
\end{array}
\label{bcWpi}
\end{equation}
We call these conditions as ``$\pi$-permuted''.

\subsection{Cardy and Pradisi-Sagnotti-Stanev's constructions}\label{sec:CPSSC}

In the standard Cardy and Pradisi-Sagnotti-Stanev(PSS) constructions,
D-branes and orientifolds in general RCFT ${\cal X}$ are expressed
as suitable linear sums of Ishibashi states which form the basis
of solutions to (\ref{bcWX}).
Here we extend this prescription to construct permutation branes
and orientifolds in ${\cal X}^N$, following the argument of \cite{Recknagel}.
Our construction of permutation orientifolds agrees with
that of \cite{Brunner-M}.

General Ishibashi states $\kket{\SB;i}$ and $\kket{\SC;i}$
in $\cal X$ are constructed as
\begin{equation}
 \kket{\SB;i}~:=~\sum_{M\in {\cal V}_i}\ket{M}\otimes \Phi\ket{M},~~~~~
 \kket{\SC;i}~:=~ e^{\pi i(L_0-h_i)}\kket{\SB;i}.
\end{equation}
Here ${\cal V}_i$ is the $i$-th highest weight representation of ${\cal A}$
spanned by an orthonormal basis $\{\ket M\}$, and $h_i$ is its
conformal weight.
$\Phi$ is the anti-unitary operator satisfying
$W_n\Phi = e^{-i\pi S_W}\Phi W^\dagger_{-n}$.
The simple products of Ishibashi states
$\kket{\SB;i_1\cdots i_N},~\kket{\SC;i_1\cdots i_N}$
satisfy the boundary or crosscap conditions (\ref{bcWpi})
in ${\cal X}^N$ with $\pi={\rm id}$.
Define an operator $R^\pi$ acting only on the left-moving ($=$ antiholomorphic)
operators and primary states as permutations
\begin{equation}
\begin{array}{rcl}
 R^\pi\widetilde{W}_n^a R^{\pi^{-1}}
 &=&   \widetilde{W}_n^{\pi(a)},\\
 R^\pi \,\cdot\,
 \ket{i_1\otimes \ti_1}_1 \cdots
 \ket{i_N\otimes \ti_N}_N
 &=& (\pm)
 \ket{i_1\otimes \ti_{\pi^{-1}(1)}}_1 \cdots
 \ket{i_N\otimes \ti_{\pi^{-1}(N)}}_N.
\end{array}
\label{Rpi}
\end{equation}
Note that, in the second equation, $R^\pi$ should be understood
to annihilate the state unless the state $\ket{i_a\otimes\ti_{\pi^{-1}(a)}}$
is contained in the Hilbert space of $\cal X$ for all $a$.
The $\pm$ sign in the right hand side of the second equation
arises if the theory ${\cal X}$ contains fermionic states and
currents.
The $\pi$-permuted Ishibashi states are then simply given by
\begin{eqnarray}
\kket{\SB^\pi;i_1\cdots i_N} &=& R^\pi\kket{\SB;i_1\cdots i_N},\nn\\
\kket{\SC^\pi;i_1\cdots i_N} &=& R^\pi\kket{\SC;i_1\cdots i_N}.
\end{eqnarray}

In the rest of this subsection we assume $\cal X$ to be an
RCFT with charge conjugation modular invariant,
so that $R^\pi$ annihilates the primary state
$\ket{i_1\otimes\bi_1}_1\cdots\ket{i_N\otimes\bi_N}_N$
unless $i_{\pi^{-1}(a)}=i_a$ for all $a$.
We also assume, for simplicity, that all the states and currents
in $\cal X$ are bosonic.
We denote the number of cycles in a given
permutation $\pi$ by $\nc{\pi}$, the $c$-th cycle of
$\pi$ as $\pi_c$ and its length by $\lc{\pi_c}$.
The $\pi$-permuted Ishibashi states can then be labelled by
$j_c~(c=1,\cdots,\nc{\pi})$ such that
\begin{equation}
  i_a=j_c~~{\rm if}~a\in \pi_c.
\label{IvsJ}
\end{equation}
So we introduce another expression for Ishibashi states:
\begin{equation}
\begin{array}{rclcl}
 \kket{\SB^\pi;j_1\cdots j_{\nc{\pi}}}
 &=& \otimes_{c=1}^{\nc\pi}\kket{\SB^{\pi_c};j_c}
 &=& \ds\sum_{i_a}\delta^{(\pi)}_{i,j}\,R^\pi\kket{\SB;i_1\cdots i_N},\\
 \kket{\SC^\pi;j_1\cdots j_{\nc{\pi}}}
 &=& \otimes_{c=1}^{\nc\pi}\kket{\SC^{\pi_c};j_c}
 &=& \ds\sum_{i_a}\delta^{(\pi)}_{i,j}\,R^\pi\kket{\SC;i_1\cdots i_N}.
\end{array}
\end{equation}
The delta symbol $\delta^{(\pi)}_{i,j}$ enforces the condition (\ref{IvsJ}).
The inner products of these Ishibashi states read
\begin{eqnarray}
 \bbra{\SB^{\tilde\pi},\tj_1\cdots\tj_{\nc{\pi}}}
 e^{\pi i\tau H}\kket{\SB^\pi;j_1\cdots j_{\nc{\pi}}}
 &=& \sum_{i,j'}\delta^{(\tilde\pi)}_{i,\tj}
     \delta^{(\pi)}_{i,j}\delta^{(\sigma)}_{i,j'}
 \prod_{c=1}^{\nc\sigma}\chi_{j'_c}(\lc{\sigma_c}\tau),
\nn\\
 \bbra{\SC^{\tilde\pi},\tj_1\cdots\tj_{\nc{\pi}}}
 e^{\pi i\tau H}\kket{\SC^\pi;j_1\cdots j_{\nc{\pi}}}
 &=& \sum_{i,j'}\delta^{(\tilde\pi)}_{i,\tj}
     \delta^{(\pi)}_{i,j}\delta^{(\sigma)}_{i,j'}
 \prod_{c=1}^{\nc\sigma}\chi_{j'_c}(\lc{\sigma_c}\tau),
\nn\\
 \bbra{\SB^{\tilde\pi},\tj_1\cdots\tj_{\nc{\pi}}}
 e^{\pi i\tau H}\kket{\SC^\pi;j_1\cdots j_{\nc{\pi}}}
 &=& \sum_{i,j'}\delta^{(\tilde\pi)}_{i,\tj}
     \delta^{(\pi)}_{i,j}\delta^{(\sigma)}_{i,j'}
 \prod_{c=1}^{\nc\sigma}
 {\cal I}^{\lc{\sigma_c}}\chi_{j'_A}(\lc{\sigma_c}\tau),
\nn\\
\lefteqn{
(\sigma\equiv \pi^{-1}\circ\tilde\pi)
}
\label{ProdI}
\end{eqnarray}
where ${\cal I}$ is an involutive operator defined to act on characters as
\begin{equation}
  {\cal I}\chi_i(\tau) = \widehat\chi_i(\tau)\equiv
  e^{-\pi i(h_i-\frac{c}{24})}\chi_i(\tau+1/2).
\label{hatch}
\end{equation}

D-branes and orientifolds are linear combinations of
Ishibashi states satisfying certain consistency conditions.
Recknagel \cite{Recknagel} constructed the permutation branes
as follows:
\begin{equation}
  \ket{\SB^\pi_{\bf J}}
 ~=~ \ket{\SB^\pi_{J_1\cdots J_{\nc\pi}}}
 ~=~ \bigotimes_{c=1}^{\nc\pi}\ket{\SB^{\pi_c}_{J_c}}
 ~=~ \bigotimes_{c=1}^{\nc\pi}\sum_{j_c}
     \frac{S_{J_cj_c}}{(S_{0j_c})^{\lc{\pi_c}/2}}
     \kket{\SB^{\pi_c};j_c}.
\label{BpiJ}
\end{equation}
In \cite{Recknagel} it was also shown that the open string spectrum
between any two such D-branes satisfies integrality.
To see this, let us consider the finest possible
decomposition of the set of $N$ letters,
$\{1,\cdots,N\}=\bigcup_b{\cal S}_b$ such that any cycle
of $\pi,\tilde\pi$ or $\sigma=\pi^{-1}\circ\tilde\pi$ is
contained in one of ${\cal S}_b$.
The annulus amplitude then becomes
\begin{eqnarray}
 \bra{\SB^{\tilde\pi}_{\bf \tilde J}}e^{-\pi H/l}\ket{\SB^\pi_{\bf J}}
 &=&
 \sum_{J'_1,\cdots,J'_{\nc\sigma}}\prod_b
 {\cal N}_b({\bf \tilde J,J,J'})
 \prod_{c=1}^{\nc\sigma}\chi_{J'_c}(il/\lc{\sigma_c}),
 \nn\\
 {\cal N}_b({\bf \tilde J,J,J'}) &=& \sum_j
 \prod_{\tilde\pi_c\in{\cal S}_b}
 \frac{S^\ast_{\tilde J_cj}}{(S_{0j})^{\lc{\tilde\pi_c}/2}}
 \prod_{\pi_c\in{\cal S}_b}\frac{S_{J_cj}}
  {(S_{0j})^{\lc{\pi_c}/2}}
 \prod_{\sigma_c\in{\cal S}_b}S_{J'_cj}.
\end{eqnarray}
The coefficient ${\cal N}_b$ always takes the form
\begin{eqnarray}
\lefteqn{
  \sum_j \frac{S_{J_1j}\cdots S_{J_{n+3}j}}{S_{0j}^{2g+n+1}}~~~
  (n\ge 0,~g\ge0)
} \nn\\ &=&
\left\{
\begin{array}{ll}
  [N_{J_2}N_{J_3}\cdots N_{J_{n+2}}]_{J_1}^{~\bar J_{n+3}}  & (g=0) \\
  \sum_{j_1,\cdots,j_g}
 {\rm Tr}[N_{J_1}N_{J_2}\cdots N_{J_{n+3}}\cdot N_{j_1}N_{\bj_1}\cdots
  N_{j_{g-1}}N_{\bj_{g-1}}]  & (g>0)
\end{array}
\right.
\label{Nb}
\end{eqnarray}
where $N_i$ is the fusion matrix whose elements are all non-negative
integers,
$$
 [N_j]_k^{~l}~=~
 N_{jk}^l    ~=~ \sum_i\frac{S_{ji}S_{ki}S^\ast_{li}}{S_{0i}}.
$$
Hence ${\cal N}_b$ is always a nonnegative integer.
The right hand side of (\ref{Nb}) has an interpretation as
the number of $(n+3)$-point conformal blocks on genus-$g$
Riemann surface.

The construction of \cite{Recknagel} can be extended to crosscap
states in a straightforward manner.
General permutation orientifold of ${\cal X}^N$ is labelled by
an involutive permutation $\pi$ and a parity
$P_I\equiv \otimes_{a=1}^N P_{I_a}$ satisfying $P_{I_{\pi(a)}}=P_{I_a}$.
Then we propose the following crosscap states,
\begin{eqnarray}
  \ket{\SC^\pi_{\bf I}}
 ~=~ \ket{\SC^\pi_{I_1\cdots I_{[\pi]}}}
 ~=~ \bigotimes_{c=1}^{\nc\pi}\ket{\SC^{\pi_c}_{I_c}}
 &=& \bigotimes_{c=1}^{\nc\pi}\sum_{j_c}
   \frac{X_{I_cj_c}}{(S_{0j_c})^{\lc{\pi_c}/2}}
   \kket{\SC^{\pi_c};j_c},
\label{CpiI} \\
    X_{I_cj_c} &=& \left\{\begin{array}{ll}
    P_{I_cj_c} & (\lc{\pi_c}=1) \\
    S_{I_cj_c} & (\lc{\pi_c}=2)
  \end{array}\right.
\nn
\end{eqnarray}
Note that the lengths of all the cycles of $\pi$ have to be
one or two for $\pi$ to be involutive.
The integrality of Klein bottle and M\"{o}bius strip amplitudes
can be checked by a direct computation.
One encounters factors of the form
\begin{equation}
  \sum_j \frac{S_{J_1j}\cdots S_{J_{m+1}j}P_{I_1j}\cdots P_{I_{2n}j}}
  {S_{0j}^{2l+m+2n-1}}~~~~~
  (m,n,l\ge0,~m+2n\ge 2),
\label{Yb}
\end{equation}
which can be rewritten in a similar way as (\ref{Nb}),
using the $N$- and $Y$-matrices
\begin{equation}
 [Y_j]_k^{~l}~=~
 Y_{jk}^{l}  ~=~ \sum_i\frac{S_{ji}P_{ki}P^\ast_{li}}{S_{0i}},
\end{equation}
whose elements are all known to be integers.
For this rewriting to be possible, the number of $P$-matrices in
(\ref{Yb}) has to be always even; this is actually the case because
we put $X_{I_cj_c}=P_{I_cj_c}$ or $S_{I_cj_c}$
depending on $\lc{\pi_c}=1$ or $2$.
To check this, let us consider the Klein bottle amplitudes
between $\pi$- and $\tilde\pi$-permuted crosscap states.
In order to write them down one needs the decomposition
$\{1,\cdots,N\}=\bigcup_b{\cal S}_b$
in the same way as for the annulus amplitudes.
The factors of the form (\ref{Yb}) are associated to each of ${\cal S}_b$.
One finds the number of $P$-matrices in (\ref{Yb}) is the sum
of the numbers of odd-length cycles of $\pi$ and those of $\tilde\pi$
contained in ${\cal S}_b$, which is always even.
The same argument applies to M\"{o}bius strip amplitudes.

In summary, for an RCFT $\cal X$ defined with charge conjugation
modular invariant, the formulae (\ref{BpiJ}) and (\ref{CpiI}) give
general $\pi$-permuted boundary and crosscap states in ${\cal X}^N$.

\subsection{Simple current orbifold}\label{sec:SCO}

Here we briefly review some basic properties of simple current
orbifolds ${\cal X}/G$ and the constructions of D-branes and
orientifolds in such theories.

Suppose a CFT $\cal X$ has a group $G$ of simple currents.
A simple current $g\in G$ is by definition a representation of
$\cal A$ which maps any representation into another unique
representation under fusion:
$$
  g\times i \to gi.
$$
It follows that $g$ induces (infinitely many) invertible maps between two
highest weight representations ${\cal V}_i, {\cal V}_{gi}$ of ${\cal A}$.
For an RCFT $\cal X$ defined with charge conjugation modular invariant,
its orbifold ${\cal X}/G$ is defined by the modular invariant
\begin{equation}
  Z^{{\cal X}/G} = \frac{1}{|G|}\sum_i\sum_{g_1,g_2\in G}
  e^{2\pi i(Q_{g_2}(i)-q(g_1,g_2))}\chi_i(\tau)\chi_{g_1\bi}(-\bar\tau).
\label{Zorb}
\end{equation}
Here $Q_g(i)$ is defined and characterized by
\begin{equation}
\begin{array}{cll}
(1) & Q_g(i)=h_i+h_g-h_{gi}&~({\rm mod}~\ZZ),\\
(2) & Q_{g}(i)+Q_{g'}(i)=Q_{gg'}(i)&~({\rm mod}~\ZZ),\\
(3) & Q_g(i)+Q_g(j)=Q_g(k) ~~{\rm if}~N_{ij}^k\ne 0&~({\rm mod}~\ZZ),
\end{array}
\label{Qgi}
\end{equation}
and $q(g_1,g_2)$ is a symmetric bilinear function of the elements
of $G$ satisfying
\begin{equation}
\begin{array}{cll}
(4) & Q_{g_1}(g_2)=2q(g_1,g_2)&~({\rm mod}~\ZZ),\\
(5) & q(g,g)=-h_g&~({\rm mod}~\ZZ).
\end{array}
\label{q}
\end{equation}
Modular invariance of $Z^{{\cal X}/G}$ follows from the above
conditions together with an important formula
\cite{Schellekens-Y, Intriligator}:
\begin{equation}
  S_{ij}e^{2\pi iQ_g(j)} = S_{gi,j}.
\end{equation}

In the RCFT terms, the sector twisted by $g\in G$ of
the orbifold theory ${\cal X}/G$ consists of the representation
spaces ${\cal V}_i\otimes{\cal V}_{g\bi}$ of ${\cal A}\otimes{\cal A}$.
The ground state in this sector has the eigenvalue
\begin{equation}
g'~=~e^{2\pi i(Q_{g'}(i)-q(g,g'))},
\label{g-tw}
\end{equation}
as can be read off from (\ref{Zorb}).
In a formal field theory terms, each term in the torus partition
function (\ref{Zorb}) of the orbifold theory ${\cal X}/G$ is given by
the path integral of the fields $\phi(z)$ on a torus
$(z\sim z+2\pi\sim z+2\pi\tau)$ with the periodicity conditions
\begin{equation}
  \phi(z)~=~ g_1^{-1}\phi(z+2\pi)g_1~=~ g_2^{-1}\phi(z+2\pi\tau)g_2.
\end{equation}

\subsubsection{The issue of doubled periodicity}\label{sec:IDP}

Although the function $Q_g(i)$ only needs to be defined modulo $\ZZ$
in constructing the modular invariant torus partition function,
we wish to have it defined modulo $2\ZZ$ for constructing boundary
or crosscap states in later sections.
In what follows we assume that $Q_g(i)$ is defined modulo $2\ZZ$
so as to satisfy the equations (2,3,4) of
(\ref{Qgi})-(\ref{q}) modulo $2\ZZ$, namely it is bilinear
in $g$ and $i$ modulo $2\ZZ$.
However, $Q_g(i)$ so defined will not always be single-valued
($=$periodic) modulo $2\ZZ$.
For example, $\prod_ag_a={\rm id}$ does not necessarily lead
to $\sum_aQ_{g_a}(i)=0$ modulo $2\ZZ$, although the equality always
holds modulo $\ZZ$.
In later sections, this kind of subtlety will be called ``doubled
periodicity''.

In constructing crosscap states in orbifolds, we will also need
to find an improvement of conformal weights
\begin{equation}
 h_i~\to~ h_i-\theta(i),
\label{hmod}
\end{equation}
by an integer-valued function $\theta(i)$ so that the equations
(1) and (5) hold modulo $2\ZZ$ as well.
Again, the function $\theta(i)$ will not in general be single valued
as a function of representation label $i$.

\subsubsection{Branes}\label{sec:SCOB}

Boundary states in orbifolds ${\cal X}/G$ are constructed by
summing over images and twists.
Pick a boundary state $\ket{\SB_J}$ in $\cal X$, and
let $H\subset G$ be the stabilizer of $J$.
Then there are boundary states in ${\cal X}/G$
in one to one correspondence with the characters $\rho$ of its
untwisted stabilizer $U\subset H$ \cite{Fuchs-KLLSW, Fuchs-HSSW},
\begin{equation}
 \ket{\SB^\rho_J}^{{\cal X}/G} ~=~
 \frac{\sqrt{|H|}}{\sqrt{|G||U|}}
 \sum_{g\in G/H,~ h\in U}g\ket{\SB_J}^h\;\rho(h),
\end{equation}
Here $\ket{\SB_J}^h$ is the boundary state in the $h$-twisted sector
and defined to satisfy
\begin{equation}
  ^h\bra{\SB_{J'}}e^{-\pi H_c/l}g\ket{\SB_J}^h ~=~
  {\rm Tr}_{J',gJ}[he^{-2\pi H_ol}],
\end{equation}
i.e. their overlaps should be proportional to the
traces over open string Hilbert space with additional weight $h$.
It is important that the twist $h$ does not run over all the
elements in $H$.
The definition of untwisted stabilizer group will be given in
Section \ref{sec:PBO}.
To construct the boundary states in orbifolds explicitly, one
therefore needs the expression for the states $\ket{\SB_J}^h$
in terms of Ishibashi states,
\begin{equation}
  \ket{\SB_J}^h ~=~ \sum_j\frac{S^{(h)}_{Jj}}{\sqrt{S_{0j}}}\kket{\SB;h(j)}^h.
\label{Smat-h}
\end{equation}
Here the matrix $S^{(h)}$ has indices $J,j$ which run only over
representations fixed by $h$, and the elements are supposed to satisfy
\begin{equation}
 S_{g(J),j}^{(h)} ~=~ S_{J,j}^{(h)}\exp 2\pi i(Q_g(j)+q(g,h)).
\end{equation}

\subsubsection{Orientifolds}\label{sec:SCO-O}

Crosscap states in ${\cal X}/G$ are constructed as sums of
crosscaps in $\cal X$.
Here we review the construction of \cite{Brunner-H1}.

Let $P_I$ be an involutive parity symmetry of $\cal X$
and $\ket{P_I}$ the corresponding crosscap state.
The parity $P_I$ maps a state in ${\cal V}_j\otimes{\cal V}_{\bj}$ to
a state in ${\cal V}_{\bj}\otimes{\cal V}_j$.
For any $g\in G$, $gP_I$ defines a parity whose action is that
of $P_I$ followed by the phase multiplication (\ref{g-tw}).
$gP_I$ is also involutive due to $gP_I=P_Ig^{-1}$ which one
can easily check.
So there are crosscaps $\ket{gP_I}$ satisfying
\begin{equation}
  \bra{g\tilde gP_I}e^{-\pi H_c/l}\ket{gP_I}
 ~=~ {\rm Tr}_{\tilde g}[g\tilde gP^{(0)}_I e^{-\pi H_c l}]
 ~=~ {\rm Tr}_{\tilde g}[gP^{(\pi)}_I e^{-\pi H_cl}],
\end{equation}
where the trace in the right hand side is over the
$\tilde g$-twisted closed string states,
and the superscripts $(0),(\pi)$ indicate the fixed points of
the parity on the circle of circumference $2\pi$.
The crosscap $\ket{P_I}^{{\cal X}/G}$ in the orbifold
is therefore described by a sum of crosscaps in ${\cal X}$,
\begin{equation}
  \ket{P_I}^{{\cal X}/G} ~=~ \frac{1}{\sqrt{|G|}}\sum_{g\in G}
  \ket{gP_I}.
\label{Corb}
\end{equation}
One can also consider the sum of crosscaps in ${\cal X}$ dressed
by characters of $G$,
\begin{equation}
  \ket{P_I^\epsilon}^{{\cal X}/G}
  ~=~ \frac{1}{\sqrt{|G|}}\sum_{g\in G}
  \ket{gP_I}\epsilon(g).
\label{Corbdr}
\end{equation}
Note here that, since $g\ket{P_I} = \ket{gP_Ig^{-1}} = \ket{g^2P_I}$
from (\ref{gketP}), the character $\epsilon$ in (\ref{Corbdr})
have to be $\ZZ_2$-valued if the crosscap states in orbifold are
made of $G$-invariant closed string states.
Such a degree of freedom arises only when $G$ contains an element
of even order, i.e. if $G/G^2$ is non-trivial.

To extend the PSS construction to orbifolds, one needs to find
the precise relation (including the normalization) between the
crosscap state $\ket{gP_I}$ corresponding to the parity $gP_I$
and the PSS state
\[
  \ket{\SC_{gI}} = \sum_j\frac{P_{gI,j}}{\sqrt{S_{0j}}}\kket{\SC;j}.
\]
From the formula for overlaps of two PSS states,
\begin{eqnarray}
  \bra{\SC_{g\tilde gI}}e^{-\pi H_c/l}\ket{\SC_{gI}}
 &=& \sum_jY_{j,gI}^{~g\tilde gI}\chi_j(il)
 \nn\\
 &=& \sum_jY_{j,I}^{~\tilde gI}\chi_j(il)
  e^{\pi i\{h_{gI}+h_{\tilde gI}-h_{g\tilde gI}-h_I-2Q_g(j)\}},
\end{eqnarray}
one finds that, for an arbitrary character $e^{i\pi\Delta(g)}$ of $G$,
the following sum of PSS states
\begin{equation}
  \frac{1}{\sqrt{|G|}}\sum_{g\in G}\ket{\SC_{gI}}
  \exp\pi i\left\{h_I-q(g,g)-h_{gI}-\Delta(g)\right\},
\label{Cpss}
\end{equation}
corresponds to a parity symmetry of the theory ${\cal X}/G$
which acts as
\[
 P_I \exp i\pi\left\{h_{gI}+q(g,g)-h_I+\Delta(g)\right\}
\]
on $g$-twisted sector.
The crosscap state (\ref{Cpss}) is $G$-invariant provided
$\Delta(g^2) = 2Q_g(I)$ modulo $2\ZZ$, as follows from the identity
\cite{Fuchs-HSSW, Huiszoon-SS2}
\begin{equation}
  e^{2\pi iQ_g(j)}P_{i,j}~=~
  P_{g^2i,j}\exp i\pi(2h_g+2h_{gi}-h_i-h_{g^2i}).
\end{equation}

We have thus found that, in order to define a parity $P_I$ and
the corresponding crosscap state in orbifold ${\cal X}/G$ from those
in ${\cal X}$, we need to choose a character $e^{i\pi\Delta}$ of $G$
satisfying $\Delta(g^2) = 2Q_g(I)$ mod $2\ZZ$.
We find it most convenient to set $\Delta(g)= Q_g(I)$ mod $2\ZZ$,
although this gives rise to some subtleties because $e^{i\pi\Delta}$
is actually not always a character of $G$.

We first notice that there exists an integer-valued function
$\theta$ on the set of representations of ${\cal A}$
with the following property\footnote{
$G$ is assumed to act on $I$ freely.}:
\begin{equation}
 h_I-q(g,g)-h_{gI} ~=~ 
 Q_g(I)+\theta(I)-\theta(gI)~~{\rm mod}~2,
\label{theta}
\end{equation}
Putting $I:={\rm id}$ and setting $\theta({\rm id})=0$, one finds
$\theta(g)=h_g+q(g,g)$.
Inserting this back into (\ref{theta}) one finds that
$\theta(I)$ can be thought of as a modification
of $h_I$ discussed at (\ref{hmod}).
Introducing $\sigma_I\equiv e^{i\pi\theta(I)}$,
the requirement that (\ref{Corb}) coincides with (\ref{Cpss}) up to
an overall sign when $\Delta(g)=Q_g(I)$ just boils down to
\begin{equation}
  \ket{gP_I} ~=~ \ket{\SC_{gI}}\sigma_{gI}.
\label{gPI-CgI}
\end{equation}
The general crosscap state in ${\cal X}/G$ is thus given by
\begin{equation}
  \ket{P_I^\epsilon}^{{\cal X}/G}~=~\frac{1}{\sqrt{|G|}}
  \sum_{g\in G}\ket{\SC_{gI}}\sigma_{gI}\cdot \epsilon(g).
\label{Corb2}
\end{equation}
The parity $P_I^\epsilon$ corresponding to this crosscap acts on
$g$-twisted sector as $P_I\epsilon(g)\sigma_I\sigma_{gI}$.

The crosscaps $\ket{gP_I}$ defined by (\ref{gPI-CgI}) satisfies the
shift relation $g\ket{P_I}=\ket{g^2P_I}$, so the crosscap state
(\ref{Corb2}) is a $G$-invariant closed string state.
However, $\ket{gP_I}$ has in general doubled periodicity because
of the doubled periodicity of $\sigma_{gI}$.
Therefore, $\epsilon$ in (\ref{Corb2}) should be chosen in such
a way that the summand in the right hand side is a single-valued function of
$g\in G$.

\subsection{Permutation branes in orbifolds}\label{sec:PBO}

In this and the next subsections we consider the permutation branes
and orientifolds in the orbifold ${\cal X}^N/{\cal G}$, where
${\cal G}$ is a subgroup of $G^N$.
For simplicity, we assume $\cal G$ is invariant under $S_N$, namely,
\begin{equation}
 g\equiv(g_1,\cdots,g_N)\in {\cal G} ~\Longrightarrow~ 
 g_\pi\equiv(g_{\pi(1)},\cdots,g_{\pi(N)})\in {\cal G}.
\end{equation}
D-branes in ${\cal X}^N/{\cal G}$ are constructed
as sums over images and twists.
The simple current $g=(g_1,\cdots,g_N)$ acts on $\pi$-permuted
boundary states $\ket{\SB^\pi_{\bf J}}$ in ${\cal X}^N$ as
\begin{equation}
  g\ket{\SB^\pi_{\bf J}} ~=~
  g\otimes_{c=1}^{\nc\pi}\ket{\SB^{\pi_c}_{J_c}} ~=~
   \otimes_{c=1}^{\nc\pi}\ket{\SB^{\pi_c}_{J'_c}},~~~~~
  J'_c = (\ts\prod_{a\in \pi_c}g_a)J_c.
\end{equation}
In particular, $g$ fixes the brane $\ket{\SB^\pi_{\bf J}}$ if
\[
  J_c = \ts\prod_{a\in \pi_c}g_a\cdot J_c~~~~~c=1,\cdots,\nc\pi.
\]
As a simple example, all the $\pi$-permuted branes
are fixed by $g$ if $\prod_{a\in \pi_c}g_a=1$ for all cycles $\pi_c$.
Let us denote by ${\cal H}\subset{\cal G}$ the stabilizer of
$\ket{\SB^\pi_{\bf J}}$.
Then the corresponding permutation brane in the orbifold takes the form
\cite{Fuchs-KLLSW,Fuchs-HSSW}
\begin{equation}
  \ket{\SB^{\pi,\rho}_{\bf J}}^{{\cal X}^N/{\cal G}}=
  \frac{|\cal H|}{\sqrt{|\cal G||\cal U|}}
  \sum_{h\in\cal U}\sum_{g\in\cal G/H}
  g\ket{\SB^\pi_{\bf J}}^h\;\rho(h),
\label{Borb}
\end{equation}
where $\ket{\SB^\pi_{\bf J}}^h$ denotes the boundary state in
$h$-twisted sector.
The twist $h$ runs over the group $\cal U\subset\cal H$ called
the untwisted stabilizer (see below for the definition) of the brane,
and $\rho$ is a character of $\cal U$.

The permutation boundary states in twisted sectors are
constructed as follows.
Since they factorize into pieces representing each cycle,
\begin{equation}
  \ket{\SB^\pi_{\bf J}}^h ~=~
  \otimes_{c=1}^{\nc\pi}\ket{\SB^{\pi_c}_{J_c}}^h,
\end{equation}
we focus on the cases where $\pi$ itself is a cyclic
permutation, $\pi=(1\,2\cdots N)$.
For such $\pi$ the boundary states in the sector twisted
by $h=(h_1,\cdots,h_N)$ are defined by
\begin{equation}
  \ket{\SB^\pi_J}^h ~=~
  \sum_j
  \frac{S^{(h_{\rm tot})}_{Jj}}{(S_{0j})^{N/2}}\kket{\SB^\pi;j}^h,
\label{BpiJh}
\end{equation}
where the matrix $S^{(h)}$ was introduced in (\ref{Smat-h}),
$h_{\rm tot}\equiv h_1h_2\cdots h_N$ and the Ishibashi states
in $h$-twisted sector are defined by
\begin{eqnarray}
&& \kket{\SB^\pi;j}^h ~\equiv~
 R^\pi\kket{\SB;j_1\cdots j_N},
 \nn\\&&~~
 j_k=h_kj_{k-1}~~~(k=1,\ldots,N;\,j_0\equiv j).
\label{twpiI}
\end{eqnarray}
Note that the Ishibashi states defined in this way depend on the choice
of the ``first'' entry in the cycle.
For more general cyclic permutation $\pi=(a_1a_2\cdots a_N)$ we
define $\kket{\SB^\pi;j}^h$ so that $\bj$ appears in the $a_1$-th
antiholomorphic sector and $h_{\rm tot}j=j$ appears in the $a_N$-th
holomorphic sector.

In order for the sum over twisted sectors to make sense,
we need to require that the $J$-label of $\ket{\SB^\pi_J}^h$
is transformed in the same way as that of $\ket{\SB^\pi_J}^{h={\rm id}}$
by simple currents:
\begin{equation}
  g\ket{\SB^\pi_J}^h ~=~
  \ket{\SB^\pi_{g_{\rm tot}(J)}}^h\omega_\pi(g,h).
~~~~(g_{\rm tot}\equiv g_1g_2\cdots g_N)
\end{equation}
The factor $\omega_\pi(g,h)$, if nontrivial, means that
$g\in{\cal G}$ not only acts on the $J$-label
of the brane $\ket{\SB^{\pi,\rho}_J}$ but also transforms
$\rho(h)$ to $\rho(h)\omega_\pi(g,h)$.
The simple current prescription gives
\begin{eqnarray}
 \omega_\pi(g,h) &=& \exp 2\pi i\left\{
-q(g_1\cdots g_N,h_1\cdots h_N) \right. \nn\\
 && \left. ~~
+q(g_1,h_1)+q(g_2,h_1^2h_2)+\cdots+q(g_N,h_1^2h_2^2\cdots h_{N-1}^2h_N)
 \right\}.
\label{omega-pi}
\end{eqnarray}
For a state $\ket{\SB^\pi_J}^h$ in $h$-twisted sector to
contribute to (\ref{Borb}), $\cal H$ should be realized trivially on it;
otherwise it would be projected out by the orbifolding procedure.
The {\it untwisted stabilizer group} $\cal U\subset {\cal H}$ of a
boundary state is formed by such $h$'s.
${\cal U}$ is therefore formed by those $h\in\cal H$
satisfying $\omega_\pi(g,h)=1$ for all $g\in{\cal H}$.

\subsubsection{Diagonal Branes}\label{sec:PBO-DB}

An interesting class of permutation D-branes are the ``diagonal
branes'' in ${\cal X}^2$ or its orbifolds, which are regarded
as wrapping the diagonal, ${\cal X}\subset {\cal X}^2$.

First, let us consider the following boundary state in the product
theory ${\cal X}^2$,
\begin{equation}
  \ket{\SB_{\rm diag}}^{{\cal X}^2} ~\equiv~
  \ket{\SB^{(12)}_0} ~=~ \sum_iR^{(12)}\kket{\SB;i,i}.
\end{equation}
Note that the modular S-matrices in the enumerator and denominator
of Recknagel's construction canceled out.
It gives the annulus partition function,
\begin{equation}
 {}^{{\cal X}^2}\bra{\SB_{\rm diag}}e^{-\pi H_c/l}
                \ket{\SB_{\rm diag}}^{{\cal X}^2} ~=~
 \sum_i\chi_i(i/l)\chi_i(i/l) ~=~
 \sum_i\chi_i(il)\chi_\bi(il) ~=~ Z^{\cal X}_{T^2}(il).
\end{equation}
Let us next consider an orbifold ${\cal X}^2/{\cal G}$.
For simplicity, we take ${\cal G}=G\otimes G=\{(g_1,g_2)|g_1,g_2\in G\}$
with $G$ acting on all the representations in the theory ${\cal X}$ freely.
The diagonal brane is invariant under the elements
$h\otimes h^{-1}\in{\cal G}$, so we consider the sum over
$h\otimes h^{-1}$-twisted sectors,
\begin{eqnarray}
  \ket{\SB_{\rm diag}}^{\rm orb}
  &=& \frac{1}{\sqrt{|\cal G|}}
  \sum_{g,h\in G}(g\otimes 1)\ket{\SB_{\rm diag}}^{h\otimes h^{-1}}
 \nn\\ &=&
  \frac{1}{\sqrt{|\cal G|}}
  \sum_{g,h\in G}\sum_i (g\otimes 1)R^{(12)}\kket{\SB;h(i),i}.
\end{eqnarray}
This diagonal brane gives the annulus partition function,
\begin{eqnarray}
\lefteqn{
 {}^{\rm orb}\bra{\SB_{\rm diag}}e^{-\pi H_c/l}
 \ket{\SB_{\rm diag}}^{\rm orb}
~=~
 \frac{1}{|G|}\sum_{g,h,i}e^{2\pi iQ_g(i)+2\pi iq(g,h)}
 \chi_{h(i)}(i/l)\chi_i(i/l)
} \hskip30mm \nn\\ &=&
 \frac{1}{|G|}
 \sum_{g,h,j}e^{2\pi iQ_{h}(j)+2\pi iq(g,h)}
 \chi_j(il)\chi_{g^{-1}(\bj)}(il)
  ~=~ Z^{{\cal X}/G}_{T^2}(il).
\label{dgtorus}
\end{eqnarray}

Let us reconsider the properties of diagonal branes in more
abstract terms.
We first consider the product theory ${\cal X}^2$
defined on a strip of width $\pi$ parametrized by
$(\sigma\in[0,\pi],t\in\RR)$.
We wish to consider what boundary condition on the fields
$\phi_{1,2}$ corresponds to the diagonal brane.
Suppose that the theory $\cal X$ on a circle ($\sigma\sim\sigma+2\pi$)
has an involutive parity symmetry $P$ which acts linearly on fields $\phi$ as
\begin{equation}
  P~:~ \phi(\sigma)~\mapsto~ \SR(P)\phi(-\sigma),
\end{equation}
where $\SR(P)$ is a matrix representation of $P$ when
$\phi$ is a vector describing the collection of fields.
Then consider the theory ${\cal X}^2$ on a strip with the following
boundary condition on fields at $\sigma=0,\pi$:
\begin{equation}
\begin{array}{rcl}
  \phi_1(0)~=~\SR(P)\phi_2(0),\\
  \phi_2(0)~=~\SR(P)\phi_1(0),
\end{array}
~~
\begin{array}{rcl}
  \phi_1(\pi)~=~\SR(P)\phi_2(\pi),\\
  \phi_2(\pi)~=~\SR(P)\phi_1(\pi).
\end{array}
\label{diagdef}
\end{equation}
One can then define a periodic field $\phi$ of the theory ${\cal X}$
on a circle of radius $2\pi$ by
\begin{equation}
\begin{array}{rcll}
 \phi(\sigma) &=& \phi_1(\sigma) &(\sigma\in[0,\pi]),\\
 \phi(\sigma) &=& \SR(P)\phi_2(2\pi-\sigma) & (\sigma\in[\pi,2\pi]).
\end{array}
\label{phidbl}
\end{equation}
The theory ${\cal X}^2$ on a strip with boundary
condition (\ref{diagdef}) is thus equivalent to the theory $\cal X$
on a periodic cylinder.
We therefore identify the fundamental diagonal branes
$\ket{\SB_{\rm diag}}, \bra{\SB_{\rm diag}}$ with the boundary
conditions (\ref{diagdef}) on fields.

Let us next consider the orbifold theory.
We first wish to show that the overlap of $\bra{\SB_{\rm diag}}$ and
$(g_1\otimes g_2)\ket{\SB_{\rm diag}}$ gives
a toroidal partition function of the theory ${\cal X}$ with
periodicity along the $\sigma$ direction twisted by $g_1^{-1}g_2^{-1}$.
In field theoretic terms, the multiplication of
$(g_1\otimes g_2)$ corresponds to the modification
of the boundary condition on fields at $\sigma=\pi$,
\begin{equation}
\begin{array}{rcl}
  g_1\phi_1g_1^{-1}~=~g_2(\SR(P)\phi_2)g_2^{-1},\\
  g_2\phi_2g_2^{-1}~=~g_1(\SR(P)\phi_1)g_1^{-1}.
\end{array}
\end{equation}
Assuming that the action of simple currents on fields
is also linear and using the notation $g^{-1}\phi g\equiv \SR(g)\phi$
it can be written as
\begin{equation}
\begin{array}{rcl}
  \SR(g_1^{-1})\phi_1~=~\SR(P)\SR(g_2^{-1})\phi_2,\\
  \SR(g_2^{-1})\phi_2~=~\SR(P)\SR(g_1^{-1})\phi_1.
\end{array}
\label{Bdgcjg}
\end{equation}
It follows that the field $\phi$ defined as in (\ref{phidbl})
satisfies the twisted periodicity, as claimed above:
\begin{equation}
 \phi(\sigma)
 ~=~ \SR(g_1g_2)\phi(\sigma-2\pi)
 ~=~ (g_1g_2)^{-1}\phi(\sigma-2\pi)g_1g_2.
\label{phitw}
\end{equation}
Second, the overlaps of diagonal boundary states in
$(h\otimes h^{-1})$-twisted sector correspond to path integral
over fields of ${\cal X}^2$ on a cylinder with the twisted periodicity
along $t$,
\begin{equation}
  \phi_1(\sigma,t)~=~h\phi_1(\sigma,t-2\pi l)h^{-1},~~~~
  \phi_2(\sigma,t)~=~h^{-1}\phi_2(\sigma,t-2\pi l)h.
\end{equation}
In terms of the field $\phi$ this is simply
\begin{equation}
  \phi(\sigma,t)~=~h\phi(\sigma,t-2\pi l)h^{-1}.
\end{equation}
From these two observations it follows that the diagonal branes of
${\cal X}^2$ sitting in twisted sectors satisfy the formula
\begin{equation}
 {}^{h\otimes h^{-1}}\bra{\SB_{\rm diag}}e^{-\pi H/l}(g_1\otimes g_2)
 \ket{\SB_{\rm diag}}^{h\otimes h^{-1}}
 ~=~ {\rm Tr}^{\cal X}_{g_1^{-1}g_2^{-1}}[he^{-2\pi Hl}].
\end{equation}
By comparing this with (\ref{dgtorus}), one can check that
the RCFT construction gives the diagonal branes with the
correct property.

We have seen in the previous subsection that the PSS prescription
allows to construct crosscaps corresponding to different parity
symmetry.
The fundamental diagonal brane we have studied above should
be associated to the fundamental parity $P$
corresponding to the crosscap $\ket{\SC_0}$.
The diagonal branes corresponding to other parities are obtained
by a similar argument as was given above.
For each representation $I$ of ${\cal A}$ satisfying the fusion rule
$I\times \bar I \longmapsto {\rm id}$, there is a
boundary state $\ket{\SB^{(12)}_I}$ in ${\cal X}^2$
\begin{equation}
  \ket{\SB^{(12)}_I} ~=~ \sum_i\frac{S_{Ii}}{S_{0i}}R^{(12)}\kket{\SB;i,i}.
\end{equation}
The fields of the two copies of ${\cal X}$ are glued
via the parity $P_I$.
The corresponding diagonal branes in the orbifold are given by
\begin{equation}
  \ket{\SB^{(12),\rho}_I}^{\rm orb}
  ~=~
  \frac{1}{\sqrt{|{\cal G}|}}
  \sum_{g,h\in G} (g\otimes 1)
  \ket{\SB^{(12)}_I}^{h\otimes h^{-1}}\rho(h),
\label{Bdg2}
\end{equation}
where $\rho(h)$ is a character of (the double cover of) $G$, and
the boundary states in twisted sectors are defined as
\begin{equation}
  \ket{\SB^{(12)}_I}^{h\otimes h^{-1}} ~=~
  \sum_i\frac{S_{Ii}}{S_{0i}}R^{(12)}\kket{\SB;h(i),i}e^{i\pi Q_h(I)},
\label{Bdgtw}
\end{equation}
where the last factor is added so that
$(g_1\otimes g_2)\ket{\SB^{(12)}_I}^{h\otimes h^{-1}}
  =\ket{\SB^{(12)}_{g_1g_2I}}^{h\otimes h^{-1}}$
is satisfied.
Note that (\ref{Bdgtw}) in general has doubled periodicity as a function
of $h$, so $\rho(h)$ in (\ref{Bdg2}) should be chosen so that
the summand of the right hand side is single valued.

\subsection{Permutation crosscaps in orbifolds}\label{sec:PCO}

Let us next construct permutation crosscaps in orbifolds\footnote{
The outline of the argument in this subsection was suggested to
us by K. Hori.}.
For an involutive permutation $\pi\in S_N$ and a $\pi$-invariant parity
$P_I$ of ${\cal X}^N$, PSS's construction gives us the crosscap state
in ${\cal X}^N$ corresponding to the parity $P_I\pi$.
To obtain crosscap states in the orbifold ${\cal X}^N/{\cal G}$,
one needs crosscaps corresponding to the parities
$gP_I\pi~(g\in{\cal G})$ which map the states of ${\cal X}^N$
as follows:
\begin{equation}
   gP_I\pi:~a_1\otimes\cdots\otimes a_N ~\to~
   (g_1P_{I_1}a_{\pi(1)})\otimes\cdots\otimes(g_NP_{I_N}a_{\pi(N)}).
\label{gPpiACT}
\end{equation}
The permutation crosscaps in ${\cal X}^N/{\cal G}$ are
sums over those in ${\cal X}^N$,
\begin{equation}
  \ket{P^{\pi,\epsilon}_I}^{{\cal X}^N/{\cal G}}
 ~=~ \frac{1}{\sqrt{|\cal G|}}\sum_{g\in{\cal G}}
  \ket{gP_I\pi}^{{\cal X}^N}\epsilon(g),
\label{Cpiorb}
\end{equation}
dressed by a character $\epsilon(g)$ of (the double cover of)
${\cal G}$ satisfying suitable periodicity conditions.
The $\cal G$-invariance of the crosscap state requires
$\epsilon(gg_\pi)=1$ for all $g\in\cal G$, but it does not necessarily
require that $\epsilon$ be $\ZZ_2$-valued.
Note also that, for the equation (\ref{Cpiorb}) to define
an involutive parity in the orbifold, $P_I$ actually does not have
to be involutive; it only has to square to an element of ${\cal G}$.

The $\pi$-permuted crosscap states should factorize into pieces
representing the cycles of $\pi$,
\begin{equation}
 \ket{gP_I\pi} ~=~ \otimes_{c=1}^{\nc\pi}\ket{g_cP_{I_c}\pi_c},
\end{equation}
where all the cycles of $\pi$ are of length one or two because
$\pi$ is involutive.
For cycles of length one we have seen the correspondence (\ref{gPI-CgI}),
so it remains to construct the crosscaps $\ket{gP_I\pi}$ for
the cyclic permutation of length two, $\pi=(1\,2)$.

We focus first on the crosscap $\ket{gP\pi}$ corresponding to
the fundamental PSS parity $P$.
The overlaps of two permutation crosscaps $\bra{gP\pi}$ and
$\ket{\tilde gP\pi}$ correspond to the theory ${\cal X}^2$ on a space
$(\sigma\in[0,\pi],t\sim t+4\pi l)$ with boundary conditions
\begin{equation}
\begin{array}{rcl}
  \phi_1(0,t)   &=& \SR(P)\SR(g_2^{-1})       \phi_2(0,t-2\pi l),\\
  \phi_2(0,t)   &=& \SR(P)\SR(g_1^{-1})       \phi_1(0,t-2\pi l),\\
  \phi_1(\pi,t) &=& \SR(P)\SR(\tilde g_2^{-1})\phi_2(\pi,t-2\pi l),\\
  \phi_2(\pi,t) &=& \SR(P)\SR(\tilde g_1^{-1})\phi_1(\pi,t-2\pi l).
\end{array}
\label{CCbc1}
\end{equation}
As states in the Hilbert space of the theory ${\cal X}^2$, the
crosscap states $\bra{gP\pi}$ and $\ket{\tilde gP\pi}$
belong to the sector twisted by
$gg_\pi^{-1}=(g_1g_2^{-1}\otimes g_2g_1^{-1})$ and
$\tilde g\tilde g_\pi^{-1}$, respectively.
Therefore, $g_1g_2^{-1}\equiv\tilde g_1\tilde g_2^{-1}$ for pairs
of crosscaps with nonzero overlaps.
By arguing in a similar way to the construction of diagonal branes,
one finds that the theory ${\cal X}^2$ with boundary conditions
(\ref{CCbc1}) is equivalent to the theory ${\cal X}$ on torus
$(\sigma\sim\sigma+2\pi,t\sim t+4\pi l)$ with periodicity,
\begin{equation}
  \phi(\sigma,t) ~=~ g_1\tilde g_1^{-1}\phi(\sigma-2\pi,t)g_1^{-1}\tilde g_1
                 ~=~ g_1g_2^{-1}\phi(\sigma,t-4\pi l)g_2g_1^{-1}.
\end{equation}
The overlaps of permutation crosscaps thus gives the torus
partition function of the theory ${\cal X}$,
\begin{eqnarray}
 \bra{gP\pi}e^{-\pi H_c/l}\ket{\tilde gP\pi} &=&
 {\rm Tr}^{\cal X}_{g_1\tilde g_1^{-1}}[g_1g_2^{-1}e^{-4\pi H_c l}].
\end{eqnarray}

We need the formula for permutation crosscaps expressed
in terms of Ishibashi states in twisted sectors,
\begin{equation}
 \ket{gP\pi} ~=~ \sum_i X_i(g_1,g_2)\kket{\SC^\pi;g_1i,g_2i}.
\end{equation}
We determine it by requiring that it has the following overlap
with the fundamental diagonal brane,
\begin{equation}
  \bra{\SB_{\rm diag}}e^{-\pi H_c/2l}\ket{gP\pi}
  ~=~ {\rm Tr}^{\cal X}_{{\cal H}_{g_2^{-1}g_1^{-1}}}
  [g_1e^{-2\pi l H - i\pi P}]
  ~=~ {\rm Tr}^{\cal X}_{{\cal H}_{g_2^{-1}g_1^{-1}}}
  [g_2^{-1}e^{-2\pi l H + i\pi P}],
\label{BdgC}
\end{equation}
where one should recall
\[
 H=L_0+\bar L_0-\frac c{12},~~~P=L_0-\bar L_0.
\]
To understand this condition, let us consider the theory
${\cal X}^2$ on a strip ($0\le\sigma\le\pi$) bounded by the diagonal brane
$\bra{\SB_{\rm diag}}$ and its image under the parity $gP\pi$.
The partition function on the M\"obius strip is calculated
by the path integral of fields $\phi_{1,2}$ of ${\cal X}^2$
with the following boundary condition at $\sigma=0$,
\begin{equation}
\begin{array}{rcl}
  \phi_1(0,t) &=& \SR(P)\phi_2(0,t),\\
  \phi_2(0,t) &=& \SR(P)\phi_1(0,t),
\end{array}
\end{equation}
and the periodicity along the $t$-direction,
\begin{equation}
\begin{array}{rcl}
  \phi_1(\sigma,t)&=&
  \SR(P)\SR(g_2^{-1})\phi_2(\pi-\sigma,t-2\pi l),\\
  \phi_2(\sigma,t)&=&
  \SR(P)\SR(g_1^{-1})\phi_1(\pi-\sigma,t-2\pi l).
\end{array}
\end{equation}
It follows that the boundary condition at $\sigma=\pi$ has to be
that of $(g_1\otimes g_2)\ket{\SB_{\rm diag}}$, (\ref{Bdgcjg}).
Thus the theory ${\cal X}^2$ on M\"obius strip is equivalent to
the theory ${\cal X}$ on the torus, with the field $\phi$
satisfying the periodicity along the spatial direction
(\ref{phitw}), and the time direction
\begin{equation}
 \phi(\sigma,t)
 ~=~ g_2^{-1}\phi(\sigma-\pi,t-2\pi l)g_2
 ~=~ g_1\phi(\sigma+\pi,t-2\pi l)g_1^{-1},
\end{equation}
hence the requirement (\ref{BdgC}).
We solve it and find
\begin{eqnarray}
 \ket{gP\pi} &=&
 \sum_i \kket{\SC^\pi;g_1i,g_2i}
  \exp \pi i\left(2q(g_1,g_1g_2)+2h_{g_1}+2h_i-h_{g_1i}-h_{g_2i}\right)
 \nn\\ &=&
 \sum_i \kket{\SC^\pi;g_1i,g_2i}\sigma_{g_1i}\sigma_{g_2i}
  \exp \pi i\left\{q(g_1g_2,g_1g_2)+Q_{g_1g_2}(i)\right\}.
\end{eqnarray}

The expression for more general permutation crosscaps $\ket{gP_I\pi}$
can be found by studying its overlap with the diagonal brane
$\ket{\SB^\pi_I}$.
Our final result reads
\begin{equation}
 \ket{gP_I\pi} ~=~
 \sum_i \frac{S_{Ii}}{S_{0i}}\kket{\SC^\pi;g_1i,g_2i}
 \sigma_{g_1i}\sigma_{g_2i}
 \exp \pi i\left\{q(g_1g_2,g_1g_2)+Q_{g_1g_2}(i)+Q_{g_1g_2}(I)\right\}.
\label{CgPpi}
\end{equation}
Note that this crosscap has the same periodicity as that of
$\sigma_{g_1g_2(I)}\sigma_{g_1g_2}$.

\subsection{Parity action on D-branes}\label{sec:PAD}

The action of parity $P_I\pi$ on branes in ${\cal X}^N$ is
read off from the relation
\begin{equation}
\bra{\SB}q_t^H\ket{P_I\pi}~=~\bra{P_I\pi}q_t^H\ket{\SB'}.
\end{equation}
When $\ket\SB$ is a $\sigma$-permuted brane gluing the $a$-th
holomorphic sector with the $\sigma(a)$-th anti-holomorphic sector,
its parity image $\ket{\SB'}$ should glue the $\pi(a)$-th antiholomorphic
sector with the $\pi\sigma(a)$-th holomorphic sector.
So $\ket{\SB'}$ has to be a $\sigma'=\pi\sigma^{-1}\pi$-permuted brane.
One then finds, using
\begin{equation}
 \bbra{\SB;i_1\cdots i_N}\!R^{\sigma^{-1}}q_t^H R^\pi\!
 \kket{\SC;j_1\cdots j_N}
 ~=~
 \bbra{\SC;j_1\cdots j_N}\!R^{\pi^{-1}}q_t^H R^{\sigma'}\!
 \kket{\SB;\ti_1\cdots \ti_N},
\label{BCCB}
\end{equation}
where $\ti_a=i_{\sigma^{-1}\pi(a)}$,
that the parity acts on boundary states as follows:
\begin{equation}
  \bra{\SB^\sigma_{\bf J}}q_t^H\ket{P_I\pi} ~=~
  \bra{P_{\bar I}\pi}q_t^H\ket{\SB^{\sigma'}_{\bf\bar J}} ~=~
  \bra{P_I\pi}q_t^H \omega\ket{\SB^{\sigma'}_{\bf\bar J}},
\end{equation}
where $\omega$ is a simple current satisfying $\omega_\pi\omega\bar I=I$.
Although there may be several $\omega$'s satisfying this, there must be
a unique $\omega$ that determines the action of parity $P_I\pi$ on D-branes.
For example, for the permutation crosscaps $\ket{gP\pi}$
made from the fundamental parity $P$ and $g=(g_1,\cdots,g_N)$, one finds
both from the M\"obius strip amplitudes of RCFT and from a formal
field theory argument that
\begin{eqnarray}
\lefteqn{
  \bra{\SB^\sigma_{\bf J}}q_t^H\ket{P\pi} ~=~
  \bra{P\pi}q_t^H \ket{\SB^{\sigma'}_{\bf\bar J}}
} \nn\\
 &\Longrightarrow&
  \bra{\SB^\sigma_{\bf J}}q_t^H\ket{gP\pi} ~=~
  \bra{gP\pi}q_t^H g\ket{\SB^{\sigma'}_{\bf\bar J}}.
\end{eqnarray}
Note here that the labels ${\bf J,\bar J}$ denote the sets of
representations $\{J_c\},\{\bar J_c\}~(c=1,\cdots,\nc\sigma)$.
$J_c$ and $\bar J_c$ are for the $c$-th cycle of $\sigma$ and $\sigma'$,
which are conjugate to each other thanks to $\pi$ being involutive.

By a similar argument one can derive the action of parity
$P^{\pi,\epsilon}_I$ (\ref{Cpiorb}) on branes in orbifold
${\cal X}^N/{\cal G}$.
We notice that (\ref{BCCB}) relates the bra Ishibashi states in the
$h$-twisted sector to the ket Ishibashi states in $h_\pi^{-1}$-twisted
sector.
The M\"obius strip amplitude of the orbifold theory,
\begin{equation}
  \vev{\SB^{\sigma,\rho}_{\bf J}|q_t^H|P^{\pi,\epsilon}_I}
 ~\sim~ \sum_{g,h} \rho^\ast(h)\cdot{}^h
  \vev{\SB^\sigma_{\bf J}|q_t^H|gP_I\pi}^h\epsilon(g)
  \cdot\delta_{h,gg_\pi^{-1}},
\end{equation}
allows us to read off the parity action on boundary states:
\begin{equation}
  P^{\pi,\epsilon}_I ~:~
  \ket{\SB^{\sigma,\rho}_{\bf J}} ~\longmapsto~
  \epsilon(\omega)
  \ket{\SB^{\sigma',\rho'}_{\bf\bar J}};
 ~~~~~\bar I=\omega I,~~\rho'(h_\pi)=\rho(h)\epsilon(h)^{-1}.
\label{Pchtr}
\end{equation}
The transformation law of $\rho(h)$ means that the parity
$P^{\pi,\epsilon}_I$ maps states in $h$-twisted sector to those in
$h_\pi$-twisted sector after multiplying $\epsilon(h)^{-1}$, a fact
which follows also from the construction of permutation parities
in orbifold.

The above expression is still somewhat ambiguous because of
the subtlety mentioned after (\ref{twpiI}): we need to specify the
first element for each cycle of $\sigma$ to define Ishibashi states
in twisted sectors unambiguously.
If $\sigma=(a\ssub1\cdots a\ssub N)$ is a single cycle and
$\pi\sigma^{-1}\pi=(a'\ssub1\cdots a'\ssub N)$, then we have to put
$i_{a_N}=\ti_{a'_N}$ in (\ref{BCCB}) and get
\begin{equation}
 \pi\sigma^{-1}\pi~=~ (a'_1\cdots a'_N) ~=~ (\pi(a_N)\cdots\pi(a_1)).
\end{equation}

\subsubsection{Parity invariant D-branes}\label{sec:PINVD}

As a future reference, we study the condition of parity-invariance
for permutation branes in more detail.
Here we give the condition on the pair $(\pi,\sigma)$ in order
for the $\sigma$-permuted brane to be invariant under $\pi$-permuted
orientifold.
\begin{PIB}\label{pib1}
Any pair of permutation $\pi,\sigma$ satisfying
$\sigma=\pi\sigma^{-1}\pi,\pi^2={\rm id}$ can be decomposed
into the following blocks,
\[
\begin{array}{cll}
(1) & \sigma=(a_1a_2\cdots a_{2n+1}),
    & \pi   =(a_1a_{2n+1})(a_2a_{2n})\cdots(a_na_{n+2}), \\
(2) & \sigma=(a_1a_2\cdots a_{2n}),
    & \pi   =(a_2a_{2n+1})(a_2a_{2n})\cdots(a_na_{n+2}), \\
(3) & \sigma=(a_1a_2\cdots a_{2n}),
    & \pi   =~~~\,(a_1a_{2n})(a_2a_{2n})\cdots(a_na_{n+1}), \\
(4) & \sigma=(a_1\cdots a_n)(a'_1\cdots a'_n),
    & \pi   =~~~\;(a_1a'_n)(a_2a'_{n-1})\cdots(a_na'_1).
\end{array}
\]
\end{PIB}

The simplest block $\sigma=\pi={\rm id}\in S_1$ is a special case of
the first type, and $\sigma=(a_1a_2),\;\pi={\rm id}\in S_2$ is the simplest
example of the second type.
The permutation $\sigma^{-1}\pi$ or its inverse appear in M\"obius
strip amplitudes as explained in (\ref{ProdI}).
Note $\sigma^{-1}\pi$ always squares to identity because of
$\sigma=\pi\sigma^{-1}\pi$, so it consists of cycles of lengths one
or two only.

In general, the spectrum of open string between identical D-branes
contains an identity representation.
The M\"obius strip amplitude for parity-invariant boundary
states, when written in the loop channel, should therefore contain
an identity character.
To check this explicitly, we need to show
\begin{equation}
  \vev{\SC^\pi|e^{-\pi H_c/4l}|\SB^\sigma}
  ~\sim~ q^{-\frac{Nc\ssub{\cal X}}{24}}+\cdots.~~~~~
  (q\equiv e^{-2\pi l})
\end{equation}
Here $-\frac{Nc\ssub{\cal X}}{24}$ is the energy
for the $SL(2,\RR)$-invariant ground state.
The amplitude can be written in the tree channel as a sum of
the following products of characters,
\[
 \prod_{\lc{\tilde\sigma_a}~{\rm even}}
 \chi_{j_a}(\tfrac{i\lc{\tilde\sigma_a}}{4l})
 \prod_{\lc{\tilde\sigma_b}~{\rm odd}}
 \hat\chi_{j_b}(\tfrac{i\lc{\tilde\sigma_b}}{4l})
 ~~~~~(\tilde\sigma\equiv\sigma^{-1}\circ\pi),
\]
where one should recall that each character $\chi_{j_a}$ or
$\widehat\chi_{j_b}$ corresponds to a cycle of $\tilde\sigma$ of
even or odd length.
One can read off the energy $E_0$ of the ground state of
the open string Hilbert space by modular transform,
\[
 E_0 ~=~
 -\sum_{\lc{\tilde\sigma_a}~{\rm even}}
  \frac{c\ssub{\cal X}}{6\lc{\tilde\sigma_a}}
 -\sum_{\lc{\tilde\sigma_b}~{\rm odd }}
  \frac{c\ssub{\cal X}}{24\lc{\tilde\sigma_a}}.
\]
This saturates the lower bound $-Nc\ssub{\cal X}/24$ iff all the
cycles of $\tilde\sigma$ have length one or two.
The four types of parity-invariant boundary states listed above
all satisfy this condition.

\section{Dirac Fermion and the Affine $U(1)_2$ Model}\label{sec:U1_2}

In this section we illustrate the construction of permutation branes and
orientifolds in the theory of $d$ Dirac fermions $\psi^{\pm,a}$.
It is pretty obvious how to construct the boundary or crosscap
states satisfying
\begin{equation}
\begin{array}{rcl}
 (\tilde\psi_n^{\pm,\pi(a)}+ i\eta\psi_{-n}^{\pm,a})\ket{\SB^\pi}
 \ssub{\SEC\eta} &=& 0, \\
 (\tilde\psi_n^{\pm,\pi(a)}+ i\eta e^{i\pi n}\psi_{-n}^{\pm,a})\ket{\SC^\pi}
 \ssub{\SEC\eta} &=& 0,
\end{array}
~~~~(\SEC=\NSNS~{\rm or}~\RARA~;~\eta=\pm)
\end{equation}
as Bogolioubov transforms of the vacuum following
\cite{Polchinski-C, Callan-LNY}.
On the other hand, one can construct the same states from
the boundary or crosscap states in the affine $U(1)_2^d$ model by
a suitable $(\ZZ_2)^d$ orbifold.
Since the affine $U(1)_2^d$ theory or its orbifold is purely bosonic,
one must assign Grassmann parity to the operators and states in a
suitable manner to reproduce the properties of fermions correctly,
as we will discuss here in detail.
The result obtained here also has a direct application to Gepner's
construction of superstring theories, where supersymmetric worldsheet
theories are constructed from purely bosonic RCFTs by the same orbifold.

The affine $U(1)_k$ symmetry is generated by the current
$J = i\sqrt{2k}\partial X$ augmented by spectral flow operators
$e^{\pm i\sqrt{2k}X}$, where $X$ is a canonically normalized chiral
scalar field.
There are $2k$ highest weight representations labelled by
a mod $2k$ integer $n$ corresponding to the collection of operators
$e^{iqX/\sqrt{2k}}~(q=n~{\rm mod}~2k)$ and their descendants.
The $U(1)$ charge and conformal weight of the operator
$e^{iqX/\sqrt{2k}}$ are $(J_0,L_0)=(q,\frac{q^2}{4k})$.
The model at level $k=2$ has four representations labelled by
an integer $s\sim s+4$.
We denote by $\psi$ the simple current satisfying the fusion rule
$\psi(s)=s+2$.

The affine $U(1)_2$ theory is related to the theory of a Dirac
fermion by the $\ZZ_2$-orbifolding.
This fact can be seen from the relation of characters:
from the characters of the affine $U(1)_2$ algebra,
\begin{equation}
  \chi_{s}(\tau,\nu)
  ~\equiv~ {\rm Tr}_{[s]}q^{L_0-1/24}z^{J_0/2}
  ~=~ \eta(\tau)^{-1} \sum_{l\in \ZZ+s/4}q^{2l^2}z^{2l},
 ~~~ (q=e^{2\pi i\tau},~z=e^{2\pi i\nu})
\end{equation}
one can construct characters of Dirac fermion model,
\begin{equation}
\begin{array}{lcrcl}
  \chi_0\pm\chi_2
  &=& \chi^{\NS\pm}(\tau,\nu)
  &=& \ds q^{-\frac{1}{24}}\prod_{m\ge 1}
      (1\pm zq^{m-\frac12})(1\pm z^{-1}q^{m-\frac12}), \\
  \chi_1\pm\chi_{-1}
  &=& \chi^{\RA\pm}(\tau,\nu)
  &=& \ds q^{\frac{1}{12}}(z^{\frac12}\pm z^{-\frac12})\prod_{m\ge 1}
      (1\pm zq^m)(1\pm z^{-1}q^m).
\end{array}
\label{chdf}
\end{equation}

The theory of $d$ Dirac fermions is obtained from the affine $U(1)_2^d$
model by orbifolding by $\Gamma_{\rm GSO}\equiv (\ZZ_2)^d$ generated
by the simple currents $\psi_a$, with the choice $q\equiv0$.
The choice $q\equiv0$ does not give a modular invariant
torus partition function because it does not satisfy (\ref{q}),
but the modular invariance is recovered by summing over four spin structures.
In RCFT terms, different spin structures arise from (i)
restricting to states for which the eigenvalues of all
$\psi_a$ are aligned, i.e. $\psi_a=1(\forall a)$ for NSNS sector
or $(-1)$ for RR sector, and
(ii) summing over twisted sectors with trivial weight or weighted by
a nontrivial character $\epsilon:\Gamma_{\rm GSO}\mapsto\ZZ_2$
such that $\epsilon(\psi_a)=-1(\forall a)$.
It is easy to see that the orbifold by $\Gamma_{\rm GSO}$
and summing over spin structures gives the same torus partition
function as the orbifold by a group $\wtilde\Gamma_{\rm GSO}=(\ZZ_2)^{d-1}$
of {\it even} monomials of $\psi_a$.
The orbifold group $\wtilde\Gamma_{\rm GSO}$ is used in
Gepner's original construction of superstring models\cite{Gepner}.

\subsection{D-branes}\label{sec:U1D}

The quartet of boundary states in Dirac fermion theory
should be obtained from those in affine $U(1)_2$ theory
by orbifolding,
\begin{equation}
\begin{array}{lclcl}
 \ket\SB\ssub{\NSNS+}
  &=& \kket{\SB;0}^{U(1)}+\kket{\SB;2}^{U(1)}
  &=& \frac{1}{\sqrt{2}}(\ket{\SB_0}^{U(1)}+\ket{\SB_2}^{U(1)}), \\
 \ket\SB\ssub{\NSNS-}
  &=& \kket{\SB;0}^{U(1)}-\kket{\SB;2}^{U(1)}
  &=& \frac{1}{\sqrt{2}}(\ket{\SB_1}^{U(1)}+\ket{\SB_{-1}}^{U(1)}), \\
 \ket\SB\ssub{\RARA+}
  &=& \kket{\SB;1}^{U(1)}+\kket{\SB;-1}^{U(1)}
  &=& \frac{1}{\sqrt{2}}(\ket{\SB_0}^{U(1)}-\ket{\SB_2}^{U(1)}), \\
 \ket\SB\ssub{\RARA-}
  &=& -i\kket{\SB;1}^{U(1)}+i\kket{\SB;-1}^{U(1)}
  &=& \frac{1}{\sqrt{2}}(\ket{\SB_1}^{U(1)}-\ket{\SB_{-1}}^{U(1)}).
\end{array}
\label{BU1}
\end{equation}
Here the Ishibashi and Cardy states of the affine $U(1)_2$ theory
are related by the standard formula
\begin{equation}
 \ket{\SB_S}^{U(1)}
 ~=~ \sum_s\frac{S_{Ss}}{\sqrt{S_{0s}}}\kket{\SB;s}^{U(1)},~~~~~
 S_{Ss}=\frac12e^{-i\pi Ss/2}.
\end{equation}

We would like to make sure that the boundary states
(\ref{BU1}) constructed from those in $U(1)_2$ theory
indeed satisfy the boundary conditions on Dirac fermions
$\psi^\pm(z),\tilde\psi^\pm(\bar z)$,
\begin{equation}
  (\tilde\psi^\pm_n+ i\eta\psi^\pm_{-n})\ket\SB\ssub{\SEC,\eta} ~=~0.
\label{BCpsi}
\end{equation}
We first notice that $\psi^\pm=e^{\pm iX}$ correspond to
nothing but the simple current $\psi$ in the affine $U(1)_2$ theory.
It induces invertible maps from ${\cal V}_s$ to ${\cal V}_{s+2}$
that square to the identity.
There are infinitely many such maps; for example the multiplication of
$(\psi^+_r+\psi^-_{-r})$ is easily seen to square to unity.
Pick an arbitrary such map and denote it by $\Psi$.
On closed string Hilbert space, one can thus consider operators
$\Psi,\wtilde\Psi$ acting on the right and left-moving sectors
respectively.
For a suitably chosen basis of orthonormal states, they satisfy
\begin{equation}
\begin{array}{rcl}
  \Psi(\ket{s,M}\otimes\ket{\tilde s,\tilde M}) &=&
    \ket{s+2,M}\otimes\ket{\tilde s,\tilde M}, \\
  \wtilde\Psi(\ket{s,M}\otimes\ket{\tilde s,\tilde M}) &=&
           \ket{s,M}\otimes\ket{\tilde s+2,\tilde M}
  (-i)(-)^{\frac{s-\tilde s}{2}}.
\end{array}
\label{Psidef}
\end{equation}
where $\ket{s,M}$ denotes the $M$-th state in the representation
$[s]$ of affine $U(1)_2$.
The phase factor in the second equation was chosen so that
the relations
$\Psi^2=\wtilde\Psi^2={\rm id},~\Psi\wtilde\Psi+\wtilde\Psi\Psi=0$
hold.
The boundary states defined in (\ref{BU1}) then satisfy
\begin{equation}
   (\wtilde\Psi\pm i\Psi)\ket\SB\ssub{\NSNS\pm}~=~ 0,~~~~
   (\wtilde\Psi\mp i\Psi)\ket\SB\ssub{\RARA\pm}~=~ 0,
\end{equation}
for any choice of $(\Psi,\wtilde\Psi)$
corresponding to the simple current $\psi$.
We regard this as corresponding to the boundary condition on
fermions (\ref{BCpsi}).

Let us try to extend the argument to general permutation branes
in the theory of $d$ Dirac fermions.
We wish to find a quartet of boundary states in the orbifold
$U(1)_2^{\otimes d}/(\ZZ_2)^d$ satisfying the boundary condition
\begin{equation}
  (\wtilde\Psi^{\pi(a)}\pm i\Psi^a)\ket{\SB}\ssub{\NSNS\pm}~=~ 0,~~~~
  (\wtilde\Psi^{\pi(a)}\mp i\Psi^a)\ket{\SB}\ssub{\RARA\pm}~=~ 0,
\label{Psipi}
\end{equation}
for any map $\Psi$ associated to the simple current in $U(1)_2$ model.
The operators $\Psi^a,\wtilde\Psi^a$ act on the states of the $a$-th
$U(1)_2$ theory as (\ref{Psidef}), but we also need to determine how to
pass them through the states of the first $(a-1)$ theories.
It should be determined in such a way that the maps
$\Psi_a$ and $\wtilde\Psi_a$ anticommute with one another.

Hereafter we work with the assignment that the state
$\ket{s,M}$ is Grassmann even when $s=0$ or $1$, and otherwise Grassmann odd.
This Grassmann parity has to be taken care of when the
states are permuted by operations such as $R^\pi$ (\ref{Rpi})
in constructing permutation branes.
In the following discussions, we denote by $R^\pi$ the permutation
operation with Grassmann parity taken into account, and by $R^\pi_\circ$
the one neglecting the Grassmann parity.
The two operations therefore differ by $\pm$ signs when action on
general states or operators.

To understand how the effect of Grassmann parity enters into
the definition of boundary states, let us consider the simplest
permutation brane in two Dirac fermion theory.
The boundary states are sums of states in the untwisted and
twisted sectors.
The untwisted part is given by
\begin{eqnarray}
\lefteqn{
 \frac12\left(\ket{\SB^{(12)}_S}^{U(1)^2}
                  +\ket{\SB^{(12)}_{S+2}}^{U(1)^2}\right)
} \nn\\
&=& \sum_{s=0,2}e^{-\frac{\pi iSs}{2}}R^{(12)}_\circ\kket{\SB;s,s}
~=~ \sum_{s=0,2}e^{-\frac{\pi iSs}{2}}(-)^{\frac s2}R^{(12)}\kket{\SB;s,s},
 \nn\\
\lefteqn{
 \frac12\left(\ket{\SB^{(12)}_S}^{U(1)^2}
                  -\ket{\SB^{(12)}_{S+2}}^{U(1)^2}\right)
} \nn\\
&=&\sum_{s=\pm1}e^{-\frac{\pi iSs}{2}}R^{(12)}_\circ\kket{\SB;s,s}
~=~\sum_{s=\pm1}e^{-\frac{\pi iSs}{2}}
   (-)^{\frac{s+1}{2}}R^{(12)}\kket{\SB;s,s}.
\label{B12utw}
\end{eqnarray}
These define two NSNS and two RR boundary states.
The sign factors $(-)^{\frac s2}$ or $(-)^{\frac{s+1}{2}}$ arise
from exchanging states by $R^{(12)}$.
The above states with $S=1$ can satisfy the boundary condition
(\ref{Psipi}) when suitable states in the twisted sector are added,
whereas the states with $S=0$ cannot.
The quartet of permutation boundary states is thus given by
\begin{eqnarray}
 \ket{\SB^{(12)}}\ssub{\NSNS\pm} &=& \frac12
 \sum_{h\in\cal H}\rho_\pm(h)\left(
  \ket{\SB^{(12)}_1}^h+\ket{\SB^{(12)}_{-1}}^h
 \right), \nn\\
 \ket{\SB^{(12)}}\ssub{\RARA\pm} &=& \pm\frac12
 \sum_{h\in\cal H}\rho_\mp(h)\left(
  \ket{\SB^{(12)}_1}^h-\ket{\SB^{(12)}_{-1}}^h
 \right),
\end{eqnarray}
where ${\cal H}=\ZZ_2$ is the stabilizer group generated by $\psi_1\psi_2$,
and $\rho_+~ (\rho_-)$ is the trivial (resp. nontrivial) character
of ${\cal H}$.
They can actually be rewritten in a simple form,
\begin{eqnarray}
 \ket{\SB^{(12)}}\ssub{\NSNS\pm}
 &=& R^{(12)}(\ket\SB\ssub{\NSNS\pm})^{\otimes2}
 ,\nn\\
 \ket{\SB^{(12)}}\ssub{\RARA\pm}
 &=& iR^{(12)}(\ket\SB\ssub{\RARA\pm})^{\otimes2} .
\end{eqnarray}

The overlap of the states $\ket{\SB^{(12)}}\ssub{\NSNS\pm}$
with the ordinary branes $\ket{\SB^{(1)(2)}}\ssub{\NSNS\pm}$ is
always given by the character of Ramond representation in the loop channel,
\begin{equation}
  \ssub{\NSNS,\epsilon}\bra{\SB^{(1)(2)}}e^{-\pi H/l}
  \ket{\SB^{(12)}}\ssub{\NSNS,\epsilon'}
 ~=~ \chi^{\NS-}(2i/l) ~=~ \chi^{\RA+}(il/2).
\end{equation}
Here the characters are those given in (\ref{chdf}) with $\nu$
set to zero.
This is easily seen to be consistent with the boundary condition
on supercurrent.

The construction of branes corresponding to cyclic permutations
of lengths $N\ge3$ goes in a similar way.
The boundary states are sums of the states
$\ket{\SB^\pi_S}^h,\ket{\SB^\pi_{S+2}}^h$ over the twists
$h\in(\ZZ_2)^{N-1}$ with suitable weights.
There are two distinguished weights for which the boundary conditions
on fermions are all appropriately aligned.
It also turns out that one has to choose $S=1$ for all spin structures
when the cycle has even length.

\subsection{Orientifolds}\label{sec:U1-O}

We start with constructing the orientifold of a Dirac
fermion theory via $\ZZ_2$ orbifold of $U(1)_2$ theory.
Since the choice $q\equiv0$ is somewhat unnatural, our starting
formula is (\ref{Cpss}).
Defining the basic parity $P$ by the action (\ref{pbas})
on Dirac fermions, one can consider the quartet of parity
symmetries $\epsilon^{F_R}\tilde\epsilon^{F_L}P$ defined by
the action on fields on a strip,
\begin{equation}
\begin{array}{rcl}
  (\epsilon^{F_R}\tilde\epsilon^{F_L}P )
  \psi^\pm(\sigma,t)(\epsilon^{F_R}\tilde\epsilon^{F_L}P )^{-1} &=&
  \tilde\epsilon e^{-i\pi/2}\tilde\psi^\pm(\pi-\sigma,t),\\ 
  (\epsilon^{F_R}\tilde\epsilon^{F_L}P )
  \tilde\psi^\pm(\sigma,t)(\epsilon^{F_R}\tilde\epsilon^{F_L}P )^{-1} &=&
  \epsilon e^{+i\pi/2}\psi^\pm(\pi-\sigma,t).
\end{array}
\end{equation}
The quartet of crosscap states is constructed by applying the formula
(\ref{Cpss}),
\begin{equation}
\begin{array}{lclcl}
 \ket{(-)^{F_L}P } &\equiv&
 \ket{\SC}\ssub{\NSNS+}
 &=& \frac{1}{\sqrt2}(\ket{\SC_0}^{U(1)}-i\ket{\SC_2}^{U(1)})
     e^{i\beta},\\
 \ket{(-)^{F_R}P } &\equiv&
 \ket{\SC}\ssub{\NSNS-}
 &=& \frac{1}{\sqrt2}(\ket{\SC_0}^{U(1)}+i\ket{\SC_2}^{U(1)})
     e^{-i\beta},\\
 \ket{P } &\equiv&
 \ket{\SC}\ssub{\RARA+}
 &=& \frac{1}{\sqrt2}(\ket{\SC_1}^{U(1)}+\ket{\SC_{-1}}^{U(1)}),\\
 \ket{(-)^{F}P } &\equiv&
 \ket{\SC}\ssub{\RARA-}
 &=& \frac{1}{\sqrt2}(\ket{\SC_1}^{U(1)}-\ket{\SC_{-1}}^{U(1)}),
\label{CQUA}
\end{array}
\end{equation}
where the PSS and crosscap Ishibashi states are related
by the standard formula
\begin{equation}
 \ket{\SC_S}^{U(1)} ~=~
 \sum_s\frac{P_{Ss}}{\sqrt{S_{0s}}}\kket{\SC;s}^{U(1)},~~~~~
 P_{Ss} ~=~ \frac{\delta_{S,s}^{(2)}}{\sqrt2}e^{-\frac{i\pi Ss}{4}}.
\end{equation}
The normalization was chosen to satisfy
\begin{equation}
\begin{array}{lclcl}
 \ket{\SC}\ssub{\NSNS\pm}
 &=& e^{i\pi(L_0\pm i\beta\mp\frac14)}\ket{\SB}\ssub{\NSNS\pm}
 &=& e^{\pm i\beta\mp\frac{i\pi}4}
     (\kket{\SC;0}^{U(1)}\pm i\kket{\SC;2}^{U(1)}), \\
 \ket{\SC}\ssub{\RARA+}
 &=& e^{i\pi(L_0-\frac18)}\ket{\SB}\ssub{\RARA+}
 &=& ~~~~\kket{\SC;1}^{U(1)}~+~\kket{\SC;-1}^{U(1)}, \\
 \ket{\SC}\ssub{\RARA-}
 &=& e^{i\pi(L_0-\frac18)}\ket{\SB}\ssub{\RARA-}
 &=& -i\kket{\SC;1}^{U(1)}+i\kket{\SC;-1}^{U(1)}.
\end{array}
\end{equation}
Note that these relations ensure that the crosscap condition on
fermions are automatically satisfied on the crosscap states.

The arbitrary phase $e^{\pm i\beta}$ in the definition of NSNS crosscaps
changes the action of NSNS parities on RR states uniformly by a factor
$e^{\pm 2i\beta}$.
Such a renormalization is important in constructing orientifolds in
superstring theory with real tension.
In the following we work with the choice
\[
 \beta=\frac\pi4,
\]
so that the NSNS crosscaps have real overlaps with the ground state.

We next construct the permutation crosscap in the orbifold
$U(1)^{\otimes2}/(\ZZ_2)^2$ by applying our general prescription
given in the previous section.
Our starting formula is an adaptation of the formula (\ref{CgPpi})
to the orbifold of $U(1)_2^2$ theory with $q\equiv 0$,
\begin{equation}
  \ket{\psi_1^{c_1}\psi_2^{c_2}P_S\pi} ~=~
  \sum_s\frac{S_{S\,s+2c_2}}{S_{0s}}R^{(12)}_\circ\kket{\SC;s+2c_1,s+2c_2}
  e^{i\pi\left(2h_{2c_1}+2h_s-h_{s+2c_1}-h_{s+2c_2}\right)},
\end{equation}
which has the correct overlap (\ref{BdgC}) with the diagonal brane
in twisted sectors,
\begin{equation}
 \ket{\SB^{(12)}_S}^{(\psi_1\psi_2)^c}
 ~\equiv~ \sum_s\frac{S_{Ss}}{S_{0s}}R^{(12)}_\circ
  \kket{\SB;s+2c,s}.~~~~(c=0,1)
\end{equation}
By summing over them weighted by appropriate characters of
$(\ZZ_2)^2$ we find
\begin{equation}
\begin{array}{rclcl}
 \ket{\SC^{(12)}}\ssub{\NSNS\pm} &=& \ds
  \frac12\sum_{c_1,c_2=0,1}\ket{\psi_1^{c_1}\psi_2^{c_2}P_0\pi}
  (\pm)^{c_1+c_2}
 &=& R^{(12)}(\ket{\SC}\ssub{\NSNS\pm})^{\otimes2}, \\
 \ket{\SC^{(12)}}\ssub{\RARA\pm}~~ &=& \ds
 \pm\frac12\sum_{c_1,c_2=0,1}\ket{\psi_1^{c_1}\psi_2^{c_2}P_1\pi}
  (\pm)^{c_1}(\mp)^{c_2}
 &=& i\,R^{(12)}(\ket{\SC}\ssub{\RARA\pm})^{\otimes2}.
\end{array}
\label{C12U1}
\end{equation}

\subsection{Parity action on states}\label{sec:U1-PAS}

The M\"obius strip amplitudes of $U(1)_2/\ZZ_2$ theory satisfy
\begin{equation}
\begin{array}{ccc}
 {}\ssub{\NSNS,\epsilon}\bra{\SB}q^{H_c}\ket{\SC}\ssub{\NSNS\pm}
 &=&
 {}\ssub{\NSNS\mp}\bra{\SC}q^{H_c}\ket{\SB}\ssub{\NSNS,\epsilon},\\
 {}\ssub{\RARA,\epsilon}\bra{\SB}q^{H_c}\ket{\SC}\ssub{\RARA\pm}
 &=&
 {}\ssub{\RARA\mp}\bra{\SC}q^{H_c}\ket{\SB}\ssub{\RARA,-\epsilon},
\end{array}
\end{equation}
from which one can read off the action of parity on some closed
string states.
The NSNS parities map $\ket{0\otimes0},\ket{2\otimes2}$ to themselves,
whereas the RR parities both map $\ket{\pm1\otimes\mp1}$ to
$\pm i\ket{\mp1\otimes\pm1}$.

The action of parity can also be found from the Klein bottle amplitudes.
For example, the eigenvalues of $(\pm)^FP$ on RR states are read from
\begin{equation}
  {}\ssub{\RARA\mp}\bra{\SC}e^{i\pi\tau H_c + i\pi\nu J_0}
    \ket{\SC}\ssub{\RARA\pm}
 ~=~  \pm i\chi^{\RA-}(\tau,\nu)
 ~=~  \pm  \chi^{\RA-}(\tau',\nu\tau').~~~~(\tau'=-1/\tau)
\label{KBU1RR}
\end{equation}
The parameter $\nu$ plays the role of a regulator to make
amplitudes nonzero.
In the tree channel description of Klein bottle, a nonzero $\nu$
makes the amplitude finite because $e^{i\pi\nu J_0}\ket{\SC}\ssub{\RARA+}$
satisfies the rotated crosscap condition,
\begin{equation}
  (\tilde\psi_n^\pm +ie^{i\pi n\pm 2\pi i\nu}\psi_{-n}^{\pm})
  e^{i\pi\nu J_0}\ket{\SC}\ssub{\RARA+} ~=~ 0.
\end{equation}
In the loop channel, $\nu$ twists the periodicity of the fermion
on the circle as
\begin{equation}
  \psi^\pm(\zeta e^{2\pi i})
 =-e^{\pm 2\pi i\nu}\psi^\pm(\zeta),~~~
  \tilde\psi^\pm(\bar\zeta e^{-2\pi i})
 =-e^{\mp 2\pi i\nu}\tilde\psi^\pm(\bar\zeta),
\end{equation}
so that their modes $\psi^\pm_r,\tilde\psi^\pm_r$ satisfy
$r\in \ZZ\mp\nu$.
This in particular resolves the degeneracy of RR ground states:
$\ket{\!\pm\!1\!\otimes\!\pm1}$ have $L_0=\bar L_0=\frac18\pm\frac\nu2$.
The one-loop partition sum in such a {\it spectral flowed} sector
should be described by characters with arguments $(\tau',\nu\tau')$.
From (\ref{KBU1RR}) one finds
\[
\begin{array}{rcl}
(\pm)^FP\ket{\!+\!1\otimes\!+\!1} &=&  \pm\ket{\!+\!1\otimes\!+\!1},\\
(\pm)^FP\ket{\!-\!1\otimes\!-\!1} &=&  \mp\ket{\!-\!1\otimes\!-\!1}.
\end{array}
\]

The action of parity thus found is summarized as follows,
\begin{equation}
\begin{array}{rcr}
P\ket{0\otimes0} &=&   \ket{0\otimes 0},\\
\star~~P\ket{0\otimes2} &=&  i\ket{2\otimes 0},\\
\star~~P\ket{2\otimes0} &=& -i\ket{0\otimes 2},\\
P\ket{2\otimes2} &=& - \ket{2\otimes 2},\\
\end{array}
~~~
\begin{array}{rcr}
P\ket{\!+\!1\otimes\!+\!1} &=&   \ket{\!+\!1\otimes\!+\!1},\\
P\ket{\!+\!1\otimes\!-\!1} &=&  i\ket{\!-\!1\otimes\!+\!1},\\
P\ket{\!-\!1\otimes\!+\!1} &=& -i\ket{\!+\!1\otimes\!-\!1},\\
P\ket{\!-\!1\otimes\!-\!1} &=&  -\ket{\!-\!1\otimes\!-\!1}.\\
\end{array}
\label{valP}
\end{equation}
The equations with $\star$ are not obtained from M\"obius strip nor
Klein bottle, and are chosen by hand so that $P\Psi P=\wtilde\Psi$
is satisfied.
The analysis of Klein bottles also determines the action of NSNS
parities and various fermion number operators on closed string states.
The fermion numbers $F_R,F_L$ and $F$ are implicitly defined by
the formulae (\ref{CQUA}).
The operators $(-)^{F_L,F_R}$ take $(+1)$ on both of
$\ket{0\otimes0}$ and $\ket{1\otimes1}$, and their values on other states
follow from the fact that $\Psi,\wtilde\Psi$ carry the corresponding
fermion number.
It also turns out that
\begin{equation}
(-)^{F_L+F_R+F} ~=~
\mbox{1 on NSNS states, $(-1)$ on RR states.}
\label{FRFLF}
\end{equation}

It is a simple exercise to check the action of permutation
parity on closed string states; the crosscaps $\ket{\SC^{(12)}}\ssub\SEC$
indeed correspond to the parity $P\pi,~\pi=(12)$ and its three cousins
dressed by fermion number operators.
In checking this, note that $\pi$ gives rise to $(\pm)$ signs when
permuting the states of two $U(1)_2$'s as (\ref{gPpiACT}).

\section{$N=2$ minimal model}    \label{sec:MM} 

In this section we study the permutation branes and orientifolds
in products of $N=2$ minimal models, which are basic building blocks
in Gepner's construction of worldsheet theories of superstring.
The $N=2$ minimal model at level $k$, which we denote by $M(k)$,
is known to be described as simple $N=2$ supersymmetric LG models
of a single chiral field $X$ with superpotential $X^{k+2}$ and a
$\ZZ_{k+2}$ symmetry,
\begin{equation}
\gamma~:~X\to e^{\frac{2\pi i}{k+2}}X.
\label{zgam}
\end{equation}
To construct boundary and crosscap states satisfying suitable
conditions on $N=2$ supercurrents, we start from the {\it rational}
minimal model or the coset model
\begin{equation}
 \left.\widehat{SU(2)}_k \otimes \widehat{U(1)}_2
 \right/\widehat{U(1)}_{k+2}.
\label{RMM}
\end{equation}
Since all the constituents are purely bosonic, the construction
of boundary or crosscap states of the section \ref{sec:PBCRCFT} applies
without any problem.
On the other hand, the above LG models (which we simply
call ``$N=2$ minimal model'')
are known to be described as different cosets,
\[
 M(k) ~\equiv~
 \left.\widehat{SU(2)}_k\otimes (\mbox{Dirac fermion})\right/
 \widehat{U(1)}_{k+2},
\]
so these two cosets are related by the same $\ZZ_2$-orbifolding
as was discussed in the previous section.

The representations of rational minimal model are labelled by three
integers $(l,m,s)$ specifying the properties under the affine
$SU(2)_k, U(1)_{k+2}$ and $U(1)_2$ respectively.
Namely they take values
\[
  0\le l\le k,~~~ m\simeq m+2(k+2),~~~s\simeq s+4.
\]
The labels are further restricted by $l+m+s\in2\ZZ$,
and subject to the {\it field identification}
$(l,m,s)\simeq(k-l,m+k+2,s+2)$.
Their conformal weight $h_{lms}$ is quadratic in $(l,m,s)$ modulo
integer,
\begin{equation}
  h_{lms}=\frac{l(l+2)-m^2}{4(k+2)}+\frac{s^2}{8}-\theta(l,m,s),~~~~~
  \theta(l,m,s)\in\ZZ.
\label{hlms}
\end{equation}
The functions $\theta(l,m,s)$ and $\sigma_{lms}\equiv e^{i\pi\theta(l,m,s)}$
are nothing but the improvement of conformal weight discussed at
Section \ref{sec:IDP} and equations (\ref{theta}), (\ref{gPI-CgI}).
See \cite{Brunner-H2} for their precise values.
They will be frequently used in constructing crosscap states.

The theory has a $U(1)$ R-symmetry, and the states
in the representation $(l,m,s)$ all have the same R-charge
modulo $2\ZZ$,
\begin{equation}
  J_0 ~=~ \frac{m}{k+2}-\frac{s}{2}~~~({\rm mod}~2\ZZ).
\end{equation}

The representations with $l\equiv0$ are simple currents $g_{m,s}$.
They simply shift the $m$ and $s$ quantum numbers when fused with
other representations.
The simple current $\psi\equiv g_{0,2}$ generates the group $\ZZ_2$,
and the orbifold of rational models by this $\ZZ_2$ (with $q\equiv0$)
gives the $N=2$ minimal models.
The simple current $\gamma\equiv g_{2,0}$, on the other hand,
generates the group $\ZZ_{k+2}$ which is identified with the phase
rotation of the LG field (\ref{zgam}).

Our aim in this section is to construct quartets of boundary or
crosscap states in minimal models and their products corresponding to
different spin structures.
In terms of the worldsheet $N=1$ supercurrent they are characterized
by
\begin{equation}
\begin{array}{rcl}
  (\tilde G_r\mp i G_{-r})\ket\SB\ssub{\SEC\pm} &=& 0,\\
  (\tilde G_r\mp ie^{i\pi r} G_{-r})\ket\SC\ssub{\SEC\pm} &=& 0,
\end{array}
~~~\left\{\begin{array}{ll}
   r\in\ZZ+\frac12 & (\SEC=\NSNS) \\
   r\in\ZZ         & (\SEC=\RARA) \end{array} \right.
\label{Gbc}
\end{equation}
The signs are flipped when the states are multiplied by the
operators $(-)^{F_R}$ or $(-)^{F_L}$.

In $N=2$ SCFTs, one can instead use the operators
$e^{i\pi J_0}$ or $e^{i\pi \tilde J_0}$ to flip the sign, where
$J_0,\tilde J_0$ are the right, left-moving R-charges.
Moreover, the NSNS and RR states are related by spectral flow.
Let us denote by $U$ a combination of left-right spectral flows
acting on the generators of two $N=2$ SCAs as
\begin{equation}
\begin{array}{l}
 UJ_n U^{-1} = J_n +\frac{\hat c}{2}\delta_{n,0}, \\
 U\tilde J_n U^{-1} = \tilde J_n -\frac{\hat c}{2}\delta_{n,0},
\end{array}
\begin{array}{l}
 UG^\pm_n U^{-1} = G^\pm_{n\pm1/2}, \\
 U\tilde G^\pm_n U^{-1} = \tilde G^\pm_{n\mp1/2},
\end{array}
\begin{array}{l}
 UL_n U^{-1} = 
       L_n +\frac12J_n +\frac{\hat c}{8}\delta_{n,0}, \\
 U\tilde L_n U^{-1} =
       \tilde L_n -\frac12\tilde J_n +\frac{\hat c}{8}\delta_{n,0}.
\end{array}
\end{equation}
$U$ maps a closed string state in
${\cal V}_{l,m,s}\otimes{\cal V}_{l,\tilde m,\tilde s}$ to a state in
${\cal V}_{l,m+1,s+1}\otimes{\cal V}_{l,\tilde m-1,\tilde s-1}$.
It is easy to see that $U$ or $Ue^{-i\pi J_0/2}$ map the NSNS solutions
of boundary or crosscap conditions to RR solutions.
We assign a phase $\varphi$ to each of the quartet states
as follows,
\begin{eqnarray}
  U\ket\SB\ssub{\NSNS+} &=&
   \ket\SB\ssub{\RARA+}e^{-i\pi\varphi(\SB)}, \nn\\
  Ue^{-i\pi J_0/2}\ket\SC\ssub{\NSNS+} &=&
   \ket\SC\ssub{\RARA+}e^{-i\pi\varphi(\SC)}.
\label{SUSYphase}
\end{eqnarray}
In type II superstring theory, the phase $\varphi$ of D-branes and
orientifolds characterizes the unbroken spacetime ${\cal N}=1$
supersymmetry.

\subsection{Boundary and crosscap states}\label{sec:MMBCS}

Boundary or crosscap states $\ket{\SB_{L,M,S}},~\ket{\SC_{0,M,S}}$
in rational minimal models are constructed from Ishibashi states
$\kket{\SB;l,m,s},\kket{\SC;l,m,s}$ in the standard way.
The $\ZZ_2$-orbifolding reorganizes them into solutions of suitable
boundary or crosscap conditions on supercurrent.
For boundary states, we define the Ishibashi states solving the
boundary conditions on supercurrents as follows,
\begin{equation}
\begin{array}{lcl}
 \kket{\SB;l,m}\ssub{\NSNS\pm} &=& \kket{\SB;l,m,0}\pm\kket{\SB;l,m,2},\nn\\
 \kket{\SB;l,m}\ssub{\RARA+} &=& \kket{\SB;l,m,1}+\kket{\SB;l,m,-1},\nn\\
 \kket{\SB;l,m}\ssub{\RARA-} &=& -i\kket{\SB;l,m,1}+i\kket{\SB;l,m,-1},
\end{array}
\label{BIc}
\end{equation}
whereas for crosscaps the appropriate combinations of Ishibashi states
are
\begin{equation}
\begin{array}{lcl}
\kket{\SC;l,m}\ssub{\NSNS\pm} &=&
 e^{\pi i(L_0-h_{lm0})}\sigma_{lm0}\kket{\SB;l,m}\ssub{\NSNS\pm}, \\
\kket{\SC;l,m}\ssub{\RARA\pm} &=&
 e^{\pi i(L_0-h_{lm1})}\sigma_{lm1}\kket{\SB;l,m}\ssub{\RARA\pm},
\end{array}
\end{equation}
where $\sigma_{lms}\equiv e^{i\pi\theta(l,m,s)}$ is given at (\ref{hlms}),
or more explicitly
\begin{equation}
\begin{array}{rcl}
\kket{\SC;l,m}\ssub{\NSNS\pm} &=&
~~~~\sigma_{lm0}\kket{\SC;l,m,0}\pm i\sigma_{lm2}\kket{\SC;l,m,2},\\
\kket{\SC;l,m}\ssub{\RARA+} &=&
~~~~\sigma_{lm1}\kket{\SC;l,m,1}+~~\sigma_{lm-1}\kket{\SC;l,m,-1}, \\
~~\kket{\SC;l,m}\ssub{\RARA-} &=&
-i\sigma_{lm1}\kket{\SC;l,m,1}+i\sigma_{lm-1}\kket{\SC;l,m,-1}.
\end{array}
\end{equation}
The D-branes and orientifolds in $N=2$ minimal model
are given by a sum over $\ZZ_2$-orbit of rational boundaries or
crosscaps \cite{Brunner-H2, Brunner-HHW}.
In terms of the above Ishibashi states they read
\begin{eqnarray}
  \ket{\SB_{L,M}}\ssub\SEC &=& \frac12
  \sum_{l,m}\frac{S_{LM}^{~lm}}{\sqrt{S_{00}^{lm}}}
  \kket{\SB;l,m}\ssub\SEC,
  \nn\\
 \ket{\SC_M}\ssub\SEC &=&
 \frac{\beta\ssub{M,\SEC}}2 \sum_{(l,m)}
 \frac{P_{0M}^{~lm}}{\sqrt{S_{00}^{~lm}}}
 \kket{\SC;l,m}\ssub\SEC
 ~=~
 \frac12 \sum_{(l,m)}
 \frac{P_{k,M+k+2}^{~lm}}{\sqrt{S_{00}^{~lm}}}
 \kket{\SC;l,m}\ssub\SEC,
  \nn\\ &&
 \beta\ssub{M,\NSNS\pm}=\mp i(-)^{\frac M2},~~
 \beta\ssub{M,\RARA\pm}=(-)^{\frac {M\pm1}2}.
\label{BCmin}
\end{eqnarray}
Here $(l,m)$ runs over integers $0\le l\le k,~ m\sim m+2k+4$.
The $S$ and $P$ matrices are twice the product of those of
$SU(2)_k$ and $(U(1)_{k+2})^\ast$ theories,
\begin{equation}
\begin{array}{ll}
S_{ll'} = \sqrt{\frac{2}{k+2}}\sin\pi\frac{(l+1)(l'+1)}{k+2}, &
S_{mm'} = \frac{1}{\sqrt{2k+4}}e^{\frac{i\pi mm'}{k+2}}, \\
P_{ll'} = \sqrt{\frac{4}{k+2}}\delta^{(2)}_{k+l+l'}
          \sin\pi\frac{(l+1)(l'+1)}{2k+4}, &
P_{mm'} = \frac{\delta^{(2)}_{k+m+m'}}{\sqrt{k+2}}
          e^{\frac{i\pi mm'}{2k+4}}.
\end{array}
\end{equation}
The coefficients $\beta\ssub{M,\SEC}$ are introduced mainly for
later convenience, but it also has some physical significances.
For one thing, they make the states $\ket{\SC_M}\ssub{\NSNS\pm}$
periodic and $\ket{\SC_M}\ssub{\RARA\pm}$ anti-periodic under $M\to M+2k+4$,
so that the shift of $M$ by $2k+4$ is regarded as the orientation flip.
It also preserves the action of simple current $\gamma$ on
crosscap states, so we have
\begin{equation}
\gamma\ket{\SB_{L,M}}\ssub\SEC ~=~ \ket{\SB_{L,M+2}}\ssub\SEC,~~~~~
\gamma\ket{\SC_{M}}\ssub\SEC ~=~ \ket{\SC_{M+4}}\ssub\SEC.
\end{equation}
The spectral flow $U$ for $N=2$ minimal models is identified
with the fusion with the simple current $g_{1,1}$.
The boundary and crosscap states of minimal model are then shown to
form the following quartets,
\begin{eqnarray}
\lefteqn{
  (1+e^{i\pi J_0})(1+e^{i\pi\varphi(\SB_{L,M})} U)
   \ket{\SB_{L,M}}\ssub{\NSNS+}
}\nn\\  &=&
     \ket{\SB_{L,M}}\ssub{\NSNS+}
    +\ket{\SB_{L,M+1}}\ssub{\NSNS-}
    +\ket{\SB_{L,M}}\ssub{\RARA+}
    +\ket{\SB_{L,M+1}}\ssub{\RARA-}  ,
\nn\\
\lefteqn{
  (1+e^{i\pi J_0})(1+e^{i\pi\varphi(\SC_{M})} Ue^{-\frac{i\pi J_0}{2}})
   \ket{\SC_M}\ssub{\NSNS+}
}\nn\\  &=&
     \ket{\SC_{M}}\ssub{\NSNS+}
    +\ket{\SC_{M+2}}\ssub{\NSNS-}
    +\ket{\SC_{M-1}}\ssub{\RARA+}
    +\ket{\SC_{M+1}}\ssub{\RARA-}  ,
\end{eqnarray}
with $\varphi(\SB_{L,M})=\frac{M}{k+2}$,~
$\varphi(\SC_M)=\frac{M-1}{2k+4}+\frac12$.

\subsubsection{Boundary states in $g_{k+2,2}$ twisted sector}\label{sec:MMBSTS}

When $k$ is even, the boundary states with $L=k/2$ are fixed by
$g_{k+2,2}\equiv\eta$.
We define the boundary states sitting in $\eta$-twisted sector
\cite{Brunner-HHW},
\begin{eqnarray}
  \ket{\wtilde\SB_{k/2,M,S}}^\eta &=& \frac12\sum_{(ms)}
  \frac{\wtilde S_{k/2\,MS}^{~k/2\,ms}}{\sqrt{S_{000}^{~k/2\,ms}}}
  \kket{\SB;\tfrac k2,m,s}^\eta,
 \nn\\
 \widetilde S_{k/2\,MS}^{~~k/2\,ms} &=&
 2S_{Mm}S_{Ss}e^{-\frac{i\pi}{2}(M-S+m-s)}.
\label{defBtw}
\end{eqnarray}
The boundary conditions on supercurrent are solved
by the following linear combinations of Ishibashi states,
\begin{equation}
\begin{array}{rcl}
 \kket{\tfrac k2,m}\ssub{\NSNS+}^\eta &=&
  \kket{\tfrac k2,m,0}^\eta-\kket{\tfrac k2,m,2}^\eta,\\
 \kket{\tfrac k2,m}\ssub{\NSNS-}^\eta &=&
  i\kket{\tfrac k2,m,0}^\eta+i\kket{\tfrac k2,m,2}^\eta,\\
 \kket{\tfrac k2,m}\ssub{\RARA+}^\eta &=&
  i\kket{\tfrac k2,m,1}^\eta-i\kket{\tfrac k2,m,-1}^\eta,\\
 \kket{\tfrac k2,m}\ssub{\RARA-}^\eta &=&
  i\kket{\tfrac k2,m,1}^\eta+i\kket{\tfrac k2,m,-1}^\eta.
\end{array}
\end{equation}
Note the sign difference in taking linear combinations
as compared to (\ref{BIc}) due to the difference in Grassmann parity.
The corresponding quartet of boundary states is given by
\begin{equation}
  \ket{\SB_{k/2,M}}^\eta\ssub\SEC ~=~
  \frac12
  \sum_m\frac{\widetilde S_{k/2\,M}^{~~~k/2\,m}}{\sqrt{S_{00}^{k/2\,m}}}
  \kket{\SB;\tfrac k2,m}^\eta\ssub\SEC,
\end{equation}
where
\begin{equation}
  \widetilde S_{k/2\,M}^{~~k/2\,m} ~=~
  2S_{Mm}e^{-\frac{i\pi}{2}(M+m)}.
\end{equation}
After the orbifold by $\ZZ_2$ is taken, there is no distinction in
labelling the twisted sector by $g_{k+2,2}$ or $g_{k+2,0}$.
We therefore use the symbol $\eta$ for the simple current $g_{k+2,0}$
in what follows.

\subsubsection{Tension and Charge}\label{sec:MMTC}

The tension and RR charges of D-brane and orientifolds are
given by the overlaps of the boundary or crosscap states with
the NSNS and RR vacua.
We denote the NSNS chiral primary states and RR ground states as,
\begin{equation}
\ket{l\ssub\NS} ~=~ \ket{(l,l,0)  \otimes(l,-l,0)},~~~
\ket{l\ssub\RA} ~=~ \ket{(l,l+1,1)\otimes(l,-l-1,-1)}.
\end{equation}
The overlaps of these states with boundary or crosscap states
read\cite{Brunner-HHW}
\begin{eqnarray}
 \langle l\ssub\RA\ket{\SB_{L,M}}\ssub{\RARA+}   &=&
 e^{\frac{i\pi M}{k+2}}\cdot ~\!
 \langle l\ssub\NS\ket{\SB_{L,M}}\ssub{\NSNS+} ~=~
 \frac{e^{\frac{i\pi M(l+1)}{k+2}}\sin\frac{\pi(L+1)(l+1)}{k+2}}
      {\sqrt{\frac{k+2}{2}\sin\frac{\pi(l+1)}{k+2}}},
 \nn\\
 \langle l\ssub\RA\ket{\SC_{2m-1}}\ssub{\RARA+}   &=&
 e^{\frac{i\pi (2m-l-1)}{2k+4}+\frac{i\pi}{2}}\cdot ~\!
 \langle l\ssub\NS\ket{\SC_{2m}}\ssub{\NSNS+}
 \nn\\ &=&
\left\{\begin{array}{l}
 \ds\langle l\ssub\RA
    \ket{\SB_{\frac{k}{2},\frac{2m+k+1-(-)^m}{2}}}\ssub{\RARA+}
 ~~ (k~~{\rm even}),\\
 \ds\langle l\ssub\RA
    \ket{\SB_{\frac{k+(-)^m}{2},\frac{2m+k+1}{2}}}\ssub{\RARA+}
 ~~ (k~~{\rm odd}).
\end{array}\right.
\label{RRMM1}
\end{eqnarray}
Tensions are therefore given by
\begin{eqnarray}
 \vev{0\ssub\NS|\SB_{L,M}}\ssub{\NSNS\pm} &=&
 T_0\sin\tfrac{\pi(L+1)}{k+2}, \nn\\
 \vev{0\ssub\NS|\SC_{2m}}\ssub{\NSNS\pm} &=&
\left\{\begin{array}{ll}
 T_0 e^{-\frac{i\pi(-)^m}{2k+4}}, & (k~{\rm even})\\
 T_0 \cos\frac{\pi}{2k+4}, & (k~{\rm odd})
\end{array}
\right.
\end{eqnarray}
where $T_0= \left(\ts\frac{k+2}{2}\sin\frac{\pi}{k+2}\right)^{-\frac12}$.

\subsubsection{Parity action on closed string states}\label{sec:MMPACSS}

Klein bottle amplitude gives a lot of information on the action
of parity on closed string states in minimal model or its orbifolds.
We take an arbitrary orbifold group $\Gamma\subset\ZZ_{k+2}$
and consider orientifolds in the orbifold,
\begin{equation}
  \ket{\SC_{M,r}}\ssub\SEC ~=~
  \frac{1}{\sqrt{|\Gamma|}}\sum_{\gamma^\nu\in\Gamma}
  \ket{\SC_{M+2\nu}}\ssub\SEC\exp\left(-\tfrac{2\pi i\nu r}{k+2}\right).
\end{equation}
The parameter $r$ labels the dressing by quantum symmetry that
multiplies phases to different twisted sectors, and $2r$ has to be
even for NSNS states and odd for RR states because of the
(anti-)periodicity of the crosscap states in $M$.
The parity $P_{M,r}$ corresponding to $\ket{\SC_{M,r}}\ssub{\RARA+}$,
as well as its cousins, are in general all non-involutive
and square to some quantum symmetry.
The action of $P_{M,r}$ on closed string states has to be of the form
\begin{equation}
  P_{M,r}\ket{(l,m,s)\otimes(l,\tilde m,\tilde s)}~=~
         \ket{(l,\tilde m,\tilde s)\otimes(l,m,s)}
         \exp\left(\tfrac{i\pi(\tilde m+m)2r+i\pi(\tilde m-m)M}{2k+4}\right)
	 p_{s,\tilde s}.
\end{equation}
The Klein bottle amplitudes show that this is indeed the case, and
moreover $p_{s,\tilde s}$ are given by
\begin{equation}
\begin{array}{lclcr}
p_{0,0} &=& p_{1,1}   &=& 1,\\
p_{2,2} &=& p_{-1,-1} &=& -1,
\end{array}
~~~
\begin{array}{lclcr}
p_{0,2} &=& p_{1,-1} &=& -i,\\
p_{2,0} &=& p_{-1,1} &=& i.
\end{array}
\end{equation}
The other three crosscaps with $Y=\NSNS\pm,~\RARA-$ are corresponding to
the parity $P_{M,r}$ combined with the fermion numbers
$(-)^{F_L,F_R,F}$ satisfying (\ref{FRFLF}).
Comparisons of various Klein bottle amplitudes determine the values of
these fermion numbers; the states with $s=\tilde s=0$ or $1$ have
$(-)^{F_R}=(-)^{F_L}=1$, and their values on other states follow
from the obvious rules.

Using these results one can derive the action of parity on boundary
states.
For those in the untwisted sector we have
\begin{equation}
\begin{array}{rcl}
 (-)^{F_L}P_{\bar M,r}\ket{\SB_{L,M}}\ssub{\NSNS\pm}
  &=& ~~~\ket{\SB_{L,\bar M-M}}\ssub{\NSNS\pm}, \\
          P_{\bar M,r}\ket{\SB_{L,M}}\ssub{\RARA\pm}~~
  &=& -\ket{\SB_{L,\bar M-M}}\ssub{\RARA\mp}.
\end{array}
\end{equation}
This agrees with the transformation law obtained
from M\"obius strip amplitudes (\ref{MS}).
The boundary states in $\eta$-twisted sector are transformed as follows:
\begin{equation}
\begin{array}{rcl}
 (-)^{F_L}P_{\bar M,r}\ket{\SB_{k/2,M}}\ssub{\NSNS\pm}^\eta
  &=& \mp ie^{i\pi r}\ket{\SB_{k/2,\bar M-M}}\ssub{\NSNS\pm}^\eta, \\
          P_{\bar M,r}\ket{\SB_{k/2,M}}\ssub{\RARA\pm}^\eta~~
  &=& ~~~~ e^{i\pi r}\ket{\SB_{k/2,\bar M-M}}\ssub{\RARA\mp}^\eta.
\end{array}
\label{Pacttw}
\end{equation}

\subsection{Permutation branes}\label{sec:MMPB}

It is straightforward to construct permutation branes in the
tensor products of $N$ minimal models.
We start by the permutation boundary states in the product
of $N$ rational minimal model and take $(\ZZ_2)^N$-orbifold.
We give the expression for those corresponding to the cyclic
permutation of length $N$, i.e. $\pi=(12\cdots N)$.
\begin{eqnarray}
  \ket{\SB^{(12\cdots N)}_{L,M}}\ssub\SEC &\equiv&
  \frac{\alpha\ssub\SEC}{2}
  \sum_{l,m}\frac{S_{LM}^{~lm}}{(S_{00}^{~lm})^{N/2}}
  R^{(12\cdots N)}\kket{\SB;(l,m)^{\otimes N}}\ssub\SEC,
\nn\\&&
\alpha\ssub{\NSNS\pm}=1,~~~~ \alpha\ssub{\RARA\pm}=i^{N-1}.
\label{BP}
\end{eqnarray}
Recalling the case of $U(1)_2$ where we have to sum over rational
boundary states of odd $S$-labels when $N$ is even,
we find that the labels $(L,M)$ obey
\[
\begin{array}{lcll}
 (N~~{\rm odd}) &\Longrightarrow&
 \begin{array}{l}
 L+M=({\rm even}) \\
 L+M=({\rm odd }) 
 \end{array}
 &
 \begin{array}{l}
 \mbox{for $\NSNS+,\RARA+$ states,}\\
 \mbox{for $\NSNS-,\RARA-$ states,}
 \end{array}
 \\
 (N~~{\rm even}) &\Longrightarrow&
 ~~L+M=({\rm odd })&
 ~~\mbox{for all states,}\\
\end{array}
\]
The simple current $\otimes_a\gamma_a^{\nu_a}$ shifts their $M$-label
by $2\sum_a\nu_a$.
In particular, the simple currents with $\sum_a\nu_a=0$ mod $(k+2)$
fix the boundary states.
The states $\ket{\SB^{(1\cdots N)}_{L,M}}\ssub{\SEC+},
\ket{\SB^{(1\cdots N)}_{L,M+N}}\ssub{\SEC-}$ form a quartet
with the phase $\varphi=\frac{M}{k+2}+\frac{1-N}{2}$.

The RR charges of permutation branes are given by
the overlaps with the states $\ket{l^{\otimes N}\ssub{\RA}}$,
\begin{equation}
     \vev{l^{\otimes N}\ssub\RA|\SB^{(12\cdots N)}_{L,M}}\ssub{\RARA+}
 ~=~ \frac{\sin\frac{\pi(L+1)(l+1)}{k+2}
           e^{\frac{i\pi M(l+1)}{k+2}+\frac{i\pi(1-N)}{2}}}
          {(\tfrac{k+2}{2})^{1-\frac{N}{2}}(\sin\frac{\pi(l+1)}{k+2})^{N/2}}.
\end{equation}
The tension is given by
\begin{equation}
 \vev{0^{\otimes N}\ssub\NS|\SB^{(12\cdots N)}_{L,M}}\ssub{\NSNS+}
 ~=~ (\tfrac{k+2}{2})^{\frac{N}{2}-1}(\sin\tfrac{\pi}{k+2})^{-\frac N2}
     \sin\tfrac{\pi(L+1)}{k+2}.
\end{equation}

\subsection{Permutation orientifolds}\label{sec:MMPO}

Here we construct the permutation crosscaps for tensor products of
two identical minimal models through the $(\ZZ_2)^2$-orbifold
procedure.
Denoting by $\psi_{1,2}$ the simple currents $g_{0,2}$ in the
two copies of minimal model, we sum over the following crosscaps
(with $\pi=(1\,2)$)
\begin{eqnarray}
 \ket{\psi_1^{c_1}\psi_2^{c_2}P_{M,S}\pi}
 &=& \frac12\sum_{l,m,s}\frac{S_{0MS}^{~lm\,s+2c_2}}{S_{000}^{~lms}}
     R^\pi_\circ\kket{\SC;(l,m,s+2c_1),(l,m,s+2c_2)}
 \nn\\ && ~~~~ \times \exp i\pi\left\{
       2h_{0,0,2c_1}+2h_{l,m,s}-h_{l,m,s+2c_1}-h_{l,m,s+2c_2} \right\}
\end{eqnarray}
with appropriate weight to obtain
\begin{equation}
\begin{array}{lll}
 \ket{\SC^\pi_M}\ssub{\NSNS\pm} &\hskip-3mm= \npm\ds
 \frac12\sum_{c_i=0,1}
 \ket{\psi_1^{c_1}\psi_2^{c_2}P_{M,0}\pi}(\pm)^{c_1+c_2}
 &= \ds\frac12\sum_{l,m}\frac{S_{0M}^{~lm}}{S_{00}^{~lm}}
     R^\pi\kket{\SC;(l,m)^{\otimes2}}\ssub{\NSNS\pm}
 \nn\\
 \ket{\SC^\pi_M}\ssub{\RARA\pm} &\hskip-3mm= \ds
 \pm\frac12\sum_{c_i=0,1}
 \ket{\psi_1^{c_1}\psi_2^{c_2}P_{M,1}\pi}(\pm)^{c_1}(\mp)^{c_2}
 &= \ds \frac i2\sum_{l,m}\frac{S_{0M}^{~lm}}{S_{00}^{~lm}}
     R^\pi\kket{\SC;(l,m)^{\otimes2}}\ssub{\RARA\pm}.
\end{array}
\end{equation}
Note that $M$ is even for NSNS states and odd for RR states.
One can furthermore consider the parities
$\gamma_1^{\nu_1}\gamma_2^{\nu_2}P_{M,S}$ which are non-involutive
for general $\nu_{1,2}$.
The corresponding crosscap states are obtained by applying
the formula (\ref{CgPpi}),
\begin{equation}
 \ket{\SC^\pi_{M+2\nu_1,M+2\nu_2}}\ssub\SEC
 ~=~ \frac{\alpha\ssub\SEC}{2}
     \sum_{l,m}\frac{S_{0,M+\nu_1+\nu_2}^{~l,m+\nu_1+\nu_2}}
     {S_{00}^{~lm}}
     R^\pi\kket{\SC;(l,m+2\nu_1),(l,m+2\nu_2)}\ssub\SEC.
\label{Cmin2}
\end{equation}
Here $\alpha\ssub{\NSNS\pm}=1,~\alpha\ssub{\RARA\pm}=i$.

We thus constructed the crosscap states $\ket{\SC^{(12)}_{M_1,M_2}}$
for different spin structures; the labels $M_{1,2}$ are both even
and periodic under $(2k+4)$-shift for NSNS crosscaps, while they are
both odd and anti-periodic for RR crosscaps.
The simple current $\gamma_1^{\nu_1}\gamma_2^{\nu_2}$ shifts
both of the labels $M_1,M_2$ by $2\nu_1+2\nu_2$.
They are organized into quartets satisfying
\begin{eqnarray}
\lefteqn{
 (1+e^{i\pi J_0})(1+e^{i\pi\varphi}Ue^{-i\pi J_0/2})
 \ket{\SC^{(12)}_{M_1,M_2}}\ssub{\NSNS+}
} \nn\\ &=&
 \ket{\SC^{(12)}_{M_1,M_2}}\ssub{\NSNS+}
+\ket{\SC^{(12)}_{M_1+2,M_2+2}}\ssub{\NSNS-}
+\ket{\SC^{(12)}_{M_1-1,M_2-1}}\ssub{\RARA+}
+\ket{\SC^{(12)}_{M_1+1,M_2+1}}\ssub{\RARA-},~~
\end{eqnarray}
with $\varphi = \frac{M_1+M_2-2}{2k+4}-\frac 12$.
The RR charges and tension are given by
\begin{eqnarray}
     \vev{l\ssub{\RA}^{\otimes2}|\SC^{(12)}_{M-1,M-1}}\ssub{\RARA+}
 ~=~ \vev{l\ssub{\RA}^{\otimes2}|\SB^{(12)}_{0,M-1}}\ssub{\RARA+}
 &=& e^{\frac{i\pi(M-1)(l+1)}{k+2}-\frac{i\pi}{2}},
\nn\\
     \vev{0\ssub{\NS}^{\otimes2}|\SC^{(12)}_{M,M}}\ssub{\NSNS+}
 ~=~ \vev{0\ssub{\NS}^{\otimes2}|\SB^{(12)}_{0,M}}\ssub{\NSNS+}
 &=& 1.
\label{RRMM2}
\end{eqnarray}
The permutation crosscaps with $M_1\ne M_2$ are tensionless, but they
have nonzero overlaps with RR vacua sitting in twisted sectors.
Let us define
\begin{equation}
  \ket{l\ssub\RA^{\rm tw}}~\equiv~
  \ket{(l,l+1,1)\otimes(l,l+1,1)}.
\end{equation}
Then one finds
\begin{equation}
  \vev{(k-l)\ssub\RA^{\rm tw}\otimes l\ssub\RA^{\rm tw}|
       \SC^{(12)}_{M,M+2l+2}}\ssub{\RARA+}
 ~=~ -i.
\end{equation}

\section{Gepner Models}\label{sec:GM} 

We apply the results of the preceding sections to the construction
of permutation D-branes and orientifolds in Gepner models, which are
type II superstring theories defined from orbifolds of products of
$N=2$ minimal models and affine $U(1)_2$ models \cite{Gepner}.

Gepner's original construction of the models starts with a
product of $r$ rational minimal models and $d$ copies of
affine $U(1)_2$ models, and then takes its orbifold by
a group of simple currents.
A subgroup $\tilde\Gamma_{\rm GSO}\simeq(\ZZ_2)^{r+d-1}$ of this orbifold is
formed by even monomials of the simple currents $\psi_1,\cdots,\psi_{r+d}$
discussed in previous sections that shift the $s$ quantum numbers by two.
As we have reviewed in detail in the previous sections, this is
equivalent to taking the product of $r$ $N=2$ minimal models and $d$
Dirac fermions and then summing over spin structures.
For constructing D-branes and orientifolds, this just amounts
to taking the product of boundary or crosscap states with the sector
index $\SEC$ aligned.
In this way one can focus on the $r$ minimal models describing
the internal manifold separately from the part describing
the noncompact spacetime.

It only remains to explain the ``rest'' of the Gepner's orbifold group.
Gepner models describe the CFT on certain Calabi-Yau $D$-folds
at special points in the moduli space in terms of orbifolds
of products of $r$ minimal models.
The central charges of constituent minimal models therefore add up to $3D$,
\begin{equation}
 \sum_{a=1}^r\frac{3k_a}{k_a+2} = 3D.
\end{equation}
We also assume without losing generality that
\begin{equation}
r-D ~=~ {\rm even},
\label{r-d ev}
\end{equation}
since we can add minimal models with $k=0$.
The product of minimal models is orbifolded by
$\Gamma=\ZZ_H~(H\equiv{\rm lcm}(k_a+2))$ generated
by $\gammaA\equiv\prod_{a=1}^r\gamma_a$ to ensure the
integrality of R-charge.
The orbifold is taken according to the standard simple current
prescription of Section \ref{sec:SCO} with
\[
 q(\gamma_a,\gamma_b) = \frac{\delta_{ab}}{k_a+2}.
\]

Gepner model $\otimes_{a=1}^rM(k_a)/\Gamma$ is mirror to
a different orbifold $\otimes_{a=1}^rM(k_a)/\Gamma_{\rm mir}$, where
\begin{equation}
  \Gamma_{\rm mir} ~\equiv~ \left\{{\ts\prod_a}\gamma_a^{m_a};~
                    \sum_a\frac{m_a}{k_a+2}\in\ZZ\right\}.
\label{Gmir}
\end{equation}
In particular, B-branes (B-type orientifolds) in the original Gepner
model are mirror of the A-branes (A-type orientifolds) in the
mirror Gepner model and vise versa.

\paragraph{Examples.}
We denote various Gepner models by the set of integers $(k_a+2)$.
Two main examples of Gepner models we discuss in this paper
are the model $(55555)$ corresponding to a quintic hypersurface
in $\CC\PP^4$, and $(88444)$ corresponding to an octic hypersurface
in weighted projective space ${\mathbb W}\CC\PP^4_{1,1,2,2,2}$.
These models have been extensively studied because of small $h_{1,1}$
of the corresponding Calabi-Yau spaces.

\paragraph{}
We describe the D-branes or orientifolds in superstring theory by
suitable linear combinations of quartet states of the worldsheet
CFT,
\begin{equation}
\begin{array}{rcl}
  2\ket\SB &=&
   \ket\SB\ssub{\NSNS+}
  -\ket\SB\ssub{\NSNS-}
  +\ket\SB\ssub{\RARA+}
  -\ket\SB\ssub{\RARA-}, \\
  2\ket\SC &=&
 -i\ket\SC\ssub{\NSNS+}
 +i\ket\SC\ssub{\NSNS-}
  +\ket\SC\ssub{\RARA+}
  -\ket\SC\ssub{\RARA-}.
\end{array}
\end{equation}
Here the quartet states are given by the products of the states
from the internal and spacetime CFTs,
\begin{equation}
  \ket\SB\ssub\SEC ~=~
  \ket\SB\ssub\SEC^{\rm int}\otimes
  \ket\SB\ssub\SEC^{\rm st},~~~~
  \ket\SC\ssub\SEC ~=~
  \ket\SC\ssub\SEC^{\rm int}\otimes
  \ket\SC\ssub\SEC^{\rm st}.
\end{equation}
The spacetime parts $\ket\SB\ssub\SEC^{\rm st},\ket\SC\ssub\SEC^{\rm st}$
contain the fields for $\RR^{2d+2}$ as well as ghosts
\cite{Polchinski-C, Recknagel-S},
and are normalized to produce consistent one-loop amplitudes.
In particular, they satisfy
\[
 (-)^{F_L}\ket\SB\ssub{\SEC+} ~=~ \ket\SB\ssub{\SEC-},~~~~
 (-)^{F_L}\ket\SC\ssub{\SEC+} ~=~ \ket\SC\ssub{\SEC-},
\]
\begin{equation}
 \ket\SC\ssub\SEC ~=~
 2^{d+1}\exp i\pi(L_0-h\ssub\SEC^{\rm st})\ket\SB\ssub\SEC.~~~~
 (h^{\rm st}\ssub\NS=-\tfrac12,~~
  h^{\rm st}\ssub\RA=\tfrac{d-4}{8})
\end{equation}
The normalization of the internal parts are fixed from the
integrality of various one-loop amplitudes.
Alternatively, it is determined by requiring that the NSNS states
$\ket\SB\ssub{\NSNS\pm}^{\rm int},\,
 \ket\SC\ssub{\NSNS\pm}^{\rm int}$ have real overlaps
with the ground state of the internal CFT.
Such overlaps appear as coefficients of the dilaton tadpole
and are regarded as the tensions of D-branes or orientifolds.
The overlaps with various RR ground states measure the RR charges.
The sign flip of the RR part of $\ket\SB$ or $\ket\SC$ therefore
gives anti-D-branes or anti-orientifolds.

One can compute cylinder, M\"obius strip and Klein bottle amplitudes
between various D-branes and orientifolds as overlaps of the states
$\ket\SB$ and $\ket\SC$.
In doing this, remember that the simple dagger of a ket state for a D-brane
or orientifold gives a bra state for anti-D-brane or anti-orientifold.

\paragraph{Tadpole cancellation.}

Consistent configurations of D-branes $\SB_i$ and orientifold $\SC$ in
superstring theory must be free of RR
tadpoles\cite{Polchinski-C, Callan-LNY},
namely, the tadpole state
\begin{equation}
 \ket\ST ~=~ \ket\SC + \sum_i\ket{\SB_i},
\end{equation}
must not have any overlaps with massless RR scalar states.
The non-vanishing tadpoles of massless NSNS scalars do not lead
to inconsistency\cite{Fischler-S}.
However,  the absence of RR tadpoles automatically guarantees
that NSNS tadpoles also vanish if the configuration of D-branes
and orientifolds preserves a spacetime supersymmetry.
The spacetime ${\cal N}=2$ supersymmetry is related to worldsheet
spectral flows in the left and right-moving sectors, and the phase
$\varphi$ (\ref{SUSYphase}) determines the $N=1$ supersymmetry unbroken by
the branes or orientifolds.
So $\ket{\ST}$ preserves spacetime supersymmetry if all the boundary
and crosscap states in $\ket\ST$ are labelled
by one and the same phase $\varphi$.

The absence of NSNS tadpoles for supersymmetric tadpole-free
configurations is shown by noticing that the massless NSNS
and RR states are related to the chiral primary and RR ground
states in the internal CFT, and are therefore paired up by
spectral flow.
For each of such pairs we can show
\begin{eqnarray}
\frac{\vev{l\ssub\RA|\SB_i}}
     {\vev{l\ssub\NS|\SB_i}} &=&
\exp i\pi\varphi, \nn\\
\frac{\vev{l\ssub\RA|\SC}}
     {\vev{l\ssub\NS|\SC}} &=&
i\exp i\pi\left[
 \varphi-\tfrac12J_0^{\rm int}(l\ssub\NS)
 -\{L_0^{\rm st}(l\ssub\NS)-(h_0^{\rm st})\ssub\NS\}
  \right]
 ~=~ \exp i\pi\varphi.
\end{eqnarray}
Here we used $\frac12J_0^{\rm int}+L_0^{\rm st}=L_0^{\rm int+\,st}=0$
for the state $l\ssub\NS$ of our interest, and chose a suitable
normalization for $l\ssub\RA$.
It immediately follows from this that
\begin{equation}
  {}\ssub\RARA\vev{l|\ST} ~=~ e^{i\pi\varphi}{}\ssub\NSNS\vev{l|\ST},
\end{equation}
for tadpole states $\ket\ST$ preserving spacetime supersymmetry
characterized by the phase $\varphi$.

\paragraph{Remark.} in our convention (\ref{bcWX}) of boundary or
crosscap conditions, the $N=2$ supercurrents $G^\pm$ are glued to
$\tilde G^\pm$ along the A-branes or A-type orientifolds though
they are usually called B-type conditions.

\subsection{Permutation D-branes in Gepner Models}\label{sec:PDGM}

We turn to construct and classify permutation branes in Gepner models.
They were constructed in \cite{Recknagel} and studied in
\cite{Ashok-DD, Brunner-G, Enger-RR, Caviezel-FG, Brunner-GK}.
Here we give a construction of them based on the simple current
orbifold prescription, paying particular attention to those
labelled by $L=k/2$ which require a special care.
We study the A-type branes first, and then study the B-type branes
using the mirror description.

\subsubsection{A-branes}\label{sec:PDGMA}

A-branes in Gepner models are labelled by a permutation
$\pi$ and $(L_c,M_c)$ with $c=1,\cdots,\nc\pi$, where
$\nc\pi$ denotes the number of cycles in $\pi$ and $\lc{\pi_c}$
the length of the cycle $\pi_c$.
The branes with trivial stabilizer group are simply given by
summing over $\ZZ_H$-images,
\begin{equation}
  \ket{\SB^{A,\pi}_{\bf L,M}}
   ~=~
  \frac{1}{\sqrt H}\sum_{\nu\in\ZZ_H}
  \ket{\SB^{\pi}_{{\bf L},\gammaA^\nu({\bf M})}}
   ~\equiv~
  \frac{1}{\sqrt H}\sum_{\nu\in\ZZ_H}
  \otimes_{c=1}^{[\pi]}\ket{\SB^{\pi_c}_{L_c,M_c+2\nu\lc{\pi_c}}}.
\label{BApi}
\end{equation}
Here and in the following the index for spin structure will be
suppressed whenever possible.
The label $({\bf L,M})$ contains some redundancy because different
values of ${\bf M}$ related by $\ZZ_H$-shifts label the same D-brane,
and the following change of the label $({\bf L,M})$
\begin{equation}
\SF_c~:~(\cdots L_c\cdots ~;~\cdots M_c\cdots)~\to~
                (\cdots k_c-L_c\cdots~;~\cdots M_c+k_c+2\cdots),
\label{FIB}
\end{equation}
maps $\ket{\SB^{A,\pi}_{\bf L,M}}$ to its anti-brane.

Some A-branes with special choices of $\pi$ or ${\bf L}$ have nontrivial
stabilizer groups.
The boundary state (\ref{BApi}) are invariant under
$\gammaA^{H'}~(H'<H)$ if
\begin{equation}
 \frac{H'\lc{\pi_c}}{k_c+2}\in\ZZ ~~\mbox{for all}~~c.
\label{gamma^H'}
\end{equation}
Such branes should be defined as sums over twists as well as over images.
Moreover, if $H'$ is even, the boundary states are invariant also
under $\gammaA^{H'/2}$ if
\begin{equation}
 L_c =\frac{k_c}{2}~~\mbox{for all $c$ such that}~~
 w'_c\equiv \frac{H'\lc{\pi_c}}{k_c+2}~~\mbox{is odd}.
\label{permshort}
\end{equation}
These D-branes are generalization of short-orbit branes
discussed in detail in \cite{Brunner-HHW}.
To see how the enhancement of the stabilizer occurs, note first that
$\gammaA^{H'/2}$ shifts $M_c$ by $k_c+2$ when 
$w'_c$ is odd, and acts trivially on other $M_c$'s.
Therefore, with the help of the maps $\SF_c$, $\gammaA^{H'/2}$ maps
the brane satisfying (\ref{permshort}) to itself or its antibrane
depending on how many of $w'_c$ are odd.
Since there are always an even number of odd $w'_c$ under the
condition (\ref{r-d ev}) the branes satisfying (\ref{permshort})
are always mapped to themselves by $\gammaA^{H'/2}$.

To write down the branes with nontrivial stabilizers,
we first introduce the boundary states in twisted sectors of
the product of $N$ minimal models following (\ref{BpiJh}) and
(\ref{twpiI}),
\begin{eqnarray}
 \ket{\SB^{(12\cdots N)}_{L,M}}^{(\mu)}\ssub\SEC
 &=& \frac{\alpha\ssub\SEC}2\sum_{l,m}\frac{S_{LM}^{~lm}}{(S_{00}^{lm})^{N/2}}
     \kket{\SB^{(12\cdots N)};l,m}^{(\mu)}\ssub\SEC, \nn\\
 \kket{\SB^{(12\cdots N)};l,m}^{(\mu)}\ssub\SEC
 &=& R^{(12\cdots N)}
     \kket{\SB;(l,m+2\mu)\otimes(l,m+4\mu)\otimes\cdots\otimes(l,m)}\ssub\SEC.
\label{BLMmu1}
\end{eqnarray}
Here $\alpha\ssub\SEC$ is defined in (\ref{BP}).
The label of twisted sectors $\mu$ satisfies $\mu N\in(k+2)\ZZ$.
When the level $k$ is even and $\mu N\in(k+2)(\ZZ+\frac12)$, we define
\begin{eqnarray}
 \ket{\wtilde\SB^{(12\cdots N)}_{k/2,M}}^{(\mu)}\ssub\SEC
 &=& \frac{\alpha\ssub\SEC}2
     \sum_{l,m}\frac{\wtilde S_{k/2M}^{~k/2\,m}}{(S_{00}^{k/2,m})^{N/2}}
     \kket{\wtilde\SB^{(12\cdots N)};\tfrac k2,m}^{(\mu)}\ssub\SEC, 
\label{BLMmu2}\\
 \kket{\wtilde\SB^{(12\cdots N)};\tfrac k2,m}^{(\mu)}\ssub\SEC
 &=& R^{(12\cdots N)}
     \kket{\SB;(\tfrac k2,m+2\mu)\otimes(\tfrac k2,m+4\mu)\otimes\cdots
              \otimes(\tfrac k2,m+k+2)^\eta}\ssub\SEC.
\nn
\end{eqnarray}
The tilde will be omitted in what follows unless we
need to distinguish the states (\ref{BLMmu2}) from (\ref{BLMmu1}).
The boundary states invariant under $\gammaA^h~(hH'=H)$ take the form
\begin{equation}
 \ket{\SB^{A,\pi,\rho}_{\bf L,M}} ~\equiv~
 \frac{1}{\sqrt{H}}\sum_{\nu\in\ZZ_h,\;\mu\in\ZZ_{H'}}
 \otimes_{c=1}^{[\pi]}
 \ket{\SB^{\pi_c}_{L_c,M_c+2\nu\lc{\pi_c}}}^{(\mu h)}
 \exp\left(\tfrac{2\pi i\rho\mu h}{H}\right).
\label{SOA}
\end{equation}
Here $\rho\in\ZZ_{H'}$ specifies a character of the stabilizer
group.

\paragraph{Example 1:~$(55555)$}

~\vskip2mm

The $\pi$-permuted boundary states have nontrivial stabilizer when
$\prod_{a\in \pi_c}\gamma_{a}=1$ for all cycles of $\pi$, namely,
all the cycles of $\pi$ have the lengths divisible by 5.
Therefore, $\pi=(12345)$ is up to conjugation the only case
with nontrivial stabilizer ${\cal H}=\ZZ_5$.
The untwisted stabilizer is $\cal H$ itself, so the boundary states
are sums over $\ZZ_5$-twists.

\paragraph{Example 2:~$(88444)$}

~\vskip2mm

There are D-branes with various stabilizer groups.
Generic non-permuted A-branes do not have stabilizers, while
those with $L_1=L_2=3$ are invariant under $\gammaA^4$.
Generic $\pi$-permuted A-branes are invariant under $\gammaA^4$
when $\pi$ permutes $a=1,2$.
Some of such D-branes are invariant under $\gammaA^2$ if
their $L$-labels satisfy (\ref{permshort}).
For all these cases, the untwisted stabilizer agrees with
the stabilizer itself.

\subsubsection{B-branes}\label{sec:PDGMB}

We would like to study B-branes in Gepner model using the
mirror description with the orbifold group $\Gamma_{\rm mir}$ of (\ref{Gmir}).
The label of D-branes consists of a permutation $\pi$ and
quantum numbers $(L_c,M_c)\,(c=1,\cdots,\nc\pi)$, as well as
a character of its untwisted stabilizer group.
Since the label $\bf M$ has a large redundancy due to the shifts
by elements of $\Gamma_{\rm mir}$, we sometimes use
\begin{equation}
M~\equiv~\sum_{c=1}^{\nc\pi} m_cw_c 
~~~~\left(w_c\equiv\frac{H}{k_c+2}\right),
\end{equation}
There is also a map $\SF_c$ (\ref{FIB}) that sends a brane to its antibrane.

In mirror Gepner model there are indeed branes with different
(untwisted) stabilizer groups. 
We first focus on generic permutation branes with none of
$L_c$ coinciding with $k_c/2$.
They start to have nontrivial stabilizer group as soon as $\pi$
becomes nontrivial.
If $\pi$ contains a cycle $\pi_c=(1\,2\cdots N)$, then all the
$\pi$-permuted branes are fixed by $(\ZZ_{k_c+2})^{N-1}$,
\[
 {\cal H}~\supset~(\ZZ_{k_c+2})^{N-1} \equiv
 \{\gamma_1^{\nu_1}\gamma_2^{\nu_2}\cdots \gamma_N^{\nu_N}~|~
   \sum_i\nu_i\in(k_c+2)\ZZ\}.
\]
So the generic $\pi$-permuted branes have stabilizer
${\cal H}=\otimes_{c=1}^{\nc\pi}(\ZZ_{k_c+2})^{\lc{\pi_c}-1}$.

By analyzing its action on twisted sectors using (\ref{omega-pi}),
one finds that none of the the stabilizer $(\ZZ_{k_c+2})^{N-1}$ contributes
to the untwisted subgroup ${\cal U}$ for odd $N$, while
a $\ZZ_{k_c+2}$ subgroup generated by
$\gamma_1\gamma_2^{-1}\cdots\gamma_{N}^{-1}$ contributes to ${\cal U}$
for even $N$.
As an example we list the permutation B-branes of the model
$(55555)$ with their (untwisted) stabilizers in the table below.

\begin{table}[htb]
\begin{center}
\begin{tabular}{|c||c|c|}
\hline
$\pi$ & $\cal H$ (generator) & $\cal U$ (generator) \\
\hline
$(1)(2)(3)(4)(5)$ & 1 & 1 \\
\hline
$(12)(3)(4)(5)$
 & $\ZZ_5~(\gamma_1\gamma_2^4)$ & $\ZZ_5~(\gamma_1\gamma_2^4)$ \\
\hline
$(12)(34)(5)$
 & $(\ZZ_5)^2~(\gamma_1\gamma_2^4,\gamma_3\gamma_4^4)$
 & $(\ZZ_5)^2~(\gamma_1\gamma_2^4,\gamma_3\gamma_4^4)$ \\
\hline
$(123)(4)(5)$
 & $(\ZZ_5)^2~(\gamma_1\gamma_2^4,\gamma_2\gamma_3^4)$ & 1 \\
\hline
$(123)(45)$
 & $(\ZZ_5)^3~(\gamma_1\gamma_2^4,\gamma_2\gamma_3^4,\gamma_4\gamma_5^4)$
 & $\ZZ_5~(\gamma_4\gamma_5^4)$ \\
\hline
$(1234)(5)$
 & $(\ZZ_5)^3~(\gamma_1\gamma_2^4,\gamma_2\gamma_3^4,\gamma_3\gamma_4^4)$
 & $\ZZ_5~(\gamma_1\gamma_2^4\gamma_3\gamma_4^4)$ \\
\hline
$(12345)$
 & $(\ZZ_5)^4~(\gamma_1\gamma_2^4,\gamma_2\gamma_3^4,
               \gamma_3\gamma_4^4,\gamma_4\gamma_5^4)$
 & 1 \\
\hline
\end{tabular}
\end{center}
\caption{B-branes of the model $(55555)$ and their
         stabilizer $\cal H$, untwisted stabilizer $\cal U$.}
\end{table}

The permutation branes with nontrivial untwisted stabilizers are made from
permutation boundary states $\ket{\SB^{(12\cdots N),\rho}_{L,M}}$
in the orbifold $M(k)^N/\Gamma_{\rm mir}$, where $N$ is even and
\begin{equation}
  \Gamma_{\rm mir}~=~(\ZZ_{k+2})^{N-1}~=~
  \{\gamma_1^{\nu_1}\cdots\gamma_N^{\nu_N}
  |\sum\nu_a=0~~{\rm mod}~~(k+2)\}.
\label{petitMG}
\end{equation}
The label $\rho$ specifies a character of the
untwisted stabilizer $\ZZ_{k+2}$ generated by
$\gamma_1\gamma_2^{-1}\cdots\gamma_N^{-1}$.
We find it convenient to define the boundary states in terms of
Ishibashi states as
\begin{eqnarray}
\lefteqn{
 \ket{\SB^{(12\cdots N),\rho}_{L,M}}\ssub\SEC ~=~
 \frac{1}{\sqrt{k+2}}
        \sum_\nu\exp\left(\tfrac{2\pi i\rho\nu}{k+2}\right)
        \sum_{l,m}\frac{\alpha\ssub\SEC}{2}
                  \frac{S_{LM}^{~lm}}{(S_{00}^{lm})^{N/2}}
} \nn\\
&& ~~~~~~\times
 R_{(12\cdots N)}\kket{\SB;(l,m+\nu)\otimes(l,m-\nu)\otimes\cdots
                                    \otimes(l,m-\nu)}\ssub\SEC,
\label{BPrho}
\end{eqnarray}
where $\alpha\ssub\SEC$ is defined in (\ref{BP}).
It is easy to check the following,
\begin{equation}
 \ket{\SB^{(12\cdots N),\rho}_{L,M}}~=~
 \ket{\SB^{(2\cdots N1),-\rho}_{L,M}},~~~~
 \gamma_a\ket{\SB^{(12\cdots N),\rho}_{L,M}}~=~
 \ket{\SB^{(12\cdots N),\rho}_{L,M+2}}.
\end{equation}
However, due to the non-standard definition of the Ishibashi states
in twisted sectors, $\rho$ has to be integer or half-odd integer
depending on whether $M$ is even or odd.
One also finds
\begin{equation}
\begin{array}{lcl}
 \ket{\SB^{(12\cdots N),\;\rho}_{L,M}}\ssub{\NSNS\pm} &=&
 ~~\ket{\SB^{(12\cdots N),\;\rho+\frac{k+2}{2}}_{k-L,M+k+2}}\ssub{\NSNS\pm}, \\
 \ket{\SB^{(12\cdots N),\;\rho}_{L,M}}\ssub{\RARA\pm} &=&
 -\ket{\SB^{(12\cdots N),\;\rho+\frac{k+2}{2}}_{k-L,M+k+2}}\ssub{\RARA\pm}.
\label{BPN}
\end{array}
\end{equation}
As an example, the permutation B-branes in $(55555)$ model
for $\pi=(12)(34)$ is given by
\begin{equation}
 \ket{\SB^{B,(12)(34),\,\rho,\,\rho''}_{\bf L,M}} ~=~
 \frac{1}{5}\sum_{\nu+\nu'+\nu''\in 5\ZZ}
 \ket{\SB^{(12),\,\rho}_{L,M+2\nu}}\otimes
 \ket{\SB^{(34),\,\rho'}_{L',M'+2\nu''}}\otimes
 \ket{\SB^{(5)}_{L'',M''+2\nu''}}.
\end{equation}

Next we discuss the enhancement of stabilizer
group when some of $k_c$ are even and $L_c = k_c/2$.
A permutation brane labelled by $\pi$ and
$\{L_1,\cdots,L_{\nc\pi}\}$ is invariant under the following
simple currents
\begin{equation}
\begin{array}{cll}
(i)  & \gamma_a\gamma_b^{-1}&\mbox{($a,b$ are in the same cycle)}\\
(ii) & \eta_a\eta_b         &\mbox{($a,b$ are in cycles labelled by $L=k/2$)}.
\end{array}
\label{HB}
\end{equation}
So the stabilizer group for a permutation brane gets enhanced by
$(\ZZ_2)^{n-1}$ if $n ~(\ge 2)$ cycles of $\pi$ are labelled by $L_c=k_c/2$.
The $\bf L$-label of B-branes is called {\it special} (or {\it generic})
if two or more (resp. at most one) of $L_c$ coincide with $k_c/2$.

It is a little intricate to find out the untwisted stabilizer
for these short-orbit branes.
For the D-branes with $\pi={\rm id}$ and $L_a=k_a/2$ for $a=1,\cdots,n$,
the boundary states in twisted sectors should be expressed as products
of $\ket{\SB_{k_a/2,M_a}}\ssub\SEC$ and
$\ket{\SB_{k_a/2,M_a}}\ssub\SEC^{\eta_a}$.
However, the action of $\eta=\gamma^{\frac{k+2}{2}}$ on boundary
states in the untwisted and $\eta$-twisted sectors differ by a sign,
\begin{equation}
\begin{array}{lcl}
  \eta\ket{\SB_{k/2,M}}\ssub{\NSNS\pm} &=&
  +\ket{\SB_{k/2,M}}\ssub{\NSNS\pm}, \\
  \eta\ket{\SB_{k/2,M}}\ssub{\RARA\pm} &=&
  -\ket{\SB_{k/2,M}}\ssub{\RARA\pm},
\end{array}
~~~
\begin{array}{lcl}
  \eta\ket{\SB_{k/2,M}}\ssub{\NSNS\pm}^\eta &=&
  -\ket{\SB_{k/2,M}}\ssub{\NSNS\pm}^\eta, \\
  \eta\ket{\SB_{k/2,M}}\ssub{\RARA\pm}^\eta &=&
  +\ket{\SB_{k/2,M}}\ssub{\RARA\pm}^\eta.
\end{array}
\label{eta-act}
\end{equation}
So the only states invariant under all the elements $(ii)$ of the stabilizer
group (\ref{HB}) are those in the untwisted sector and
$(\eta_1\cdots\eta_n)$-twisted sector.
The latter exists only when $n$ is even.
The untwisted stabilizer for non-permuted branes is given by
\begin{equation}
{\cal U}~=~ \left\{\begin{array}{ll}
  1     & (n~{\rm odd}) \\
  \ZZ_2=\{1,\prod_{a=1}^n\eta_a\} & (n~{\rm even})
\end{array}
\right. .
\end{equation}
Generalizing this to permutation branes, one finds the following result.
For each even-length cycle $\pi_c=(a_1,a_2,\cdots,a_{2l})$ of $\pi$,
denote by $\gamma_{\pi_c}$ the following simple current
\begin{equation}
  \gamma_{\pi_c}~=~ \gamma_{a_1}\gamma_{a_2}^{-1}\gamma_{a_3}\cdots
  \gamma_{a_{2l}}^{-1}.
\label{gammapic}
\end{equation}
Then the untwisted stabilizer for permutation branes with $L_c=k_c/2$
for more than one cycles is generated by the following:
\begin{enumerate}
\item $\gamma_{\pi_c}$, where $\pi_c$ is an even-length cycle
      labelled by $L_c\ne k_c/2$,
\item $(\gamma_{\pi_c})^2$, where $\pi_c$ is an even-length cycle
      labelled by $L_c= k_c/2$,
\item The element
\begin{equation}
 \gammaB ~\equiv~ (\prod_a\eta_a)(\prod_{L_c=k_c/2}\gamma_{\pi_c}),
\label{genU3}
\end{equation}
      where the first product is over all $a$'s belonging to odd-length
      cycles labelled by $L_c=k_c/2$, and the second is over all
      even-length cycles $\pi_c$ labelled by $L_c=k_c/2$.
      This is an element of $\Gamma_{\rm mir}$ only when
      there are even number of odd-length cycles labelled by $L_c=k_c/2$.
\end{enumerate}
Interestingly, when $L_c$'s coincide with $k_c/2$ the untwisted
stabilizer group gets reduced due to $1\to 2$ of the list,
and then enhances by $3$ of the list.

As an example, we list some of the permutation B-branes, their
stabilizers and untwisted stabilizers in the model $(88444)$.

\begin{table}[htb]
\begin{center}
{\small
\begin{tabular}{|c|c|c||c|c|}
\hline
$\pi$ & $\sharp(L_c=k_c/2)$ & ${\bf L}$ & $\cal H$ & $\cal U$ \\
\hline
$(12)(345)$
 & $2$
 & $(3,1)$
  & $\ZZ_8\times(\ZZ_4)^2\times\ZZ_2$
  & $\ZZ_4~(\gamma_1^2\gamma_2^6)$ \\
 & $\le1$
 & any
  & $\ZZ_8\times(\ZZ_4)^2$
  & $\ZZ_8~(\gamma_1\gamma_2^7)$ \\
\hline
$(1)(2)(345)$
 & $3$
 & $(3,3,1)$
  & $(\ZZ_4)^2\times(\ZZ_2)^2$
  & 1 \\
 & $2$
 & $(3,\ast,1)$
  & $(\ZZ_4)^2\times\ZZ_2$
  & $\ZZ_2~(\eta_1\eta_3\eta_4\eta_5)$ \\
 & $2$
 & $(3,3,\ast)$
  & $(\ZZ_4)^2\times\ZZ_2$
  & $\ZZ_2~(\eta_1\eta_2)$ \\
 & $\le1$
 & any
  & $(\ZZ_4)^2$
  & $1$ \\
\hline
$(12)(34)(5)$
 & $3$
 & $(3,1,1)$
  & $\ZZ_8\times\ZZ_4\times(\ZZ_2)^2$
  & $\ZZ_4\times\ZZ_2~(\gamma_1^2\gamma_2^6,\gamma_3^2\gamma_4^2)$ \\
 & $2$
 & $(3,1,*)$
  & $\ZZ_8\times\ZZ_4\times\ZZ_2$
  & $\ZZ_8\times\ZZ_2~
     (\gamma_1\gamma_2^7\gamma_3\gamma_4^3,\gamma_3^2\gamma_4^2)$ \\
 & $2$
 & $(3,*,1)$
  & $\ZZ_8\times\ZZ_4\times\ZZ_2$
  & $\ZZ_4\times\ZZ_4~(\gamma_1^2\gamma_2^6,\gamma_3\gamma_4^3)$ \\
 & $2$
 & $(*,1,1)$
  & $\ZZ_8\times\ZZ_4\times\ZZ_2$
  & $\ZZ_8\times\ZZ_2~(\gamma_1\gamma_2^7,\gamma_3^2\gamma_4^2)$ \\
 & $\le1$
 & any
  & $\ZZ_8\times\ZZ_4$
  & $\ZZ_8\times\ZZ_4~(\gamma_1\gamma_2^7,\gamma_3\gamma_4^3)$ \\
\hline
\end{tabular}
}
\end{center}
\caption{Some permutation B-branes in the model $(88444)$.}
\end{table}
Let us pick up some examples from the list and illustrate
the construction of boundary states.
We first take the case $\pi=(1)(2)(345)$, which is a rather
straightforward generalization of non-permuted branes because
all the cycles have odd length.
The $\pi$-permuted B-branes split into two when ${\bf L}=(3,*,1)$.
To describe the boundary states in $\eta_1\eta_3\eta_4\eta_5$-twisted
sector, we use the states $\ket{\SB_{k/2,M}}^\eta$
defined at Section \ref{sec:MMBSTS} and their generalization to
arbitrary odd-length cycles,
\begin{equation}
\ket{\wtilde\SB^{(12\cdots N)}_{k/2,M}}
 ~\equiv~
\ket{\SB^{(12\cdots N)}_{k/2,M}}^{\eta_1\cdots\eta_N}\ssub\SEC
 ~=~ \frac{\alpha\ssub\SEC}{2}
     \sum_{l,m}\frac{\wtilde S_{k/2,M}^{~lm}}{(S_{00}^{lm})^{N/2}}
 R_{(12\cdots N)}\kket{\SB;\otimes_{a=1}^N(l,m)^\eta}
 \ssub\SEC.
\label{BPeta}
\end{equation}

Next we study the case $\pi=(12)(34)(5)$.
The untwisted stabilizer group for $\pi$-permuted B-branes
gets smaller as the number of $L_c$'s coinciding with $k_c/2$
increases.
We wish to understood this in terms of the boundary states defined
at (\ref{BPrho}).
For generic ${\bf L}$ the branes are defined as
\begin{equation}
  \ket{\SB^{B,\pi,(\rho,\rho')}_{\bf L,M}} ~=~
  \frac{1}{4}\sum_{\nu+2\nu'+2\nu''\in 8\ZZ}
  \ket{\SB^{(12),\rho}_{L,M+2\nu}}\otimes
  \ket{\SB^{(34),\rho'}_{L',M'+2\nu'}}\otimes
  \ket{\SB^{(5)}_{L'',M''+2\nu''}},
\end{equation}
with the integers $\rho,~\rho'$ specifying a character of
the untwisted stabilizer $\ZZ_8\times\ZZ_4$.
When some $L_c$'s coincide with $k_c/2$, then the sum over orbifold
images is partially translated into the sum over shifts of $(\rho,\rho')$
due to (\ref{BPN}).
When ${\bf L}=(3,1,1)$ one can write
\begin{equation}
  \ket{\SB^{B,\pi,(\rho,\rho')}_{\bf L,M}} ~=~
  \frac{1}{4}\sum_{\nu+2\nu'+2\nu''\in 8\ZZ}
  \ket{\SB^{(12),\;\rho,+}_{3,M+2\nu}}\otimes
  \ket{\SB^{(34),\;\rho',+}_{1,M'+2\nu'}}\otimes
  \ket{\SB^{(5)}_{1,M''+2\nu''}},
\end{equation}
where we define, for any cyclic permutation $\pi$ of even length,
\begin{equation}
  \ket{\SB^{\pi,\rho,\pm}_{k/2,M}} ~\equiv~
  \frac12\left(
     \ket{\SB^{\pi,\;\rho}_{k/2,M}}
 \pm \ket{\SB^{\pi,\;\rho+(k+2)/2}_{k/2,M}}
  \right).
\label{Bpirpm}
\end{equation}
The periodicity of $\rho,\rho'$ thus becomes halved when
${\bf L}=(3,1,1)$, in accordance with the untwisted stabilizer
becoming smaller for these branes.
Note also that the states (\ref{Bpirpm}) are transformed by
$\eta_a$'s in a similar way as $\ket{\SB_{k/2,M}}$ and
$\ket{\SB_{k/2,M}}^\eta$ of (\ref{eta-act}).

When ${\bf L}=(3,1,0)$ one can write
\begin{eqnarray}
  \ket{\SB^{B,\pi,(\rho,\rho',\varepsilon)}_{\bf L,M}} &=&
  \frac14\sum_{\nu+2\nu'+2\nu''\in 8\ZZ}
  \ket{\SB^{(12),\;\rho,+}_{3,M+2\nu}}\otimes
  \ket{\SB^{(34),\;\rho',+}_{1,M'+2\nu'}}\otimes
  \ket{\SB^{(5)}_{0,M''+2\nu''}} \nn\\ && +
  \frac\varepsilon4\sum_{\nu+2\nu'+2\nu''\in 8\ZZ}
  \ket{\SB^{(12),\;\rho,-}_{3,M+2\nu}}\otimes
  \ket{\SB^{(34),\;\rho',-}_{1,M'+2\nu'}}\otimes
  \ket{\SB^{(5)}_{0,M''+2\nu''}}.
\label{Bpirre}
\end{eqnarray}
The untwisted stabilizer is twice as big as the previous case
due to the generator $\gammaB$ (\ref{genU3}).

\subsection{Permutation Orientifolds in Gepner Model}\label{POG}

We next construct and classify the permutation orientifolds
in Gepner models.
The basic building blocks are the quartets of crosscap states
$\ket{\SC_M}\ssub\SEC$ (\ref{BCmin}) or
$\ket{\SC^{(12)}_{M_1,M_2}}\ssub\SEC$ (\ref{Cmin2})
defined before.
The A-type permutation orientifolds are constructed
as sums of their products with characters of
$\Gamma_O\equiv\Gamma/(\Gamma\Gamma^\pi)$,
where $\Gamma$ is the Gepner's orbifold group
and
\[
 \Gamma\Gamma^\pi~\equiv~\{g\pi g\pi|g\in\Gamma\}.
\]
B-type orientifolds are constructed in a similar way
using the mirror description.
Below we give a general construction, and illustrate it in a few examples.

\subsubsection{A-type orientifolds}\label{sec:POGMA}

The orbifold group is $\Gamma=\ZZ_H$ and one easily finds that
\[
 \Gamma\Gamma^\pi=\Gamma^2\equiv\{g^2|g\in\Gamma\},
\]
for any models and any $\pi$.
Therefore, $\Gamma_O\equiv\Gamma/\Gamma^2$ is a $\ZZ_2$ for even $H$
and otherwise trivial.
We denote by $\ket{\SC^\pi_{\bf M}}\ssub\SEC$ the products
of crosscaps $\ket{\SC_M}\ssub\SEC$ and
$\ket{\SC^{(12)}_{M_1,M_2}}\ssub\SEC$ in minimal models.
The A-type crosscaps in Gepner models are given by their sums,
\begin{equation}
 \ket{\SC^{A,\pi,\epsilon}_{\bf M}}\ssub\SEC
  ~=~
 \frac{c\ssub\SEC}{\sqrt H }\sum_{\nu}\epsilon^\nu
 \ket{\SC^{\pi}_{\gamma^\nu({\bf M})}}\ssub\SEC
  ~\equiv~
 \frac{c\ssub\SEC}{\sqrt H }\sum_{\nu}\epsilon^\nu
 \ket{\SC^{\pi}_{\bf M+2\nu}}\ssub\SEC.
\end{equation}
with suitable normalization constants $c\ssub\NS,c\ssub\RA$.
The following crosscap states form a quartet,
\begin{eqnarray}
&&  \ket{\SC^{A,\pi,      \epsilon}_{\bf M}}\ssub{\NSNS+},~~
    \ket{\SC^{A,\pi,      \epsilon}_{\bf M+2}}\ssub{\NSNS-},~~
    \ket{\SC^{A,\pi,\tilde\epsilon}_{\bf M-1}}\ssub{\RARA+},~~
    \ket{\SC^{A,\pi,\tilde\epsilon}_{\bf M+1}}\ssub{\RARA-},
\nn\\&&
 \hskip25mm\tilde\epsilon\equiv\epsilon\cdot
      \exp\left(-\tsum_{a=1}^r\tfrac{i\pi}{k_a+2}\right).
\label{ACQ}
\end{eqnarray}
with the supersymmetry phase
\begin{equation}
 \exp i\pi\varphi
  ~=~
     \frac{c\ssub\RA}{c\ssub\NS}\exp i\pi\left(
     \sum_{a=1}^5\frac{(M_a-1)}{2k_a+4}+\frac{r+|\pi|}{2}\right).
\label{vpC}
\end{equation}
Here $|\pi|$ counts the number of cycles of length two in $\pi$.
The four possible choices of $c\ssub{\NS,\RA}$ correspond to orientifold
planes $O^\pm$ of positive or negative tension, and their anti-planes.
The label $\epsilon$ can take $\pm1$ for even $H$, while only $\epsilon=+1$
is allowed for odd $H$.

The constant $c\ssub\RA$ takes values $\pm1$, whereas the correct
values of $c\ssub\NS$ depends on the label $\epsilon$.
When $H$ is odd, the tension $T$ of the orientifold is given by $c\ssub\NS$
up to a positive proportionality constant so we should set $c\ssub\NS=\pm1$.
When $H$ is even, $T$ becomes proportional to
\[
 T~\sim~
 c\ssub\NS\left(
  e^{-i\pi\Theta_{\bf M}}
  +\epsilon e^{+i\pi\Theta_{\bf M}}
 \right),~~~~~
 \Theta_{\bf M}\equiv\sum_{c~(k_c={\rm even},\lc{\pi_c}=1)}
 \frac{(-)^\frac{M_c}{2}}{2k_c+4}.
\]
So the correct choices of $c\ssub\NS$ are
\begin{equation}
\begin{array}{lcll}
H={\rm odd},~ (\epsilon=+) &\Longrightarrow& c\ssub\NS=\pm1, &
 T\sim\pm1, \\
H={\rm even},~ \epsilon=+ &\Longrightarrow& c\ssub\NS=\pm1, &
 T\sim\pm\cos\pi\Theta_{\bf M}, \\
H={\rm even},~ \epsilon=- &\Longrightarrow& c\ssub\NS=\pm i, &
 T\sim\pm\sin\pi\Theta_{\bf M}.
\end{array}
\end{equation}

Orientifolds labelled by different $\bf M$ are related to one
another by the global symmetry generated by simple currents,
\begin{equation}
  (\otimes_{a=1}^r\gamma_a^{\nu_a})\ket{\SC^{A,\pi,\epsilon}_{\bf M}} ~=~
  \ket{\SC^{A,\pi,\epsilon}_{\bf M'}},~~~~
  M'_a\equiv M_a+2\nu_a+2\nu_{\pi(a)}.
\end{equation}
If $H$ is odd, then any ${\bf M}$ can be mapped to ${\bf M}=0$
by the global symmetry.
For even $H$ there are several choices for $\bf M$
that lead to physically inequivalent orientifolds.
An interesting fact is that, for even $H$, the involutiveness
of parity does not require $M_a=M_{\pi(a)}$.
The condition that the square of parity is an element of
$\Gamma$ implies the existence of a mod-$H$ integer $\nu$
satisfying
\begin{equation}
  M_a-M_{\pi(a)}= 2\nu~~{\rm mod}~~2(k_a+2).
\end{equation}
Since the left hand side is antisymmetric under $a\to \pi(a)$ and
the right hand side is symmetric, the only allowed $\nu$ are
$0$ or $H/2$.

\paragraph{Example 1:~$(55555)$}

~\vskip2mm

There are three involutive permutations of five elements up to
conjugation, namely $\pi={\rm id},~(12)$ or $(12)(34)$.
We denote various products of crosscap states as
\begin{eqnarray}
 \ket{\SC^{(1)(2)(3)(4)(5)}_{\bf M}}\ssub\SEC &=&
 \ket{\SC^{(1)}_{M_1}\otimes
      \SC^{(2)}_{M_2}\otimes
      \SC^{(3)}_{M_3}\otimes
      \SC^{(4)}_{M_4}\otimes
      \SC^{(5)}_{M_5}}\ssub\SEC, \nn\\
 \ket{\SC^{(12)(3)(4)(5)}_{\bf M}}\ssub\SEC &=&
 \ket{\SC^{(12)}_{M_1,M_2}\otimes
      \SC^{(3)}_{M_3}\otimes
      \SC^{(4)}_{M_4}\otimes
      \SC^{(5)}_{M_5}}\ssub\SEC, \nn\\
 \ket{\SC^{(12)(34)(5)}_{\bf M}}\ssub\SEC &=&
 \ket{\SC^{(12)}_{M_1,M_2}\otimes
      \SC^{(34)}_{M_3,M_4}\otimes
      \SC^{(5)}_{M_5}}\ssub\SEC.
\label{Cprod}
\end{eqnarray}
The A-type crosscaps in Gepner model are given by their sums.
For the parities to be involutive, we have to set $M_1=M_2$ in
the second line and $M_1=M_2, M_3=M_4$ in the third line.

Since all the levels are odd, the crosscaps with different values
of ${\bf M}$ are all related to the one with ${\bf M}={\bf 0}$
by global symmetry (simple currents).
Moreover, $\Gamma_O$ is trivial because $H$ is odd.
Therefore, there are just three physically inequivalent A-type
orientifolds in this model $\ket{\SC^{A,\pi}_{\bf 0}}$ labelled by
three different permutations.
The same argument apply to all other Gepner models with odd $H$.

\paragraph{Example 2:~$(88444)$}

~\vskip2mm

In this model there are four inequivalent permutations up
to conjugation, namely $\pi={\rm id}, (12), (34)$ or $(12)(34)$.
The orientifolds are also labelled by a character of $\Gamma_O=\ZZ_2$.
In order for the orientifold $\ket{\SC^{A,\pi,\epsilon}_{\bf M}}$
to correspond to an involutive parity, the $M$
labels have to satisfy $M_3=M_4$ if $\pi$ contains a cycle $(34)$,
and $M_1=M_2$ or $M_1=M_2+8$ if $\pi$ contains $(12)$.
Different values of ${\bf M}$ are related by the actions of global
symmetry, but this time there remain several choices for $\bf M$ leading
to inequivalent orientifolds.
The physically inequivalent choices of labels $(\pi,{\bf M})$ are
as listed below:
\begin{equation}
\begin{array}{lrcl}
\pi = {\rm id}, &
{\bf M}&=& (00000),~(02000),~(22000),\\
       &&& (00002),~(02002),~(22002), \\
\pi = (12), &
{\bf M}&=& (00000),~(00002),~(08000),~(08002),~ \\
\pi = (34), &
{\bf M}&=& (00000),~(02000),~(22000),~ \\
\pi = (12)(34), &
{\bf M}&=& (00000),~(08000).
\end{array}
\label{15diff}
\end{equation}

The crosscaps containing $\ket{\SC^{(12)}_{M_1,M_1+8}}$
are supported only on closed string Hilbert space in the
$\gamma_1^4\gamma_2^4$-twisted sector, so they are in particular
tensionless.
On the other hand, they do have nonzero overlaps with RR ground states
in $\gamma_1^4\gamma_2^4$-twisted sector.

\subsubsection{B-type permutation orientifolds}\label{sec:POGMB}

We study the B-type permutation orientifolds in Gepner models
as A-types in the mirror.
The orientifolds are given by summing the crosscap states
$\ket{\SC^\pi_{\bf M}}\ssub\SEC$ of the product theory over an orbit
of $\Gamma_{\rm mir}$ weighted by various characters of
$\Gamma_O\equiv\Gamma_{\rm mir}/(\Gamma_{\rm mir}\Gamma^\pi_{\rm mir})$,
\begin{eqnarray}
  \ket{\SC^{B,\pi,\rho}_{\bf M}}\ssub\SEC &=&
  \frac{c\ssub\SEC}{\sqrt{|\Gamma_{\rm mir}|}}
  \sum_{\gamma=\otimes_a\gamma_a^{\nu_a}\in\Gamma_{\rm mir}}
  \ket{\SC^{\pi}_{\gamma({\bf M})}}\ssub\SEC\rho(\vec\nu),
\label{CBY}
\end{eqnarray}
where $\gamma({\bf M})\equiv (M_1+2\nu_1,\cdots,M_r+2\nu_r)$
for $\gamma=\otimes_a\gamma_a^{\nu_a}\in\Gamma_{\rm mir}$.
Then the following quartet of states defines a B-type
orientifold of Gepner model,
\begin{eqnarray*}
&&
 \ket{\SC^{B,\pi,      \rho}_{\bf M}  }\ssub{\NSNS+},~~
 \ket{\SC^{B,\pi,      \rho}_{\bf M+2}}\ssub{\NSNS-},~~
 \ket{\SC^{B,\pi,\tilde\rho}_{\bf M-1}}\ssub{\RARA+},~~
 \ket{\SC^{B,\pi,\tilde\rho}_{\bf M+1}}\ssub{\RARA-},
\\&&\hskip25mm
 \tilde\rho(\vec\nu) \equiv
 \rho(\vec\nu)\exp\left(-\tsum_{a=1}^r\tfrac{i\pi \nu_a}{k_a+2}\right).
\end{eqnarray*}
Here $\rho$ is a character of $\Gamma_O$, whereas $\tilde\rho(\vec\nu)$
is anti-periodic in any of $\nu_a\to\nu_a+k_a+2$.

The label $\bf M$ is highly redundant because it has meanings only
up to shifts by $\Gamma_{\rm mir}$.
There is also a global $\ZZ_H$ symmetry of the mirror Gepner model
that relates orientifolds with different ${\bf M}$.

Let us discuss the properties of the characters $\rho$
of the group $\Gamma_O$ in some detail.
By definition, $\rho$ is a character of the group $\Gamma_{\rm mir}$
that takes trivial value on the subgroup
$\Gamma_{\rm mir}\Gamma^\pi_{\rm mir}$.
The elements of this subgroup are given by $\vec\nu$ satisfying
\begin{eqnarray*}
&({\rm i})  & \sum_{a=1}^r\frac{\nu_a}{k_a+2}\in\ZZ, \\
&({\rm ii}) & \nu_a=\nu_{\pi(a)}, \\
&({\rm iii})& \mbox{$\nu_a$ is even for all $a$ labelled by even $k_a$
                    and fixed by $\pi$.}
\end{eqnarray*}
Characters of $\Gamma_{\rm mir}$ taking trivial value at such $\vec\nu$'s
are given by
\begin{equation}
 \rho(\vec\nu) ~=~
 \prod_{c\;(\pi_c=(a_cb_c))}
 e^{-\frac{2\pi i r_c}{k_c+2}(\nu_{a_c}-\nu_{b_c})} \cdot
 \prod_{c\;(\lc{\pi_c}=1,k_c={\rm even})}\epsilon_c^{\nu_c}.
\label{rhoB}
\end{equation}
Here $r_c\in\ZZ_{k_c+2}$ is associated to the cycle $\pi_c$ of length
2 labelled by $k_c$, and the sign $\epsilon_c$ is associated to
the length-one cycle $\pi_c$ labelled by an even level $k_c$.
Sometimes the conditions (i)--(iii) accidentally imply that some more
$\nu_a$ have to be even, and $\rho$ depends upon additional $\pm$
signs (see the Example 2 below).
Finally, some of the parameters $(r_c,\epsilon_{c'})$ are redundant
because of the equivalence $\rho(\vec\nu)\simeq
\rho(\vec\nu)\exp\left(\sum_a\frac{2\pi i\nu_a}{k_a+2}\right)$ 
that follows from (i).

Recall that $\ket{\SC^{B,\pi,\rho}_{\bf M}}$ is constructed by
summing the crosscap states sitting in different twisted sectors.
In the formula (\ref{rhoB}) for characters, the parameters $r_c$
assign different weights to different twisted sectors so that they
express the dressings by quantum symmetry of the mirror Gepner model.
Such symmetry are known to map to the global symmetry of the
original Gepner model.
In other words, $r_c$'s can be absorbed by a suitable redefinition
of the LG fields $X_1,\ldots,X_r$.
On the other hand, different signs $\epsilon_c$ give physically
inequivalent orientifolds since they cannot be gauged away in such
a way.
In particular, the tension and supersymmetry phase $\varphi$ of
orientifolds do depend on $\epsilon$'s in a non-trivial manner.

\paragraph{Example 1:~$(55555)$}

~\vskip2mm

There are three inequivalent choices of permutations,
$\pi={\rm id}, (12), (12)(34)$.
For each choice of $\pi$ there is a unique choice for $\bf M$
up to shifts by $\Gamma_{\rm mir}$ and the global $\ZZ_5$ symmetry of
the mirror Gepner model.
The tension is given by $c\ssub\NS$ up to some positive proportionality
constant, and the supersymmetry phase $\varphi$ is given by (\ref{vpC}).

The group $\Gamma_{\rm mir}/(\Gamma_{\rm mir}\Gamma^\pi_{\rm mir})$
and the allowed character $\rho$ for various choices of permutation
are given by the following table (we denote
$\omega_n\equiv\exp\frac{2\pi i}{n}$),
\begin{center}
\begin{tabular}{|l||c|cl|}
\hline
$\pi$ &
$\Gamma_{\rm mir}/(\Gamma_{\rm mir}\Gamma^\pi_{\rm mir})$ &
$\rho(\vec\nu)$ & \\
\hline
${\rm id}$ &
$\{1\}$ &
$1$ & \\
\hline
$(12)$ &
$\ZZ_5$ &
$\omega_5^{-r(\nu_1-\nu_2)}$ & $r\in\ZZ_5$ \\
\hline
$(12)(34)$ &
$(\ZZ_5)^2$ &
$\omega_5^{-r(\nu_1-\nu_2)-r'(\nu_3-\nu_4)}$ & $r,r'\in\ZZ_5$ \\
\hline
\end{tabular}
\end{center}
The orientifolds labelled by different $r,r'$ are related
by quantum symmetries, so they are physically equivalent.
We thus found three inequivalent B-type orientifolds of this
model corresponding to three different choices of $\pi$.

\paragraph{Example 2:~$(88444)$}

~\vskip2mm

The orbifold group is $\Gamma_{\rm mir}=\ZZ_8\times(\ZZ_4)^3$, and
there are four inequivalent choices for the permutation,
$\pi={\rm id}, (12), (34)$ and $(12)(34)$.
For each choice of $\pi$ there are two inequivalent values for
the label ${\bf M}$ up to shifts by $\Gamma_{\rm mir}$ and
global symmetry of the mirror model,
\[
 {\bf M}=(00000)~~{\rm or}~~(20000).
\]
The orientifolds are also labelled by the character of the group
$\Gamma_O\equiv\Gamma_{\rm mir}/(\Gamma_{\rm mir}\Gamma^\pi_{\rm mir})$.
We determine the general form of the character following
the argument given above ($\omega_n\equiv\exp\frac{2\pi i}{n}$),
\begin{equation}
\begin{array}{lrclcl}
\pi={\rm id} &
\rho_{\epsilon_1,\epsilon_2,\epsilon_3,\epsilon_4,\epsilon_5}(\vec\nu)
&=& \epsilon_1^{\nu_1}\epsilon_2^{\nu_2}\epsilon_3^{\nu_3}
    \epsilon_4^{\nu_4}\epsilon_5^{\nu_5}
&\simeq&
    \rho_{-\epsilon_1,-\epsilon_2,\epsilon_3,\epsilon_4,\epsilon_5}(\vec\nu),
\\
\pi=(12) &
\rho_{r,\epsilon_1,\epsilon_3,\epsilon_4,\epsilon_5}(\vec\nu)
&=& \omega_8^{-r(\nu_1-\nu_2)}\epsilon_1^{\nu_1}\epsilon_3^{\nu_3}
    \epsilon_4^{\nu_4}\epsilon_5^{\nu_5}
&\simeq&
    \rho_{r+2,-\epsilon_1,-\epsilon_3,-\epsilon_4,-\epsilon_5}(\vec\nu),
\\
\pi=(34) &
\rho_{r,\epsilon_1,\epsilon_2,\epsilon_5}(\vec\nu)
&=& \omega_4^{-r(\nu_3-\nu_4)}\epsilon_1^{\nu_1}\epsilon_2^{\nu_2}
    \epsilon_5^{\nu_5}
&\simeq&
    \rho_{r,-\epsilon_1,-\epsilon_2,\epsilon_5}(\vec\nu),
\\
\pi=(12)(34) &
\rho_{r,r',\epsilon_1,\epsilon_5}(\vec\nu)
&=& \omega_8^{-r(\nu_1-\nu_2)}\omega_4^{-r'(\nu_3-\nu_4)}
    \epsilon_1^{\nu_1}\epsilon_5^{\nu_5}
&\simeq&
    \rho_{r+2,r'+2,-\epsilon_1,-\epsilon_5}(\vec\nu).
\end{array}
\label{rho2p}
\end{equation}
In the second and fourth cases above, we have one more $\pm$ sign as
compared to the formula (\ref{rhoB}) due to the accidental effect
explained there.
For example, for $\pi=(12)$ the elements of $\Gamma\Gamma^\pi$
are given by those $\vec\nu$ satisfying
\begin{equation}
 \nu_1+\nu_2+2(\nu_3+\nu_4+\nu_5)\in 8\ZZ,~~~~
 \nu_1=\nu_2,~~~~
 \nu_{3,4,5}\in 2\ZZ.
\end{equation}
These conditions imply that $\nu_1$ is also even, so we get
an additional parameter $\epsilon_1$ in the second line of (\ref{rho2p}).
Although $\nu_2$ is also even, we do not introduce $\epsilon_2^{\nu_2}$
because $\epsilon_2^{\nu_2}=\omega_8^{4(\nu_1-\nu_2)}\epsilon_1^{\nu_2}$.

Let us compute the tension of the orientifolds we have listed, focusing
on the dependence on $\epsilon$-labels.
We use various symmetry to set ${\bf M}=(00000)$ or $(20000)$,
$\epsilon_1=1$ and $r,r'=0$.
The tension of various orientifolds then becomes,
\begin{eqnarray}
 T(\SC^{B,{\rm id},+\epsilon_2\epsilon_3\epsilon_4\epsilon_5}
      _{{\bf M}=(00000)})
 &=& c\ssub\NS T_0^{(\rm id)}
     (\cos\tfrac\pi8)^{4-\alpha}
     (-i\sin\tfrac\pi8)^{\alpha},
 \nn\\
 T(\SC^{B,{\rm id},+\epsilon_2\epsilon_3\epsilon_4\epsilon_5}
      _{{\bf M}=(20000)})
 &=& c\ssub\NS T_0^{(\rm id)}
     (\cos\tfrac\pi8)^{3-\alpha}
     (-i\sin\tfrac\pi8)^{\alpha}
      \delta_{\epsilon_2,+},
 \nn\\
 T(\SC^{B,(12),+\epsilon_3\epsilon_4\epsilon_5}
      _{{\bf M}=(00000)})
 &=& c\ssub\NS T_0^{(12)}
     (\cos\tfrac\pi8)^{3-\alpha}
     (-i\sin\tfrac\pi8)^{\alpha},
 \nn\\
 T(\SC^{B,(12),+\epsilon_3\epsilon_4\epsilon_5}
      _{{\bf M}=(20000)})
 &=& 0,
 \nn\\
 T(\SC^{B,(34),+\epsilon_2\epsilon_5}
      _{{\bf M}=(00000)})
 &=& c\ssub\NS T_0^{(34)}
     (\cos\tfrac\pi8)^{2-\alpha}
     (-i\sin\tfrac\pi8)^{\alpha},
 \nn\\
 T(\SC^{B,(34),+\epsilon_2\epsilon_5}
      _{{\bf M}=(20000)})
 &=& c\ssub\NS T_0^{(34)}
     (\cos\tfrac\pi8)^{1-\alpha}
     (-i\sin\tfrac\pi8)^{\alpha}
      \delta_{\epsilon_2,+},
 \nn\\
 T(\SC^{B,(12)(34),+\epsilon_5}
      _{{\bf M}=(00000)})
 &=& c\ssub\NS T_0^{(12)(34)}
     (\cos\tfrac\pi8)^{1-\alpha}
     (-i\sin\tfrac\pi8)^{\alpha},
 \nn\\
 T(\SC^{B,(12)(34),+\epsilon_5}
      _{{\bf M}=(20000)})
 &=& 0.
\end{eqnarray}
Here $\alpha$ denotes the number of $\epsilon_a$'s taking $(-)$ sign,
and $T_0^\pi$ are some positive definite constants.
In order to make the tension real, one therefore have to put
$c\ssub\NS=\pm i^\alpha$.
A useful relation is $(-)^\alpha=c\ssub\NS^2=\rho(\vec\nu=\vec1)$.

\section{Some String Theory Problems}\label{sec:SSTP}

In this section we wish to study some more properties of
permutation branes and orientifolds in Gepner models.
One important problem is to find out the spectrum of massless
open string modes.
Here we will restrict our attention to the gauge fields on D-brane
worldvolumes and study what gauge group is realized on coincident
D-branes, by analyzing the action of parity on D-branes and open strings.
Another important problem is to solve the tadpole cancellation
condition and find supersymmetric tadpole-free configurations.
The tadpole cancellation in general simply amounts to the cancellation
of D-brane charges against the charge of orientifold.
It becomes more and more difficult to solve it as the dimension of
charge lattice gets larger.
For type IIA case, we will analyze in a similar way as in
\cite{Brunner-HHW} and find a few solutions involving permutation
orientifolds using the simple relations between the charges of
D-branes and orientifolds in minimal models.
For type IIB we see that the charges of D-branes and orientifolds
are summarized into simple polynomials as was discussed in
\cite{Brunner-DLR, Recknagel, Brunner-G, Becker-BVW}.

\subsection{Parity action on D-branes}\label{sec:STPAD}

We would like to find out here the action of various orientifolds
of Gepner model on D-branes from M\"obius strips amplitudes.
We begin with the M\"obius strips in the product
of $r$ minimal models,
\begin{eqnarray}
 \ssub{\NSNS+}\bra{\SB^\sigma_{\bf L,M}}
 q^H\ket{\SC^\pi_{2{\bf m}}}\ssub{\NSNS\pm}
 &=& 
 \ssub{\NSNS\mp}\bra{\SC^\pi_{2{\bf m}}}
 q^H\ket{\SB^{\sigma'}_{{\bf L},{\bf M}'}}
 \ssub{\NSNS+},
 \nn\\
 \ssub{\RARA-}\bra{\SB^\sigma_{\bf L,M}}
 q^H\ket{\SC^\pi_{2{\bf m}-{\bf 1}}}\ssub{\RARA\pm}
 &=& 
 \ssub{\RARA\mp}\bra{\SC^\pi_{2{\bf m}-{\bf 1}}}q^H
 \ket{\SB^{\sigma'}_{{\bf L},{\bf M}''}}
 \ssub{\RARA+}
 (-)^{r+|\sigma|+|\pi|}.
\end{eqnarray}
Here $\sigma'\equiv\pi\sigma^{-1}\pi$, and
${\bf M}',{\bf M}''$ have the following components,
\[
 M'_c  = \sum_{a\in\sigma'_c}2m_a - M_c,~~~~~
 M''_c = \sum_{a\in\sigma'_c}(2m_a-1) - M_c.
\]
The minus signs in the second line come from the coefficients
$\beta\ssub{M,\SEC}, \alpha\ssub\SEC$ in (\ref{BCmin}), (\ref{BP})
and (\ref{Cmin2}).
By taking the sum over orbits of the orbifold groups $\ZZ_H$ or
$\Gamma_{\rm mir}$ we find the M\"obius strip amplitudes
for A-type crosscaps,
\begin{eqnarray}
 \ssub{\NSNS+}\bra{\SB^{A,\sigma}_{\bf L,M}}
 q^H\ket{\SC^{A,\pi,\epsilon}_{2{\bf m}}}\ssub{\NSNS\pm}
 &=& 
 \ssub{\NSNS\mp}\bra{\SC^{A,\pi,\epsilon}_{2{\bf m}+{\bf 2}}}
 q^H\ket{\SB^{A,\sigma'}_{{\bf L},{\bf M}'}}
 \ssub{\NSNS+}\cdot
 \epsilon c\ssub\NS^2
 \nn\\
 \ssub{\RARA-}\bra{\SB^{A,\sigma}_{\bf L,M}}
 q^H\ket{\SC^{A,\pi,\tilde\epsilon}_{2{\bf m}-{\bf 1}}}\ssub{\RARA\pm}
 &=& 
 \ssub{\RARA\mp}\bra{\SC^{A,\pi,\tilde\epsilon}_{2{\bf m}+{\bf 1}}}q^H
 \ket{\SB^{A,\sigma'}_{{\bf L},{\bf M}''}}
 \ssub{\RARA+}
 \cdot
 \tilde\epsilon
 (-)^{r+|\sigma|+|\pi|}.
\end{eqnarray}
Here $\tilde\epsilon$ was defined in (\ref{ACQ}).
Recalling that $c\ssub\NS$ was determined so that
$\epsilon c\ssub\NS^2=1$, we conclude
\begin{equation}
 \SC^{A,\pi,\epsilon}_{\bf 2m}~:~
 \SB^{A,\sigma}_{\bf L,M} ~\mapsto~
 \SB^{A,\sigma'}_{\bf L,M'}\cdot(-)^{\frac12(r+D)+|\pi|+|\sigma|}\epsilon,
\label{PBA}
\end{equation}
where $\sigma,{\bf M}'$ are defined above and the $\pm$ sign
distinguishes the brane and antibranes, i.e. $-\SB$ denotes
the antibrane of $\SB$.
The rule for B-type orientifolds is similar,
\begin{equation}
 \SC^{B,\pi,\rho}_{\bf 2m}~:~
 \SB^{B,\sigma}_{\bf L,M} ~\mapsto~
 \SB^{B,\sigma'}_{\bf L,M'}\cdot(-)^{\frac12(r+D)+|\pi|+|\sigma|}
 \rho(\vec\nu=\vec1),
\label{PBB}
\end{equation}
where $\rho(\vec\nu)$ specifies a character of $\Gamma_O$ (\ref{CBY}).
So the condition for a brane to be parity-invariant is
$\pi\sigma^{-1}\pi=\sigma$ and
$({\bf L,M})=({\bf L},{\bf M}')$ up to shifts of $\bf M$
by orbifold elements and an even or odd times of brane identification
$\SF_c$ (\ref{FIB}) depending on the sign in the above formulae.

\vskip2mm

For later use, we study the pairs of brane and orientifold satisfying
the condition $({\bf L,M})=({\bf L,M'})$ up to $\SF_c$ by decomposing
into blocks.

\begin{PIB}\label{pib2}
If a brane $\SB^\sigma_{\bf L,M}$ in the theory $\otimes_aM(k_a)$ is
invariant under $\SC^\pi_{\bf\bar M}$~$(M_a=M_{\pi(a)})$, the pair
$(\pi,\sigma)$ decomposes into the blocks listed in {\rm PIB \ref{pib1}}.
For each block of type {\rm (1)--(3)} of {\rm PIB \ref{pib1}},
\[
\begin{array}{cll}
(1) & \sigma_c=(a_1a_2\cdots a_{2n+1}),
    & \pi     =(a_1a_{2n+1})(a_2a_{2n})\cdots(a_na_{n+2}), \\
(2) & \sigma_c=(a_1a_2\cdots a_{2n}),
    & \pi     =(a_2a_{2n+1})(a_2a_{2n})\cdots(a_na_{n+2}), \\
(3) & \sigma_c=(a_1a_2\cdots a_{2n}),
    & \pi     =~~~\,(a_1a_{2n})(a_2a_{2n})\cdots(a_na_{n+1}), \\
\end{array}
\]
the labels $(L_c,M_c)$ have to satisfy
\[
\begin{array}{rcll}
& ({\rm I}) & L_c={\rm any}, &  M_c=\tfrac12\bar M_c^{(\rm tot)} ~~{\rm or}~~
                                   \tfrac12\bar M_c^{(\rm tot)}+k_c+2,\\
{\rm or} &
({\rm II}) & L_c=\frac k2 , &  M_c=\tfrac12\bar M_c^{(\rm tot)}
                                \pm\tfrac{k_c+2}2,
\end{array}
~~~\left(\bar M_c^{\rm (tot)}\equiv\sum_{a\in\sigma_c}\bar M_a\right)
\]
and for each block of type {\rm (4)} of the list,
\[
 (4)~~\sigma_c\sigma_{c'}=(a_1\cdots a_n)(a'_1\cdots a'_n),
    ~~             \pi   =(a_1a'_n)(a_2a'_{n-1})\cdots(a_na'_1),
\]
the labels $(L_c,M_c),(L_{c'},M_{c'})$ have to satisfy
\[
\begin{array}{rll}
({\rm III}) & L_c=L_{c'},  & M_c+M_{c'}=\bar M^{(\rm tot)}, \\
{\rm or}~~
({\rm IV})  & L_c+L_{c'}=k,& M_c+M_{c'}=\bar M^{(\rm tot)}+k_c+2.
\end{array}
   \left(\bar M^{\rm (tot)}
   \equiv\sum_{a\in\sigma_c}\bar M_a
   \equiv\sum_{a\in\sigma_{c'}}\bar M_a \right)
\]
Thus the pair $(\SB^\sigma_{\bf L,M},\SC^\pi_{\bf\bar M})$ decomposes
into eight different kinds of blocks,
\[
 (1)_{\rm I},~~
 (1)_{\rm II},~~
 (2)_{\rm I},~~
 (2)_{\rm II},~~
 (3)_{\rm I},~~
 (3)_{\rm II},~~
 (4)_{\rm III},~~
 (4)_{\rm IV}.
\]
\end{PIB}

\paragraph{Parity and supersymmetry.}
The action of parity on D-branes obtained above is such that
the parity reversal of a supersymmetric configuration is again
supersymmetric.
Namely, if $\ket\SB$ preserves the same supersymmetry as $\ket\SC$,
so does $P_\SC\ket\SB$.
To see this, recall the supersymmetry phases for A-type branes and crosscaps,
\begin{eqnarray}
 \exp i\pi\varphi(\SB^{A,\sigma}_{\bf L, M}) &=&
 \exp i\pi\left(\tsum_{c=1}^{[\sigma]}
                \tfrac{M_c}{k_c+2}
               -\tfrac{\lc{\sigma_c}-1}2\right), \nn\\
 \exp i\pi\varphi(\SC^{A,\pi,\epsilon}_{\bf M}) &=&
 \tfrac{c\ssub\RA}{c\ssub\NS}
 \exp i\pi\left(\tsum_{a=1}^r
                \tfrac{M_a-1}{2k_a+4}
               +\tfrac{|\pi|+r}2\right),
\end{eqnarray}
where $\lc{\sigma_c}$ denotes the length of the $c$-th cycle of $\sigma$,
and $|\pi|$ denotes the number of cycles of length 2 in $\pi$.
Similar expressions hold also for B-types.
Also, recall that $c\ssub\RA=\pm1$, and that $c\ssub\NS$ is determined
from the group character as follows,
\begin{equation}
  c\ssub\NS^2\epsilon            = 1~~~~(\mbox{A-type})~~;~~~~
  c\ssub\NS^2\rho(\vec\nu=\vec1) = 1~~~~(\mbox{B-type}).
\end{equation}
Combining these together with (\ref{PBA}) or (\ref{PBB}) one can show that,
for any pair of an orientifold $\SC$ and a D-brane $\SB$,
\begin{equation}
 \varphi(P_\SC\SB)~=~2\varphi(\SC)-\varphi(\SB)~~~({\rm mod}~~2).
\end{equation}

\vskip4mm

The formulae (\ref{PBA}) and (\ref{PBB}) determine the action
of orientifolds on all the {\it long-orbit branes}, or branes
with trivial untwisted stabilizer group $\cal U$.
We need some more work to find out the action of orientifolds
on {\it short-orbit branes} which have non-trivial $\cal U$
and are therefore labelled by additional label specifying a character of
$\cal U$.

\subsubsection{Parity action on short-orbit A-branes}\label{sec:STPASA}

Short-orbit A-branes are made from permutation boundary states
in twisted sectors, $\ket{\SB^{(1\cdots N)}_{L,M}}^{(\mu)}$ and
$\ket{\wtilde\SB^{(1\cdots N)}_{k/2,M}}^{(\mu)}$, in the product
of $N$ identical minimal models $M(k)^{N}$ defined in
(\ref{BLMmu1}) and (\ref{BLMmu2}).
They satisfy the basic transformation laws (here
$\omega\equiv e^{\frac{2\pi i}{k+2}}$)
\begin{equation}
\begin{array}{rcl}
 \ket{\SB^{(2\cdots N1)}_{L,M}}^{(\mu)}\ssub{\NSNS\pm}
 &=& +\ket{\SB^{(12\cdots N)}_{L,M}}^{(\mu)}\ssub{\NSNS\pm}\omega^{M\mu},
      \nn\\
 \ket{\wtilde\SB^{(2\cdots N1)}_{k/2,M}}^{(\mu)}\ssub{\NSNS\pm}
 &=& \mp\ket{\wtilde\SB^{(12\cdots N)}_{k/2,M}}^{(\mu)}\ssub{\NSNS\pm}
        \omega^{M\mu},
\end{array}
\label{BTWA1}
\end{equation}
\begin{equation}
\begin{array}{rcl}
 \gamma_a\ket{\SB^{(12\cdots N)}_{L,M}}\ssub\SEC^{(\mu)}
 &=& +\ket{\SB^{(12\cdots N)}_{L,M+2}}\ssub\SEC^{(\mu)}\omega^{\mu(2a-1)},
      \nn\\
 \gamma_a\ket{\wtilde\SB^{(12\cdots N)}_{k/2,M}}\ssub\SEC^{(\mu)}
 &=& -\ket{\wtilde\SB^{(12\cdots N)}_{k/2,M+2}}\ssub\SEC^{(\mu)}
      \omega^{\mu(2a-1)}.
\end{array}
\label{BTWA2}
\end{equation}

We study the action of NS parity $\SC^\pi_{\bf \bar M}$ on these
boundary states.
It maps the $\sigma$-permuted boundary states to $\sigma'$-permuted
boundary states, where
\begin{equation}
 \sigma  = (1\,2\cdots N)~~\Longrightarrow~~
 \sigma' = \pi\sigma^{-1}\pi = (\pi(N)\;\pi(N-1)\cdots \pi(1)).
\label{sigma'}
\end{equation}
The NS parity acts on Ishibashi states as
\begin{eqnarray}
(-)^{F_L}P^\pi_{\bf \bar M}
     \kket{\SB^\sigma;l,m}^{(\mu)}\ssub{\NSNS\pm}
 &=& \otimes_a\gamma_a^{\bar M_a/2}\cdot
     \kket{\SB^{\sigma'};l,-m}^{(\mu)}\ssub{\NSNS\pm}, \nn\\
(-)^{F_L}P^\pi_{\bf \bar M}
     \kket{\wtilde\SB^\sigma;l,m}^{(\mu)}\ssub{\NSNS\pm}
 &=& \otimes_a\gamma_a^{\bar M_a/2}\cdot
     \kket{\wtilde\SB^{\sigma'};l,k+2-m}^{(\mu)}\ssub{\NSNS\pm}\cdot(\pm i).
\end{eqnarray}
Therefore the boundary states are transformed as,
\begin{eqnarray}
     (-)^{F_L}P^\pi_{\bf \bar M}
     \ket{\SB^\sigma_{L,M}}\ssub{\NSNS\pm}^{(\mu)}
 &=& \otimes_a\gamma_a^{\bar M_a/2}\cdot
     \ket{\SB^{\sigma'}_{L,-M}}\ssub{\NSNS\pm}^{(\mu)},
 \nn\\
     (-)^{F_L}P^\pi_{\bf \bar M}
     \ket{\wtilde\SB^\sigma_{L,M}}\ssub{\NSNS\pm}^{(\mu)}
 &=& \otimes_a\gamma_a^{\bar M_a/2}\cdot
     \ket{\wtilde\SB^{\sigma'}_{L,-M}}\ssub{\NSNS\pm}^{(\mu)}\cdot(\mp i).
\end{eqnarray}

The above formula can be directly applied to the parity action
on short-orbit A-branes in Gepner models.
A general permutation A-brane with stabilizer group $\ZZ_{H'}~(H'=H/h)$
takes the form (\ref{SOA}),
\begin{equation}
 \ket{\SB^{A,\sigma,\rho}_{\bf L,M}}\ssub\SEC ~=~ \frac{1}{\sqrt H}
 \sum_{\nu\in \ZZ_h}\sum_{\mu\in \ZZ_{H'}}
 \gammaA^\nu\bigotimes_{c=1}^{[\sigma]}
 \ket{\SB^{\sigma_c}_{L_c,M_c}}\ssub\SEC^{(\mu h)}
 \exp\left(\tfrac{2\pi i\mu h\rho}{H}\right).
\end{equation}
The orientifolds $\SC^{A,\pi,\epsilon}_{\bf\bar M}$ maps the brane
$\ket{\SB^{A,\sigma,\rho}_{\bf L,M}}$ to
$\otimes_a\gamma_a^{\bar M_a/2}\cdot\ket{\SB^{A,\sigma',\rho'}_{\bf L,-M}}$.
The permutations $\sigma$ and $\sigma'$ are related cycle by
cycle as follows,
\begin{equation}
 \sigma_c=(a_1\cdots a_{n}) ~\Longleftrightarrow~
 \sigma'_c=(\pi(a_n)\cdots \pi(a_1)).
\end{equation}
The mod-$H'$ integer $\rho$ gets shifted according to the following
rules:
\begin{enumerate}
\item $\rho$ gets shifted by $\frac H2=\frac{hH'}2$
      if $H$ is even and the orientifold has $\epsilon=(-)$.
\item $\rho$ gets shifted by $\frac{nH'}2$ if the boundary state in
      $\gammaA^h$-twisted sector contains $2n$ tilded boundary states.
\end{enumerate}

As an application, let us find out the condition for an A-brane
$\SB^{A,\sigma,\rho}_{\bf L,M}$ to be invariant under the A-type
orientifold $\SC^{A,\pi,\epsilon}_{\bf\bar M}$.
For simplicity, we assume their labels are chosen in such a way
that the pair $(\SB^\sigma_{\bf L,M},\SC^\pi_{\bf\bar M})$ satisfy
the condition PIB \ref{pib2}.
The problem is then how the label $\rho$ is transformed under the parity.
Besides the possible shifts of $\rho$ listed above, it gets shifted when
we use the formula (\ref{BTWA1}), (\ref{BTWA2}) or the identification
$\SF_c$ to transform the labels $(\pi\sigma^{-1}\pi,{\bf L,M'})$ into
$(\sigma,{\bf L,M})$.
A detailed analysis shows
\begin{enumerate}
\setcounter{enumi}{2}
\item $\rho$ gets shifted by $\frac{nH'}2$ if the boundary state in
      $\gammaA^h$-twisted sector contains $n$ tilded boundary states
      of type $(1)_{\rm II}, (2)_{\rm II}$ or $(3)_{\rm II}$.
\item $\rho$ gets shifted by $(1+\frac{\bar M_{a_1}}2)\frac{H'}{2}$ or
      $(h+1+\frac{\bar M_{a_1}}2)\frac{H'}{2}$ if the boundary state in
      $\gammaA^h$-twisted sector contains a tilded boundary states
      of type $(2)_{\rm I}$ or $(2)_{\rm II}$.
\end{enumerate}
In any case, the action of orientifold on $\rho$ of
the brane $\SB^{A,\sigma,\rho}_{\bf L,M}$ is at most a half
period shift, and it only occurs when $\bf L$ is special so that
the tilded boundary states are involved in its construction.
The parity action on the label $\rho$ is thus determined from the expression
of boundary state in $\gammaA^h$-twisted sector.
Whether $\rho$ is invariant or shifted by half-period is
determined by the following sign (where the notation should be
obvious from the above explanation),
\begin{equation}
 \lambda ~\equiv~
 \epsilon^h(-)^{\frac12\sharp(\tilde B)-\sharp(\tilde B_{(\rm II)})}
 \prod_{\tilde B:(2)_{\rm  I}}(-)^{\frac{\bar M_{a_1}}{2}+1}
 \prod_{\tilde B:(2)_{\rm II}}(-)^{\frac{\bar M_{a_1}}{2}+1+h}.
\label{lambdaA}
\end{equation}

\subsubsection{Parity action on short-orbit B-branes}\label{sec:STPASB}

B-branes in Gepner model with nontrivial untwisted stabilizer $\cal U$
are made of permutation boundary states $\ket{\SB_{L,M}^{\sigma,\rho}}$
defined at (\ref{BPrho}), with $\sigma=(12\cdots N)$ a cycle of even length.
Each of $\ket{\SB_{L,M}^{\sigma,\rho}}$ contributes a factor of
$\ZZ_{k+2}$ or $\ZZ_{(k+2)/2}$ to $\cal U$, depending on whether
$L$ is generic or coincides with $k/2$.
For the D-branes whose untwisted stabilizer contain the generator
$\gammaB$ of (\ref{genU3}), we construct the boundary states in
$\gammaB$-twisted sector using
$\ket{\wtilde\SB^{\sigma}_{k/2,M}}$ and
$\ket{\SB^{\sigma,\rho,-}_{k/2,M}}$
defined at (\ref{BPeta}) and (\ref{Bpirpm}).

The $\sigma$-permuted short orbit B-branes are therefore labelled by
the half integers $(\rho_c)$, and also by a sign $\varepsilon$ if
${\cal U}$ contains the element $\gammaB$.
Each $\rho_c$ is associated to an even-length cycle $\sigma_c$, and has
period $k_c+2$ or $(k_c+2)/2$ depending on whether $L_c$ is generic or not.
The sign $\varepsilon$ appears in the expression for boundary states
as follows,
\begin{eqnarray}
 \ket{\SB^{B,\sigma,(\rho,\varepsilon)}_{\bf L,M}}
 &\sim&
 \sum_{\otimes\gamma_a^{\nu_a}\,\in\,\Gamma_{\rm mir}/{\cal H}}
 \otimes_a\gamma_a^{\nu_a}\cdot
 \bigotimes_{\rm  odd}^{(L_c\ne \frac{k_c}{2})}
 \ket{\SB^{\sigma_c}_{L_c,M_c}}
 \bigotimes_{\rm even}^{(L_c\ne \frac{k_c}{2})}
 \ket{\SB^{\sigma_c,\rho_c}_{L_c,M_c}}
 \nn\\ && \times\left\{
 \bigotimes_{\rm  odd}
 \ket{\SB^{\sigma_c}_{\frac{k_c}2,M_c}}
 \bigotimes_{\rm even}
 \ket{\SB^{\sigma_c,\rho_c,+}_{\frac{k_c}2,M_c}}
 + \varepsilon
 \bigotimes_{\rm  odd}
 \ket{\wtilde\SB^{\sigma_c}_{\frac{k_c}2,M_c}}
 \bigotimes_{\rm even}
 \ket{\SB^{\sigma_c,\rho_c,-}_{\frac{k_c}2,M_c}}
  \right\}.~~~
\end{eqnarray}
An example is the boundary state (\ref{Bpirre}) for a B-brane
in the $(88444)$ model.

We wish to find out the action of various B-type (NSNS) parities on
B-branes, in particular how the labels $(\rho_c,\varepsilon)$ are
transformed.
We consider the parity $P^{B,\pi,\vec r}_{\bf\bar M}$
corresponding to a general B-type orientifold,
\[
 \ket{\SC^{B,\pi,\vec r}_{\bf\bar M}} ~=~ \frac{1}{\sqrt{|\Gamma_{\rm mir}|}}
 \sum_{\otimes_a\gamma_a^{\nu_a}\in\Gamma_{\rm mir}}
 \ket{\SC^{\pi}_{{\bf\bar M}+2\vec\nu}}\exp\left(
 -\tsum_a\tfrac{2\pi ir_a\nu_a}{k_a+2}  \right).
\]
Actually the transformation law of $\{\rho_c\}$ is obtained simply
by applying the general formula (\ref{Pchtr}), thanks to the fact
that the boundary states in twisted sector is essentially unique
unlike the case with A-branes (cf. equation (\ref{BTWA1})).
To illustrate this, let us work out the condition on $\rho$-labels
for a B-brane $\SB^{B,\sigma,(\rho_c,\epsilon)}_{\bf L,M}$ to be
invariant under the orientifold $\SC^{B,\pi,\vec r}_{\bf\bar M}$.

\begin{PIB}\label{pib3}
Take a pair $(\SB^\sigma_{\bf L,M},\SC^\pi_{\bf\bar M})$ satisfying
the condition PIB \ref{pib2}.
Then the B-type orientifold $\SC^{B,\pi,\vec r}_{\bf\bar M}$
acts on the $\rho$-labels of the B-brane
$\SB^{B,\sigma,\rho}_{\bf L,M}$ in a non-trivial manner.
By analyzing the condition of parity invariance on $\rho$ block by block
one finds the following:
\begin{enumerate}
\item the blocks of type $(1)$ do not contain $\rho$-labels.
\item in a block of type $(2)$, the boundary state
  $\SB^{\sigma_c,\rho_c}_{L_c,M_c}~(\sigma_c=(a_1a_2\cdots a_{2n}))$
  has the label $\rho_c$ which transform under parity as
\[
 \rho_c~\mapsto~~\rho_c+r^{(\rm tot)}, ~~~
 r^{(\rm tot)}\equiv r_{a_1}-r_{a_2}+\cdots -r_{a_{2n}}.
\]
  It follows from the involutiveness of parity that $r^{(\rm tot)}=0$
  or $\frac{k_c+2}{2}$ mod $k_c+2$. If the latter is the case $L_c$
  has to equal $k_c/2$, but there arise no condition on $\rho_c$.
\item in a block of type $(3)$ the parity transform the $\rho$-label
  as $\rho_c~\mapsto~-\rho_c-r^{(\rm tot)}$,
  where $\rho_c,r^{(\rm tot)}$ are defined similarly to the previous case.
  The parity invariance requires
\[
\begin{array}{rcll}
 ({\rm A}) &:& \rho_c=-\frac12r^{(\rm tot)}~~{\rm mod}~~ \frac{k_c+2}{2},&
  L_c={\rm any},
\\ {\rm or}~~~
 ({\rm B}) &:& \rho_c=-\frac12r^{(\rm tot)}+\frac{k_c+2}{4}~~{\rm mod}~~
 \frac{k_c+2}{2},&
 L_c=\frac{k_c}{2}.
\end{array}
\]
\item in a block of type $(4)$, we take
 $\sigma_c\circ\sigma_{c'}=(a_1\cdots a_{2n})\circ(a'_1\cdots a'_{2n})$ and
 consider the boundary state
 $\SB^{\sigma_c,   \rho_c   }_{L_c   ,M_c   }\otimes
  \SB^{\sigma_{c'},\rho_{c'}}_{L_{c'},M_{c'}}$.
 The parity acts on the labels $\rho_c,\rho_{c'}$ as
\[
\begin{array}{rcl}
  \rho_c    &\mapsto& -r^{(\rm tot)}-\rho_{c'},\\
  \rho_{c'} &\mapsto& -r^{(\rm tot)}-\rho_c   ,
\end{array}
~~~
\begin{array}{rcl}
 r^{(\rm tot)}&\equiv& r_{a_1 }-r_{a_2 }+r_{a_3 }\cdots -r_{a_{2n}} \\
               &=&     r_{a'_1}-r_{a'_2}+r_{a'_3}\cdots -r_{a'_{2n}}.
\end{array}
\]
The parity-invariant blocks of type $(4)_{\rm III}$ or $(4)_{\rm IV}$
have to satisfy
\[
\begin{array}{rcl}
 ({\rm III}) &:& \rho_c+\rho_{c'}+r^{(\rm tot)}=0
  ~~{\rm mod}~{k_c+2},  \\
 ({\rm IV }) &:& \rho_c+\rho_{c'}+r^{(\rm tot)}=\frac{k_c+2}{2}
  ~~{\rm mod}~{k_c+2}.
\end{array}
\]
\end{enumerate}
The pair $(\SB^{B,\sigma,\rho}_{\bf L,M}, 
           \SC^{B,\pi,\vec r}_{\bf\bar M})$ therefore decomposes
into blocks of 10 different kinds,
\[
 (1)_{\rm I},~
 (1)_{\rm II},~
 (2)_{\rm I},~
 (2)_{\rm II},~
 (3)_{\rm IA},~
 (3)_{\rm IB},~
 (3)_{\rm IIA},~
 (3)_{\rm IIB},~
 (4)_{\rm III},~
 (4)_{\rm IV}.
\]

\end{PIB}

\paragraph{Parity action on $\varepsilon$.}

A naive application of the formula (\ref{Pchtr}) does not work for
determining the action of parity on $\varepsilon$ because we have been
making no distinction between $\gamma^{\frac{k+2}{2}}$-twisted sector
and $\psi\gamma^{\frac{k+2}{2}}$-twisted sector of minimal models.
Here we focus on short-orbit B-branes
$\SB^{B,\sigma,(\rho,\varepsilon)}_{\bf L,M}$ satisfying the condition
PIB\ref{pib3} discussed above and ask what is the relation between
$\varepsilon$ and $\varepsilon'$ in the formula:
\[
  (-)^{F_L}P^{B,\pi,\vec r}_{\bf\bar M}~:~
  \SB^{B,\sigma,(\rho,\varepsilon)}_{\bf L,M}~\mapsto
  \SB^{B,\sigma,(\rho,\varepsilon')}_{\bf L,M}.
\]
The result is summarized as
\begin{equation}
  \frac{\varepsilon'}{\varepsilon} ~=~
        (-)^{\sharp(1)_{\rm II}+\sharp(2)_{\rm II}
            +\sharp(3)_{\rm IIA}+\sharp(3)_{\rm IIB}}
  \cdot (-)^{\sharp(3)_{\rm IB}+\sharp(3)_{\rm IIB}}
  \cdot \hskip-3mm
  \prod_{\sigma_c~{\rm odd,}~L_c=k_c/2}
  (-i)\cdot(-)^{\sum_{a\in\sigma_c}r_a},
\label{lambdaB}
\end{equation}
where $\sharp(\cdots)$ counts the number of blocks of each type.
The factors in the right hand side arise from the following reason.
The first sign $(-)^{\sharp(1)_{\rm II}+\sharp(2)_{\rm II}
                    +\sharp(3)_{\rm IIA}+\sharp(3)_{\rm IIB}}$
arises because the states $\ket{\wtilde\SB^{\sigma}_{k/2,M}}$,
$\ket{\SB^{\sigma,\rho,-}_{k/2,M}}$ are odd under the shift $M\to M+k+2$.
The second sign $(-)^{\sharp(3)_{\rm IB}+\sharp(3)_{\rm IIB}}$
is from the states $\ket{\SB^{\sigma,\rho,-}_{k/2,M}}$ which are odd
under the shift $\rho\to \rho+\frac{k+2}{2}$.
The last factor arises from the odd-length cycles $\sigma_c$
labelled by $L_c=k_c/2$.
A $(-i)$ is due to the parity action
\[
 (-)^{F_L}P^{\pi}_{\bf\bar M}\ket{\wtilde B^{\sigma}_{k/2,M}}\ssub{\NSNS\pm}
 ~=~\mp i\ket{\wtilde B^{\sigma}_{k/2,\bar M_{\rm tot}-M}}\ssub{\NSNS\pm}.
\]
The $r_a$-dependent sign arises from the action of quantum symmetry
labelled by $\vec r$ on states sitting in
$(\eta_{a_1}\cdots\eta_{a_n})$-twisted sector.

\subsection{Gauge group}\label{sec:STGG}

If a brane $\SB$ is invariant under the orientifold $\SC$,
then the corresponding M\"obius strip amplitude shows a massless gauge
boson running along the strip.
The parity eigenvalue of the gauge boson determines whether the
gauge group is $O$ or $Sp$.
We read off the eigenvalues of NS parities $(-)^{F_L}P_\SC$ or
$(-)^{F_R}P_\SC$ for the orientifold $\SC$ from the amplitudes
\[
    \mp i\;\ssub{\NSNS+}\vev{\SB|q^H|\SC}\ssub{\NSNS\pm}
 ~=~\mp i\;\ssub{\NSNS\mp}\vev{\SC|q^H|\SB}\ssub{\NSNS+}.
\]
We regard $\mp i$ as the value of NS parities for
open string NS ground state.
Since NS parities square to fermion number, it follows that the
NS tachyon (and all the NS states that are projected out by GSO
projection) has {\it odd} fermion number, and the remaining states
have eigenvalues $\pm1$ of the NS parities.
The gauge group is $O$ or $Sp$ depending on the gauge boson having
eigenvalues $-1$ or $1$ of NS parities.

We compute the eigenvalues of NS parities by decomposing the
M\"obius strip amplitudes into parts.
The spacetime part of the amplitude reads
\[
    \mp i\;\ssub{\NSNS+\!}^{~~~~~~\rm st}
    \vev{\SB|e^{-\pi H_c/4l}|\SC}\ssub{\NSNS\pm}^{\rm st}
 ~ \sim ~ \mp i\cdot q^{-\frac{c_{\rm st}}{24}-\frac12}
   \{\widehat\chi_0(q)\mp i\widehat\chi_2(q)\}^d
    e^{\pm\frac{i\pi d}{4}}~~~~
    (q\equiv e^{-2\pi l})
\]
where $\chi_s$ are characters of $U(1)_2$ and the hat operation is
defined in (\ref{hatch}).
The spacetime part therefore contributes $-e^{\frac{\pm i\pi d}{4}}$
to the eigenvalue of $(-)^{F_{L,R}}P_\SC$ on gauge boson.
The internal part, if the brane is parity invariant, can be
studied by decomposing them into blocks as explained in section
\ref{sec:PINVD}.
Let us forget about the orbifolding for the moment and first
consider M\"obius strip of a single minimal model,
\begin{eqnarray}
\lefteqn{
  \ssub{\NSNS+}\bra{\SB_{L,M}}e^{-\frac{\pi H_c}{4\ell}}
  \ket{\SC_{\bar M}}\ssub{\NSNS\pm}
}\nn\\
  &=& \sum_{l=0}^{{\rm min}(L,k-L)}
      \left\{
      (-)^{l+L-\frac{\bar M}{2}}e^{\mp\frac{i\pi}{4}}
      \widehat\chi^{\NS\mp}_{2l,2M-\bar M}(q)
  +   e^{\pm\frac{i\pi}{4}}
      \widehat\chi^{\NS\mp}_{k-2l,2M-\bar M-k-2}(q)
      \right\},
\end{eqnarray}
where $\widehat\chi^{\NS\pm}_{l,m}$ are linear combinations
of hatted characters in minimal model,
\begin{equation}
 \widehat\chi^{\NS\pm}_{l,m} ~\equiv~
  \sigma_{lm0}\widehat\chi_{l,m,0}
 \pm i\sigma_{lm2}\widehat\chi_{l,m,2},
\end{equation}
and $\sigma_{lms}=e^{i\pi\theta(l,m,s)}$ was defined at (\ref{hlms}).
From the coefficient of $\widehat\chi^{\NS\pm}_{0,0}$ one finds
the value of NS parities on the ground state,
\begin{equation}
\begin{array}{llllll}
  L={\rm any},& M=\frac{\bar M}2~{\rm or}~
                  \frac{\bar M}2+k+2 & \Rightarrow &
 (-)^{F_{L,R}}P_{\bar M} &=& e^{\mp\frac{i\pi}4}, \\
  L=\frac k2,   & M=\frac{\bar M}2\pm\frac{k+2}2& \Rightarrow &
 (-)^{F_{L,R}}P_{\bar M} &=& e^{\pm\frac{i\pi}4}.
\end{array}
\end{equation}

We generalize this analysis to the pairs of a permutation brane
$\SB^\sigma_{\bf L,M}$ and orientifold $\SC^{\pi}_{\bf\bar M}$
in tensor products of minimal models.
We again assume $\bar M_a=\bar M_{\pi(a)}$ for simplicity.
We decompose them into blocks satisfying the condition PIB\ref{pib2}
and compute the values of NS parities block by block.
\vskip1mm\noindent
\underline{\bf NS Parity eigenvalue formula}
\begin{equation}
\begin{array}{lcl}
  (1)_{\rm I}   &:& (-)^{F_{L}}P = e^{-\frac{i\pi}{4}}, \\
  (1)_{\rm II}  &:& (-)^{F_{L}}P = e^{+\frac{i\pi}{4}}, \\
  (2)_{\rm I}   &:& (-)^{F_{L}}P = 1, \\
  (2)_{\rm II}  &:& (-)^{F_{L}}P = - i(-)^{\frac{\bar M_{a_1}}{2}}, \\
\end{array}
~~~
\begin{array}{lcl}
  (3)_{\rm I}   &:& (-)^{F_{L}}P = - i, \\
  (3)_{\rm II}  &:& (-)^{F_{L}}P = 1, \\
  (4)_{\rm III} &:& (-)^{F_{L}}P = 1, \\
  (4)_{\rm IV}  &:& (-)^{F_{L}}P = 1.
\end{array}
\end{equation}

To determine the gauge group on D-branes in Gepner model, one
has to combine the NS parity eigenvalue from all the blocks together
with the overall coefficient of the crosscap $c\ssub\NS$,
and then sum over orbifold images.

Let us start with type IIA and consider a brane
$\SB^{A,\sigma,\rho}_{\bf L,M}$ invariant under the
orientifold $C^{A,\pi,\epsilon}_{\bf\bar M}$.
The M\"obius strip amplitude is given by the sum over orbifold
orbit,
\begin{eqnarray}
 \ssub\SEC\bra{\SB^{A,\sigma,\rho}_{\bf L,M}}q^H
 \ket{\SC^{A,\pi,\epsilon}_{\bf\bar M}}\ssub{\SEC'}
 &=&
 \frac{1}{|{\cal H}|}\sum_{\gamma\in\Gamma}
 {}\ssub\SEC\bra{\SB^\sigma_{\bf L,M}}q^H
 \ket{\SC^\pi_{\gamma({\bf\bar M})}}\ssub{\SEC'}\epsilon(\gamma)c\ssub\NS
 \nn\\ &\equiv&
 \frac{1}{|{\cal H}|}\sum_{\gamma\in\Gamma}{\SM}(\gamma),
\label{GAMo}
\end{eqnarray}
where $\epsilon(\gamma)\equiv\epsilon^\nu$ when
$\gamma({\bf \bar M})={\bf \bar M}+{\bf 2}\nu$,
and ${\cal H}\subset\Gamma$ is the stabilizer group of the brane.
In the sum in the right hand side, there are $|{\cal H}|$ terms
satisfying the condition PIB\ref{pib2} and therefore contributing to
the NS parity eigenvalue.
However, for generic ${\bf L}$ the sum is trivial so that it simply
removes the factor $1/|{\cal H}|$ in front.
If ${\bf L}$ is such that the enhancement of the stabilizer group
occurs, the sum boils down to an average of two terms with $\gamma$
being identity or the generator $\gammaA^h$ of the stabilizer
group.
Expanding $\SM({\rm id})$ and $\SM(\gamma^h)$ as power series
in the loop-channel modular parameter, the coefficients of the
leading term gives the eigenvalues of operators $(-)^{F_L}P$
and $(-)^{F_L} \gammaA^hP$ on ground state.
The value of $\gammaA^h$ on open string ground state obtained in
this way should coincide with $\lambda$ at (\ref{lambdaA}).

Let us next consider type IIB case and take a brane
$\SB^{B,\sigma,(\rho,\varepsilon)}_{\bf L,M}$ invariant under
the orientifold $C^{B,\pi,\vec r}_{\bf\bar M}$.
The parity eigenvalue of NS ground state on the brane can be computed
by summing the M\"obius strips $\SM(\gamma)$ in the product of
minimal models satisfying the condition PIB\ref{pib3}.
When $\sigma$ contains a cycle $\sigma_c$ of even length,
this involves summing $\SM(\gamma)$ over orbits generated by
the elements $\gamma_{\sigma_c}\in{\cal U}$ defined at (\ref{gammapic}).
This not only enforces the condition PIB\ref{pib3} on $\rho_c$ but
moreover projects out the terms containing blocks of type
$(2)_{\rm II}, (3)_{\rm IB}$ and $(3)_{\rm IIA}$.
The terms which survive this averaging are therefore those consisting
only of the blocks
\[
 (1)_{\rm I}, (1)_{\rm II}, (2)_{\rm I}, (3)_{\rm IA}, (3)_{\rm IIB},
 (4)_{\rm III}, (4)_{\rm IV}.
\]
The non-trivial part of averaging thus amounts to the sum over
$\gamma\in (\ZZ_2)^{p-1}\subset\Gamma_{\rm mir}$,
where $p$ is the number of odd-length cycles $\sigma_c$
labelled by $L_c=\frac{k_c}2$ and $(\ZZ_2)^{p-1}$ is the group of
even-order monomials of $\eta_c^{(\rm tot)}\equiv\prod_{a\in\sigma_c}\eta_a$.
Including the spacetime part and other factors, the NS parity
eigenvalue of gauge bosons finally becomes
\begin{eqnarray}
 (-)^{F_L} P
 &=& -c\ssub\NS
     (-i)^{\frac12\left\{\sharp(1)_{\rm I}-\sharp(1)_{\rm II}-d\right\}
           +\sharp(3)_{\rm IA}}\times
 \nn\\ &&\times
 {\rm Re}\left(
  2^{-[p/2]}(-i)^{\sharp(1)_{\rm II}}
  \cdot\hskip-3mm
  \prod_{\sigma_c~{\rm odd},\,L_c=k_c/2}\hskip-8mm
  \left(1+i(-)^{\sum_{a\in\sigma_c}r_a}\right)
  \right).
\label{NSPB}
\end{eqnarray}

\subsubsection{Example 1: $(55555)$}\label{sec:STGGQ}

Let us study the gauge group on A-branes $\ket{\SB^{A,\sigma}_{\bf L,M}}$
in the model $(55555)$ which are invariant under the orientifold
$\ket{\SC^{A,\pi}_{\bf \bar M}}$.
We put $c\ssub\NS=-1$ and set ${\bf M}={\bf \bar M}={\bf 0}$ for simplicity.
For each of the allowed $\sigma$'s we compute the supersymmetry phase
of the brane $\ket{\SB^\sigma_{\bf L,0}}$ and the eigenvalue of
corresponding NS parity $\wtilde P$ and summarize them in the Table
\ref{table:gAqui} below.
Because $H$ is odd, the parity eigenvalue are computed simply
by multiplying the contributions from blocks.
\begin{table}[htb]
\begin{center}
{\small
\begin{tabular}[t]{|c||c|c|}
\multicolumn{3}{c}{$\pi={\rm id},~\varphi(\SC)=0$} \\
\hline
\hline
$\sigma$ & $\varphi(\SB)$ & $\wtilde P$ \\
\hline
${\rm id}$ & $0$        & $-1$ \\
$(12)    $ & $-\frac12$ & $-i$ \\
$(12)(34)$ & $-1$       & $+1$ \\
\hline
\end{tabular}
~~
\begin{tabular}[t]{|c||c|c|}
\multicolumn{3}{c}{$\pi=(12),~\varphi(\SC)=\frac12$} \\
\hline
\hline
$\sigma$ & $\varphi(\SB)$ & $\wtilde P$ \\
\hline
${\rm id} $ & $0$        & $-i$ \\
$(12)     $ & $-\frac12$ & $-1$ \\
$(34)     $ & $-\frac12$ & $+1$ \\
$(12)(34) $ & $-1$       & $-i$ \\
$(123)    $ & $-1$       & $-i$ \\
$(123)(45)$ & $-\frac32$ & $+1$ \\
\hline
\end{tabular}
~~
\begin{tabular}[t]{|c||c|c|}
\multicolumn{3}{c}{$\pi=(12)(34),~\varphi(\SC)=1$} \\
\hline
\hline
$\sigma$ & $\varphi(\SB)$ & $\wtilde P$ \\
\hline
${\rm id} $ & $0$        & $+1$ \\
$(12)     $ & $-\frac12$ & $-i$ \\
$(345)    $ & $-1$       & $+1$ \\
$(12)(34) $ & $-1$       & $-1$ \\
$(13)(24) $ & $-1$       & $+1$ \\
$(1234)   $ & $-\frac32$ & $-i$ \\
$(12)(345)$ & $-\frac32$ & $-i$ \\
$(13542)  $ & $-2$       & $+1$ \\
\hline
\end{tabular}
}
\end{center}
\caption{Parity eigenvalue of gauge boson on various D-branes
         of the model $(55555)$.}
\label{table:gAqui}
\end{table}

When the eigenvalue of $\tilde P$ is pure imaginary, the gauge
boson has $(-)^F=-1$ and is therefore GSO projected out.
This is in consistency with that the brane $\SB$ is mapped to
its anti-brane under an orientifold $\SC$ when
$\varphi(\SB)-\varphi(\SC)=\frac12$ (mod $\ZZ$), as the table shows.

Since nontrivial stabilizer group or summing over orbifold images
do not affect the computation of parity eigenvalue, the analysis
for B-type branes and orientifolds is essentially the same and
the result summarized in table \ref{table:gAqui} applies
also to B-types.

\subsubsection{Example 2: $(88444)$}\label{sec:STGG2P}

We take this model to discuss the gauge group on branes with special
${\bf L}$-labels.
We first present some type IIA examples:
\begin{itemize}

\item Consider a non-permuted brane $\SB^{A, \sigma=\rm id}_{\bf L,M}$
   invariant under the orientifold $\SC^{A,{\pi=\rm id},+}_{\bf \bar M}$.
   When $c\ssub\NS=-1$, the branes with generic ${\bf L}$ support
   $O(N)$ gauge group.
   If $L_1=L_2=3$ the branes split into two short-orbit branes
   exchanged to each other by orientifold because $\lambda$ of
   (\ref{lambdaA}) takes $-1$, and
   the short-orbit brane supports a unitary gauge group.

\item Consider a pair ($\SB^{A,(12)(345)}_{\bf L,M}$,
   $\SC^{A,(12)(34),-}_{\bf\bar M}$) with the latter normalized
   as $c\ssub\NS=-i$.
   Assume the pair
  ($\SB^{\sigma}_{\bf L,M},\SC^\pi_{{\bf\bar M}+{\bf 2}\nu}$) satisfy
   the condition $(3)_{\rm I}\times(1)_{\rm I}$ of PIB\ref{pib2},
   namely
\begin{eqnarray*}
 M_{12}  &=& \frac12(\bar M_1+\bar M_2)+2\nu~~{\rm mod}~~8, \\
 M_{345} &=& \frac12(\bar M_3+\bar M_4+\bar M_5)+3\nu~~{\rm mod}~~4.
\end{eqnarray*}
   The gauge group on branes with generic $\bf L$ is either $Sp$
   or $O$ depending on whether $\nu$ is even or odd.
   For special $\bf L$, namely $(L_{12}=3,~L_{345}=1)$ they break
   into short-orbit branes supporting a unitary gauge group.
\end{itemize}
We next consider some type IIB examples:
\begin{itemize}
\item Consider a non-permuted brane $\SB^{B,{\rm id}}_{\bf L,M}$
 invariant under the orientifold
 $\SC^{B,{\rm id},\epsilon_1\cdots\epsilon_5}_{\bf\bar M}$.
We normalize the orientifold by setting
$c\ssub\NS ~=~ -i^\alpha$, where $\alpha$ is the number of
$\epsilon_a$'s taking minus sign.
The $\bf L$-label of branes is called generic if $L_a=k_a/2$
for at most one $a$.
If a brane $\SB^{B,{\rm id}}_{\bf L,M}$ with generic $\bf L$ is invariant
under the orientifold
$\SC^{B,{\rm id},\epsilon_1\cdots\epsilon_5}_{\bf\bar M}$,
then there is a set of integer $\{\nu_a\}$
such that $\SB^{\rm id}_{\bf L,M}$ and $\SC^{\rm id}_{\bf\bar M+2\vec\nu}$
satisfy the condition PIB\ref{pib2}.
The NS parity eigenvalue is then given by
\begin{equation}
 (-)^{F_L}P
  ~=~ -i^{\alpha+\sharp(1)_{\rm II}}\prod_a\epsilon_a^{\nu_a}
  ~=~ -i^{\alpha+\sharp(1)_{\rm II}}
       \prod_a\epsilon_a^{L_a}\cdot\prod_a\epsilon_a^{\bar M_a/2}.
\end{equation}
Here we used that $L_a+M_a$ and $\frac{k_a+2}{2}$ are even for all $a$.
Note also that $\alpha+\sharp(1)_{\rm II}$ is always even if the brane
and orientifold preserve the same supersymmetry.
The branes with $p(\ge2)$ of $L_a$'s coinciding with $\frac{k_a}{2}$
are special.
The NS parity eigenvalue for such branes is determined by applying
the general formula (\ref{NSPB}),
\begin{equation}
 (-)^{F_L}P
  ~=~ -{\rm sgn}\left[{\rm Re}\left(i^\alpha(1+i)^p\right)\right]
       \prod_a\epsilon_a^{L_a}\cdot\prod_a\epsilon_a^{\bar M_a/2}.
\end{equation}
We thus recover the result of Tables 9,10 of \cite{Brunner-HHW}.
The gauge group is unitary when $p$ is even and $\alpha+\frac p2$
is an odd integer.

\end{itemize}

\subsection{Tadpole Cancellation}\label{sec:TC}

Here we discuss the RR tadpole cancellation condition and its
solutions.
The formula relating the charges of crosscaps and boundary states
in minimal models allows us to find a set of D-branes cancelling
the RR-charge of any given orientifold.
It is more difficult to find the set of D-branes preserving
a spacetime supersymmetry.
In principle we have to deal with a system of coupled linear
equations with integer coefficients, and the complexity of the
problem depends on the number of linear equations which equals
the dimension of the RR-charge lattice.

\subsubsection{Type IIA on $(55555)$}\label{sec:TCQ}

There are three physically inequivalent orientifolds,
$\SC^{\rm id}_{\bf 0}$, $\SC^{\rm (12)}_{\bf 0}$ and
$\SC^{\rm (12)(34)}_{\bf 0}$.
We only consider those with negative tension ($O^-$-planes).
These three orientifolds have supersymmetry phase $\varphi=0,1/2,1$
respectively.
The simplest tadpole-free configurations for these orientifolds
are obtained by wrapping four D-branes of the like charge, same
supersymmetry phase on top of the orientifolds.
Such configurations are described by the tadpole states,
\begin{equation}
 \ket{\SC^{\rm id}_{\bf 0}}+4\ket{\SBPLM{\rm id}{22222}{22222}},~~~
 \ket{\SC^{\rm (12)}_{\bf 0}}+4\ket{\SBPLM{(12)}{0222}{9222}},~~~
 \ket{\SC^{\rm (12)(34)}_{\bf 0}}+4\ket{\SBPLM{(12)(34)}{002}{992}}.
\end{equation}
These will be all interpreted as four D6-branes on top of orientifold
plane wrapping an $\RR\PP^3$ \cite{Brunner-H2}, and supporting $O(4)$
gauge theory with various matters.

\subsubsection{Type IIA on $(88444)$}\label{sec:TC2P}

We have found 30 physically inequivalent orientifolds
labelled by different choices of $(\pi,{\bf M})$ (\ref{15diff}) and
a sign $\epsilon$.
The choice
\[
 c\ssub\RARA=-1,~~~~
 c\ssub\NSNS=-1~(\epsilon=1)~~{\rm or}~~-i~(\epsilon=-1)
\]
ensures the negative semi-definiteness of the tension for
all choices of $(\pi,{\bf M})$ in the list.
For 12 of them labelled by $\pi={\rm id}$, one finds the
expressions the RR-charges in terms of those of D-branes\cite{Brunner-HHW},
\begin{equation}
\begin{array}{lcr}
\charge{\SC^{{\rm id},\pm}_{(00000)}}
+2\charge{\SBPLM{\rm id}{33111}{33111}}
\mp 2\charge{\SBPLM{\rm id}{33111}{33111}}&=&0, \\
\charge{\SC^{{\rm id},\pm}_{(00002)}}
+2\charge{\SBPLM{\rm id}{33111}{33113}}
\mp 2\charge{\SBPLM{\rm id}{33111}{33111}}&=&0, \\
\charge{\SC^{{\rm id},\pm}_{(02000)}}
+2\charge{\SBPLM{\rm id}{33111}{35111}}
\mp 2\charge{\SBPLM{\rm id}{33111}{33111}}&=&0, \\
\charge{\SC^{{\rm id},\pm}_{(02002)}}
+2\charge{\SBPLM{\rm id}{33111}{35113}}
\mp 2\charge{\SBPLM{\rm id}{33111}{33111}}&=&0, \\
\charge{\SC^{{\rm id},\pm}_{(22000)}}
+2\charge{\SBPLM{\rm id}{33111}{55111}}
\mp 2\charge{\SBPLM{\rm id}{33111}{33111}}&=&0, \\
\charge{\SC^{{\rm id},\pm}_{(22002)}}
+2\charge{\SBPLM{\rm id}{33111}{55113}}
\mp 2\charge{\SBPLM{\rm id}{33111}{33111}}&=&0.
\end{array}
\end{equation}
Note that each of the D-brane charges appearing above equalities expresses
the sum of the charges of two short-orbit branes labelled by ${\bf L,M}$
(recall that the non-permuted branes with $L_1=L_2=3$ are fixed under
 $\gamma^4$).
These relations immediately give RR tadpole free configurations,
which are however not supersymmetric except for those in the first line.
In \cite{Brunner-HHW}, some supersymmetric tadpole-free
configurations were found by rewriting these equations using
the relations between D-brane charges in minimal models,
\begin{equation}
 \charge{\SB_{L,M}} ~=~
 \charge{\SB_{0,M-L}}+
 \charge{\SB_{0,M-L+2}}+\cdots+
 \charge{\SB_{0,M+L}}.
\end{equation}

For some of the other 18 orientifolds, we found the following
equalities for the RR charges,
\begin{equation}
\begin{array}{rcl}
\charge{\SC^{(12),\pm}_{(00000)}}
   +2\charge{\SBPLM{(12)}{0111}{3333}}
\mp 2\charge{\SBPLM{(12)}{0111}{1333}}&=&0, \\
\charge{\SC^{(12),\pm}_{(00002)}}
   +2\charge{\SBPLM{(12)}{0111}{3335}}
\mp 2\charge{\SBPLM{(12)}{0111}{1333}}&=&0, \\
\charge{\SC^{(34),\pm}_{(00000)}}
   +2\charge{\SBPLM{(34)}{3301}{5533}}
\mp 2\charge{\SBPLM{(34)}{3301}{5513}}&=&0, \\
\charge{\SC^{(34),\pm}_{(02000)}}
  + 2\charge{\SBPLM{(34)}{3301}{5733}}
\mp 2\charge{\SBPLM{(34)}{3301}{5513}}&=&0, \\
\charge{\SC^{(34),\pm}_{(22000)}}
   +2\charge{\SBPLM{(34)}{3301}{7733}}
\mp 2\charge{\SBPLM{(34)}{3301}{5513}}&=&0, \\
\charge{\SC^{(12)(34),\pm}_{(00000)}}
   +2\charge{\SBPLM{(12)(34)}{001}{333}}
\mp 2\charge{\SBPLM{(12)(34)}{001}{113}}&=&0.
\end{array}
\end{equation}
Applying recombination to some of them, we found
the following supersymmetric tadpole-free configurations,
\begin{eqnarray}
&&\ket{\SC^{(12),-}_{(00000)}}
   +2\sum_\epsilon\ket{\SBPLM{(12),\epsilon}{1111}{2111}}, \nn\\
&&\ket{\SC^{(34),-}_{(00000)}}
   +2\sum_\epsilon\ket{\SBPLM{(34),\epsilon}{3311}{5523}}, \\
&&\ket{\SC^{(34),+}_{(22000)}}
   +2\sum_\epsilon\ket{\SBPLM{(34),\epsilon}{3301}{7733}}
   +2\sum_\epsilon\ket{\SBPLM{(34),\epsilon}{3321}{5553}}.\nn
\end{eqnarray}
Here $\epsilon$ specifies the characters
of the stabilizer group $\ZZ_2$ of short-orbit D-branes.

The remaining 6 orientifolds all involve the permutation
orientifold $\ket{\SC^{(12)}_{M,M+8}}$ of the first two minimal models.
The crosscap states are made of closed string states sitting
in $\gammaA^4$-twisted sector, and are in particular tensionless.

\subsubsection{Type IIB}\label{sec:TCB}

In type IIB Gepner models, the tadpole-free condition can be solved more
easily because the charge of D-branes span a lattice of relatively
low dimension.

Let us first focus on the charges arising from the untwisted sector
(in the mirror description).
In mirror Gepner model labelled by $(k_1\cdots k_r)$ and
$H\equiv {\rm l.c.m.}(k_a+2)$, the relevant RR ground states
are labelled by a mod-$H$ integer $\nu$ which is not multiple of
any of $(k_a+2)$.
They take the form
\begin{equation}
 \ket{\nu}\ssub\RARA ~=~ i^{-r}
 \bigotimes_{a=1}^r\ket{(l_a,l_a+1,1)\otimes(l_a,-l_a-1,-1)}\cdot(-)^{d_a},
\label{RRGUN}
\end{equation}
where $(l_a,d_a)$ is a unique pair of integers satisfying
$\nu=d_a(k_a+2)+l_a+1$.
Counting the allowed $\nu$'s one finds the dimension of RR charge
lattice spanned by the ground states in the untwisted sector, which is
4 for $(k_a+2)=(5,5,5,5,5)$ and 6 for $(k_a+2)=(8,8,4,4,4)$.
Since the dimension agrees with the known value of $2h_{1,1}+2$ for
both cases, there are no RR-charges from twisted sectors
for these two theories.

The boundary states $\ket{\SB^{\sigma,\rho}_{\bf L,M}}\ssub{\RARA+}$
are shown to have the following overlaps,
\begin{equation}
  {}\ssub\RARA\vev{\nu|\SB^{\sigma,\rho}_{\bf L,M}}\ssub{\RARA+}
 ~=~ \frac{1}{2^{[p/2]}\sqrt H}
     \frac{\prod_{c=1}^{[\sigma]}F_{L_c,M_c}(\omega^{\nu w_c})
           (k_c+2)^{\delta_c}}
          {\prod_{a=1}^r|1-\omega^{\nu w_a}|^{1/2}}.
\end{equation}
Here we denoted $\omega\equiv e^{\frac{2\pi i}{H}}$,
$w_a\equiv \frac{H}{k_a+2}$ and
\begin{eqnarray}
 F_{L,M}(x) &\equiv& x^{\frac12(M+L+1)}-x^{\frac12(M-L-1)}, \nn\\
 \delta_c   &\equiv& {\rm max}([\tfrac{|\sigma_c|-1}{2}],0), \nn\\
 p          &\equiv& (\mbox{number of odd-length cycles labelled by
                      $L=k/2$}).
\end{eqnarray}
The powers of $(k_c+2)$ and the factor $2^{[p/2]}$ arise from
the order of the stabilizer group and its untwisted subgroup.
The RR charge of B-branes are thus expressed conveniently
by the polynomial,
\begin{equation}
 [\SB^\sigma_{\bf L,M}] (x)~\equiv~
 2^{-[p/2]}\prod_{c=1}^{[\sigma]}F_{L_c,M_c}(x^{w_c})(k_c+2)^{\delta_c}.
\end{equation}
In particular, if the argument $x$ of the polynomials is assumed
to satisfy
\begin{equation}
 1-x^H ~=~ 1+x^{w_a}+x^{2w_a}+\cdots x^{w_a(k_a+1)}~=~ 0,
\label{ch-x}
\end{equation}
one can rewrite every polynomial in terms of a finite
number of monomials.
The number of monomials required is the same as the dimension
of the (untwisted) RR-charge lattice.
So $[\SB^\sigma_{\bf L,M}](x)$ are naturally identified
with vectors on the RR-charge lattice \cite{Brunner-DLR}.
As an application of this formula, the intersection number
of D-branes is computed by the index,
\begin{eqnarray}
 I(\SB^{\sigma',\rho'}_{\bf L',M'},\SB^{\sigma,\rho}_{\bf L,M})
 &\equiv&
  {}\ssub{\RARA+}
    \bra{\SB^{\sigma',\rho'}_{\bf L',M'}}e^{-i\pi J_0}q^H
    \ket{\SB^{\sigma ,\rho }_{\bf L ,M }}\ssub{\RARA+}
 \nn\\ &=&
   \sum_\nu
  {}\ssub{\RARA+}
    \vev{\SB^{\sigma',\rho'}_{\bf L',M'}|e^{-i\pi J_0}|\nu}\ssub{\RARA+}\cdot
  {}\ssub{\RARA+}
    \vev{\nu|\SB^{\sigma ,\rho }_{\bf L ,M }}\ssub{\RARA+}
 \nn\\ &=&
     \frac1H\sum_\nu
     \frac{[\SB^{\sigma'}_{\bf L',M'}](\omega^{-\nu})
           [\SB^{\sigma }_{\bf L ,M }](\omega^{\nu})
         }{\prod_{a=1}^r(1-\omega^{\nu w_a})}.
\end{eqnarray}

The polynomials $[\SB^\sigma_{\bf L,M}](x)$ satisfy various relations
under the assumption (\ref{ch-x}).
For example, for the model $(55555)$ one finds relations
among RR-charges of various permutation branes by a
repeated use of the formula
$(x^{\frac12}-x^{-\frac12})^{-1}
 =\frac15(x^{-\frac32}+2x^{-\frac12}-2x^{\frac12}-x^{\frac32})$.
\begin{eqnarray}
  [\SB^{(12)}_{{\bf 0},M}] &=& \frac15\left(
    [\SB^{\rm id}_{{\bf 0},M-3}]
  +2[\SB^{\rm id}_{{\bf 0},M-1}]
  -2[\SB^{\rm id}_{{\bf 0},M+1}]
   -[\SB^{\rm id}_{{\bf 0},M+3}] \right),
 \nn\\ ~~
  [\SB^{(12)(34)}_{{\bf 0},M}]  ~=~
  \frac15[\SB^{(123)}_{{\bf 0},M}] &=& \frac15\left(
    [\SB^{\rm id}_{{\bf 0},M-4}]
  -2[\SB^{\rm id}_{{\bf 0},M}]
   +[\SB^{\rm id}_{{\bf 0},M+4}]  \right),
\end{eqnarray}
where we used the label $M\equiv\sum_cM_c$ (mod 10) instead of $\bf M$.

It is straightforward to express the RR charge of orientifolds
in terms of similar polynomials, using the relations
(\ref{RRMM1}) and (\ref{RRMM2}).
For the model $(55555)$ one has simple relations
\begin{eqnarray}
  [\SC^{\rm id   }_{(00000)  }] &=&
 -4[\SB^{\rm id  }_{(22222),0}],\nn\\ ~
  [\SC^{(12)     }_{(00000)  }] &=&
 -4[\SB^{(12)    }_{( 0222),5}],\\ ~
  [\SC^{(12)(34) }_{(00000)  }] &=&
 -4[\SB^{(12)(34)}_{( 0 02),0}]. \nn
\end{eqnarray}
This agrees with the result of \cite{Becker-BVW} using the
(twisted) Landau-Ginzburg description \cite{Hori-W}.
For the model $(88444)$, there are orientifolds
labelled by $(\pi,{\bf M})$ as well as $\epsilon$'s and $r$'s as
explained in Example 2 of Section \ref{sec:POGMB}.
Restricting to those with $r=r'=0$, the RR-charges are given by
the following polynomials:
\begin{eqnarray}
 [\SC^{\rm id,\epsilon_1\epsilon_2\epsilon_3\epsilon_4\epsilon_5}_{(00000)}](x)
 &=& -[\SB^{\rm id}_{(33111),-4}](x)\cdot
  (1+\epsilon_1\epsilon_2x)(1+\epsilon_3x)(1+\epsilon_4x)(1+\epsilon_5x),
 \nn \\ ~
 [\SC^{\rm id,\epsilon_1\epsilon_2\epsilon_3\epsilon_4\epsilon_5}_{(20000)}](x)
 &=& -[\SB^{\rm id}_{(33111),-2}](x)\cdot
  (1+\epsilon_1\epsilon_2)(1+\epsilon_3x)(1+\epsilon_4x)(1+\epsilon_5x),
 \nn \\ ~
 [\SC^{(12)  ,\epsilon_1\epsilon_3\epsilon_4\epsilon_5}_{(00000)}](x)
 &=& -[\SB^{(12)}_{(0111),5}](x)\cdot
      (1+\epsilon_1\epsilon_3x)(1+\epsilon_1\epsilon_4x)
      (1+\epsilon_1\epsilon_5x),
 \nn \\ ~
 [\SC^{(12)  ,\epsilon_1\epsilon_3\epsilon_4\epsilon_5}_{(20000)}](x)
 &=& 0,
 \nn \\ ~
 [\SC^{(34)  ,\epsilon_1\epsilon_2\epsilon_5}_{(00000)}](x)
 &=& -2[\SB^{(34)}_{(3301),6}](x)\cdot(1+\epsilon_1\epsilon_2x)(1+\epsilon_5x),
 \nn \\ ~
 [\SC^{(34)  ,\epsilon_1\epsilon_2\epsilon_5}_{(20000)}](x)
 &=& -2[\SB^{(34)}_{(3301),8}](x)\cdot(1+\epsilon_1\epsilon_2)(1+\epsilon_5x),
 \nn \\ ~
 [\SC^{(12)(34),\epsilon_1\epsilon_5}_{(00000)}](x)
 &=& -2[\SB^{(12)(34)}_{(001),-1}](x)\cdot(1+\epsilon_1\epsilon_5 x)
 \nn \\ ~
 [\SC^{(12)(34),\epsilon_1\epsilon_5}_{(20000)}](x)
 &=& 0.
\end{eqnarray}

\paragraph{RR charges from twisted sectors.}

Finally we briefly discuss the case where the RR charge lattice is
not entirely spanned by the states in the untwisted sector.
We take as an example the model $(44666),~H=12$.
The RR charge lattice is known to be 14 dimensional, of which
8 arise from the states $\ket{\nu}\ssub\RARA$ in the untwisted sector
defined at (\ref{RRGUN}).
The values $\nu=0,4,6,8$ (mod 12) are excluded, but for $\nu=4,8$
there are RR vacua of the form
\begin{eqnarray}
\ket{\mu,\tilde\nu}\ssub\RARA &=&
 \ket{(\mu-1, \mu, 1)\otimes(\mu-1, \mu, 1)}\otimes
 \ket{(\mu-1,-\mu,-1)\otimes(\mu-1,-\mu,-1)}
 \nn\\ &&\otimes
 \prod_{a=3}^5\ket{(\tilde\nu-1,\tilde\nu,1)\otimes
                   (\tilde\nu-1,-\tilde\nu,-1)},
 \nn\\ && \hskip30mm
 \tilde\nu\equiv\nu~\mbox{mod}~6 ~=~ 4~{\rm or}~2,~~~
 \mu~=~ 1,2,3.
\end{eqnarray}
These 6 RR states from twisted sectors complete the full set of
RR charges.
They are sitting in the $(\gamma_1^\mu\gamma_2^{-\mu})$-twisted
sector of the mirror Gepner model.

The B-type permutation orientifolds of the model $(44666)$ have
twisted RR-charges if $\pi$ permutes 1 and 2.
The permutation B-branes have twisted RR-charges if their
untwisted stabilizer group contains elements $\gamma_1^\mu\gamma_2^{-\mu}$.
The RR-charges of these branes and orientifolds are again
conveniently expressed by polynomials of
$(y\equiv e^{\frac{2\pi i\mu}{4}},\;z\equiv e^{\frac{2\pi i\tilde\nu}{6}})$
which therefore satisfy
\[
 1+y+y^2+y^3 ~=~ 1+z+z^2 ~=~ 0.
\]
The branes carrying the twisted RR-charges are
\begin{eqnarray}
 [\SB^{(12),\rho}_{L,M}\otimes\SB^{\sigma,\rho'}_{\bf L',M'}]
 &=&
 [\SB^{(12)}_{L,2\rho}](y)[\SB^{\sigma}_{\bf L',M'}](z),
 \nn\\ ~
 [\SB^{(1)(2),\pm}_{{\bf L}=(11),{\bf M}}\otimes\SB^{\sigma,\rho'}_{\bf L',M'}]
 &=&
 (1-y+y^2-y^3)[\SB^{\sigma}_{\bf L',M'}](z).
\end{eqnarray}
In the second line, none of $L'_c$ equals 2 because otherwise the
untwisted stabilizer of the brane would not contain $\eta_1\eta_2$.
The orientifolds carrying the twisted RR-charges are
\[
\begin{array}{lcrcl}
 \SC^{B,(12),\rho}_{\bf M} &:&
 \rho_{r,\epsilon_1,\epsilon_3,\epsilon_4,\epsilon_5}(\vec\nu) &=&
 \omega_4^{-r(\nu_1-\nu_2)}
 \epsilon_1^{\nu_1}\epsilon_3^{\nu_3}\epsilon_4^{\nu_4}\epsilon_5^{\nu_5}
 \\
 \SC^{B,(12)(34),\rho}_{\bf M} &:&
 \rho_{r,r',\epsilon_1,\epsilon_5}(\vec\nu) &=&
 \omega_4^{-r(\nu_1-\nu_2)}\omega_6^{-r'(\nu_3-\nu_4)}
 \epsilon_1^{\nu_1}\epsilon_5^{\nu_5}.
\end{array}
\]
We restrict to those with ${\bf M}=(00000)$ or $(20000)$ and
$\epsilon_1=+1$ since all the others are related to them by
symmetries.
Their twisted RR-charges are expressed by the polynomials
\begin{eqnarray}
 [\SC^{B,(12),\rho}_{(M0000)}] &=&
 -2\left(      [\SB^{(12)}_{0,-2r}](y)
   +(-)^{M/2}[\SB^{(12)}_{0,-2r+4}](y) \right)
 \nn\\ && \times \frac14 [\SB^{\rm id}_{(222),6}](z)
   (1+\epsilon_3z^2)(1+\epsilon_4z^2)(1+\epsilon_5z^2), \nn\\ ~
 [\SC^{B,(12)(34),\rho}_{(M0000)}] &=&
 -2\left(      [\SB^{(12)}_{0,-2r}](y)
   +(-)^{M/2}[\SB^{(12)}_{0,-2r+4}](y) \right)
 \nn\\ && \times \frac12 [\SB^{(34)}_{(02),1}](z)(1+\epsilon_5z^2).
\end{eqnarray}

\section{Concluding Remarks}\label{sec:CONC}

In this paper we discussed the construction of permutation orientifolds
in general RCFTs and then studied those in Gepner models.
Although our analysis was limited to the Gepner point,
it will serve as a starting point to explore a new
class of four-dimensional string vacua.
It will be interesting to see how various properties of permutation
orientifolds continue in moduli space to large volume.
In doing this, it will be useful to switch from the description
in terms of coset CFTs to those in terms of Landau-Ginzburg orbifolds
or linear sigma models.
A number of works along this path have appeared
recently\cite{Hori-W, Becker-BVW, Brunner-M}.

\section*{Acknowledgment}

This work is a continuation of the author's previous work
in collaboration with I. Brunner, K. Hori and J. Walcher;
it is a preasure to thank them for discussions and correspondences.
The author especially thanks K. Hori for collaboration at an early stage,
and I. Brunner for informing about the work on closely related problems.

\end{document}